%% file: master_manuscript_humans.tex
\documentclass[aps, prb,
 amsmath,amssymb,
 reprint,
 longbibliography
]{revtex4-2}

\usepackage{dcolumn}
\usepackage{bm}
\usepackage[dvipsnames]{xcolor}
\usepackage{soul}
\usepackage{amssymb,mathrsfs}
\usepackage{mathtools}
\usepackage[normalem]{ulem}
\usepackage{physics}
\usepackage{comment}
\definecolor{refkey}{gray}{.75}
\definecolor{labelkey}{gray}{.75}
\usepackage{graphicx}
\usepackage{tikz}
\usetikzlibrary{decorations.pathmorphing,arrows.meta}
\usetikzlibrary{patterns}
\usepackage{pgfplots}
\pgfplotsset{compat=1.16}

\usepackage[utf8]{inputenc}
\usepackage[T1]{fontenc}
\usepackage{etoolbox}
\usepackage{placeins}

\makeatletter
\def\@email#1#2{%
 \endgroup
 \patchcmd{\titleblock@produce}
  {\frontmatter@RRAPformat}
  {\frontmatter@RRAPformat{\produce@RRAP{*#1\href{mailto:#2}{#2}}}\frontmatter@RRAPformat}
  {}{}
}%
\makeatother
\newcommand{\pref}[1]{(\ref{#1})}
\newcommand{\epref}[1]{Eq.~(\ref{#1})}
\newcommand{\figref}[1]{Fig.~\ref{#1}}

\newcommand{\appref}[1]{Appendix~\ref{#1}}
\newcommand{\secref}[1]{Sec.~\ref{#1}}
\newcommand{\iu}{\mathrm i}
\newcommand{\ie}{\textit{i.e.}}

\newcommand{\eg}{\textit{e.g.}}

\newcommand{\me}{m_{\mathrm e}}

\newcommand{\Up}{U_{\mathrm p}}
\newcommand{\alphav}{{\bm{\alpha}_0^{(+)}}}
\newcommand{\Qv}{\vb Q}
\newcommand{\WR}{\Omega_{\mathrm R}}
\newcommand{\EE}{\mathcal E}
\newcommand{\EEv}{\bm{\mathcal E}}
\newcommand{\wL}{\omega_{\mathrm L}} 
\newcommand{\Ap}{\vb{A}^{(+)}}
\newcommand{\SF}{\mathscr S_{\mathrm F}}
\newcommand{\SFc}{\mathscr S_{\mathrm{F,c}}}
\newcommand{\Pbar}{\bar{\mathcal P}}
\newcommand{\chiaux}{\chi_{\mathrm{aux}}}
\newcommand{\ImH}{\operatorname{Im}_{H}}
\begin{document}

\title{Floquet density response in laser-assisted fast-electron scattering from solids}

\author{Richard Korytár$^*$}
\email{richard.korytar@matfyz.cuni.cz}
\affiliation{Department of Condensed Matter Physics, Faculty of Mathematics and Physics, Charles University, Ke Karlovu 5, 121 16, Praha 2, Czech Republic}
\author{Sascha Schäfer}
\affiliation{Institute of Experimental and Applied Physics, University of Regensburg, Universitätstraße 31, 93053 Regensburg, Germany}
\author{Ferdinand Evers}
\affiliation{Institute of Theoretical Physics, University of Regensburg and Halle-Berlin-Regensburg Cluster of Excellence CCE, Universitätstraße 31, 93053 Regensburg, Germany}

\begin{abstract}
We extend the Bethe--Floquet formalism of Joachain and
coworkers, originally developed for laser-assisted electron--atom
collisions, to inelastic scattering of fast electrons from a
many-body condensed-matter target in a time-periodic light field. At
first order in the projectile--target interaction, the general
Floquet--Fourier cross section separates into exact laser-dressed
projectile kernels and a matrix-valued Floquet density response of the
target. For spatially structured or translationally non-invariant
systems, the latter is the bi-momentum Floquet structure factor
$\SF^{nn'}(\vb q_1,\vb q_2;\omega)$; its momentum
diagonal defines the Floquet generalization
$S_{\mathrm F}^{nn'}(\Qv,\omega)$ of the dynamic structure factor,
constrained by Hermiticity, positivity, and sum rules. As a controlled
realization of the general projectile kernel, an eikonal--Volkov
approximation for slowly varying inhomogeneous fields yields a
finite-momentum-resolution convolution of the target response. In the
homogeneous-field dipole limit, the projectile kernel reduces to
Bessel-function sidebands and the cross section takes the familiar
Bethe--Floquet form. The Floquet density correlator separates exactly
into the outer product of the pump-induced coherent mean density and
connected fluctuations. In the straight-trajectory nonrecoil limit,
the coherent, target-elastic sector defines the weak-coupling PINEM
amplitude \(g_{\rm ind}\); repeated coherent insertions generate the
PINEM Bessel ladder, with an explicit no-double-counting prescription
for combining this channel with connected losses. As worked examples,
we evaluate the connected, target-changing cross section analytically
for a metal in the Drude and diffusive
limits, recovering Bessel-weighted plasmon-loss and diffusive combs,
and for a driven two-level model that exhibits off-diagonal Floquet
coherences through Bessel-channel interference.
\end{abstract}

\maketitle

\section{Introduction}
\label{sec:intro}
Inelastic scattering of fast electrons from matter is one of the most
informative spectroscopies of condensed-phase excitations, from plasmons and
excitons to phonons and core-level transitions~\cite{Egerton2011}. The
cornerstone of its quantitative interpretation is the theory of
Bethe~\cite{Bethe1930,Inokuti1971}: in the first Born approximation and at
high projectile energy the inelastic cross section factorizes into a Coulomb
form factor and the \emph{dynamic structure factor} $S(\vb q,\omega)$ of the
target -- an intrinsic many-body property of the sample.

A growing family of experiments now combines such fast-electron probes with
intense optical fields. Photon-induced near-field electron microscopy
(PINEM)~\cite{Barwick2009,deAbajo2010} and its descendants have demonstrated
that a free-electron wave function can be coherently dressed into a frequency
comb of sidebands by an optical field, and that this dressing carries
information about the optical near-field of the sample. On the theoretical
side, the parent atomic problem -- laser-assisted electron--atom collisions
-- was developed in detail in the 1990s by Joachain, Dörr, Potvliege,
Vučić, and coworkers, who formulated a Born--Floquet $S$-matrix theory of
$(e,e')$ in a strong monochromatic
field~\cite{Dorr1994,Vucic1995,Joachain2012}.

\begin{figure*}
\centering
\includegraphics[width=\textwidth]{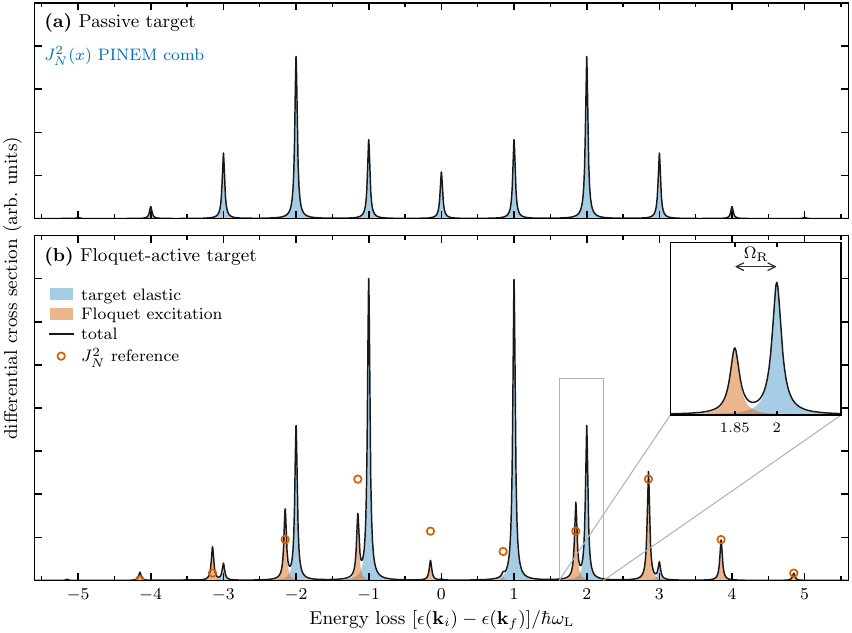}
\caption{Laser-assisted EELS spectrum illustrating projectile dressing
and target Floquet dynamics.
(a)~For a passive target, projectile dressing produces the usual
target-elastic PINEM comb with squared-Bessel sideband weights.
(b)~For the driven two-level target, the blue comb is the first-Born
target-elastic coherent term, which now also contains the induced-dipole
harmonics of the driven target; it is not the all-orders localized-field
PINEM resummation. An additional target-changing comb (orange) appears at
$N\hbar\wL+\hbar(\wL-\WR)$.
Its sideband weights deviate from a squared-Bessel distribution because
different Floquet channels interfere coherently; open circles show the
corresponding Bessel reference with the same integrated weight.
The inset highlights the displacement from the
neighboring target-elastic sideband, given by the Rabi frequency $\WR$.
Parameters are Bessel argument $x=|\Qv\cdot\boldsymbol{\alpha}_0|=3$,
mixing angle $\theta=\pi/3$, $\WR=0.15\,\wL$. Each line is broadened
into a Lorentzian of half width at half maximum $0.02\,\hbar\wL$.
The relative strengths are illustrative:
equal transition and static form factors, $q_{\rm dip}|F|=|\varrho_{gg}|$,
and a real, positive ratio of the induced-dipole to the static
target-elastic amplitude (\secref{sec:two-level}).}
\label{fig:eels-spectrum}
\end{figure*}

Figure~\ref{fig:eels-spectrum} previews the central spectroscopic
signature of the theory: target Floquet dynamics add a displaced,
non-Bessel sideband comb to the familiar PINEM spectrum, as derived
for the driven two-level model in \secref{sec:two-level}.

The conceptual step taken in the present work is to upgrade the target from
a single field-dressed atom to a generic many-body solid with its own
Floquet structure. The formalism is built around the possibility that
the light field dresses both the projectile (via the Volkov mechanism)
and the interacting target (via the Floquet eigenstates of
$\hat H_1(t)$), and that the cross section can in principle probe the
\emph{correlations} between the two dressings (\figref{fig:geometry}).
The scope of the framework -- including the closed-target idealization
and the projectile-only limit for the connected target response used in the metallic worked example
of \secref{sec:plasmon} -- is set in \secref{sec:scope}.
We work consistently to first
order in the electron--target interaction $\hat V$ but to all orders in
the laser--electron coupling. At first Born order, and without invoking
an eikonal, Volkov, homogeneous-field, or slowly-varying-envelope
approximation, the cross section separates into exact laser-dressed
projectile kernels and the bi-momentum Floquet density-correlation
matrix of the target, as expressed in \pref{eq:dsigma-bimomentum};
this separation is the principal theoretical result of the paper. The
projectile dynamics enter only through the kernel that contracts with
this target response. Translational invariance of the target reduces
the bi-momentum response to $S_{\mathrm F}^{nn'}(\Qv,\omega)$, while the
eikonal--Volkov reduction \pref{eq:dsigma-inhom} and the homogeneous
Volkov construction are subsequent projectile-side specializations that
evaluate the kernel; they are not the origin of the Floquet matrix or
of its bi-momentum structure.

\paragraph*{Relation to existing PINEM theories.}
Standard PINEM theory describes projectile-side Bessel dressing by a
prescribed optical near
field~\cite{Park2010,ParkZewail2014,deAbajo2010,deAbajoRMP2010,Feist2015,Vanacore2018,Gaida2024}.
The present work incorporates the projectile dressing in an external
field and describes the target through the \emph{full} unsymmetrized
density correlator: $S(\vb q,\omega)$ in equilibrium,
$S_{\mathrm F}^{nn'}(\vb q,\omega)$ for a Floquet-dressed target, and
$\SF^{nn'}(\vb q_1,\vb q_2;\omega)$ when target-side
translational symmetry is broken.  Its connected part describes
target-changing density fluctuations, whereas its coherent
mean-density outer product describes target-elastic PINEM.  Optical
sidebands can therefore carry either genuine target excitations or
coherent electron gain/loss without a change of target state.  As
shown in \secref{sec:pinem-bridge}, the first-Born coherent amplitude
is the weak-coupling PINEM vertex, while exponentiating its repeated
action produces the familiar PINEM Bessel ladder.

\paragraph*{Division of roles between the worked examples.}
The metallic examples of \secref{sec:plasmon} benchmark the
projectile-only limit for the connected target response, in which the optical field dresses the
projectile while the target is represented by its equilibrium density
response. They therefore demonstrate how established condensed-matter
loss functions enter the general scattering framework, but they do not
constitute calculations of a nontrivial Floquet response matrix of a
driven metal. The driven two-level example of \secref{sec:two-level}
is used separately to isolate the off-diagonal Floquet-coherence
mechanism in closed form.

\subsection{Scope and assumptions}
\label{sec:scope}
Five assumptions and approximations underlie the present analysis.

(i)~First Born in $\hat V$ -- valid for sufficiently fast projectiles,
thin samples, and screened momentum transfers, i.e.\ in the regime
where the single-scattering amplitude is perturbative. At very small
$Q$, or in thick samples, the long-range Coulomb interaction is no
longer perturbative and dielectric screening or multiple scattering
must be incorporated explicitly; that regime lies outside the present
treatment.

(ii)~Nonrelativistic projectile -- the kinetic Hamiltonian \pref{eq:H0}
omits the relativistic terms.
In
the aloof, small-recoil, spin-insensitive PINEM regime the leading
relativistic correction is the projectile-side kinematic substitution
of \pref{eq:rel-mass-substitution}, together with the relativistic
kinematic conversion $p=\hbar k=\gamma_{\mathrm p}m_{\mathrm e}v$ and
$\epsilon_{\rm k}=(\gamma_{\mathrm p}-1)m_{\mathrm e}c^{2}$ between beam
energy, projectile momentum, and velocity, leaving the Bessel sideband
structure intact~\cite{fn:rel-catalogue,ParkZewail2012}.
For quantitative comparison with $100$--$300\,\mathrm{keV}$ TEM
experiments the kinematics should be evaluated relativistically and, in
low-loss dielectric, Cherenkov, large-angle, or core-loss regimes, the
Coulomb kernel should be replaced by a retarded electromagnetic
interaction coupling to charge and current response; the formalism
developed here is the nonrelativistic scalar-response core of a more
general relativistic Bethe--Floquet theory. The scalar Coulomb
collision kernel is expected to be reliable in the quasistatic,
longitudinal, non-radiative regime of small-angle scattering at finite
momentum transfer, $Q\gg\omega/c$. This is a characterization of the
controlled regime rather than a sharply universal inequality: near the
light line, in Cherenkov-active media, and for interfaces, guided
modes, or aloof-beam geometries the corresponding retarded
electromagnetic response is required.

(iii)~Dipole approximation for the light field -- strictly requires
$\lambda_{\text{light}}\gg L_{\text{target}}$, the standard condition
for optical or near-IR drives of mesoscopic samples. This is relaxed
in Secs.~\ref{sec:projdyn-eikonal}--\ref{sec:dsigma-inhom}
to a slowly varying spatial envelope $\vb A(\vb r)$ via the eikonal
(semi-classical) approximation $kL\gg 1$.

(iv)~Linear-Volkov truncation -- the second-harmonic dressing of the
projectile, generated by the $\hat A^2$ part of \pref{eq:H0}, is
dropped. This is a controlled approximation in the regime of ponderomotive energy
$\Up\ll\hbar\wL$ where the light frequency is denoted by $\wL$,
relevant to fast-electron probes.

(v)~Closed (unitary) target -- three levels should be distinguished.
The general first-Born bilinear factorization represented by
\pref{eq:dsigma-bimomentum} does \emph{not} require a pure target
state. The bi-momentum structure factor defined in
\pref{eq:SF-bimomentum}, however, is the closed-system pure-state
Floquet--Lehmann realization used in the present work, as are
\pref{eq:SF}, \pref{eq:SF-pure} and
\pref{eq:SF-pure-app}--\pref{eq:chiF}; these assume unitary target
evolution under $\hat H_1(t)$. For a periodically driven open steady
state, \pref{eq:SF-bimomentum} must be replaced by the corresponding
unsymmetrized Floquet density correlator evaluated with the periodic
stationary density matrix. Absorptive driven solids therefore require
the open-system (Floquet--Keldysh / kinetic) treatment discussed in
\secref{sec:plasmon-scope} and
\secref{sec:outlook}; the metallic worked examples of
\secref{sec:plasmon} avoid this by using the projectile-only limit
for the connected target response,
$S_{\mathrm{F,c}}^{nn'}\to\delta_{n,0}\delta_{n',0}\,S_{\mathrm c}$, with the Drude
$\gamma$ and diffuson $1/\tau$ supplied as phenomenological parameters
of the equilibrium dissipative response.

None of the algebra of Secs.~\ref{sec:1Born}--\ref{sec:floquet}
requires (iii) or (iv); they enter only in the Volkov construction of
Secs.~\ref{sec:projdyn}--\ref{sec:applications}.

\paragraph*{Organization of the paper.}
Secs.~\ref{sec:hamilton}--\ref{sec:floquet} establish the
Hamiltonian, first-Born matrix element, exact Floquet--Fourier
factorization, bi-momentum cross section, and coherent/connected
decomposition, including its bridge to PINEM.
\secref{sec:projdyn} constructs the Volkov and eikonal--Volkov
projectile states, and \secref{sec:dsigma-inhom} evaluates the
projectile kernel and cross section for a slowly varying structured
field.
\secref{sec:applications} introduces $S_{\mathrm F}^{nn'}(\Qv,\omega)$
in the homogeneous limit. \secref{sec:floquet-response} develops the
response-function interpretation and the structural identities.
Secs.~\ref{sec:plasmon} and~\ref{sec:two-level} treat the metallic
Drude/diffusive and driven two-level worked examples.
\secref{sec:outlook} closes with the
open-system extension. Appendices collect the Floquet density response
and its coarse graining, derive the $f$-sum rule, collect
the classical limits and literature comparison, treat elastic scattering
and the PINEM angular transform in an inhomogeneous field, validate the
straight-line eikonal trajectory and the slowly varying
envelope (SVE) approximation.

\input{figures/fig_geometry}

\section{Hamiltonian}
\label{sec:hamilton}
The scattering system is described by the Hamiltonian with three terms,
\begin{equation}
\label{eq:H}
\hat H(t) \coloneqq \hat H_0(t) + \hat H_1(t) + \hat V,
\end{equation}
governing the (fast) electron projectile, the target many-body system,
and the collision potential. The light field enters $\hat H(t)$ as a classical field,
which we describe first.

\subsection{Description of the light field}
\label{sec:hamiltonian-light-field}
Throughout this work, \(\vb A(\mathbf r,t)\) denotes a prescribed external laser vector potential. 
It does not include electromagnetic fields induced by the target charge or current response; such induced fields are not
fed back into either the projectile Hamiltonian or the target-driving Hamiltonian.
We adopt the Coulomb gauge,
$\nabla\!\cdot\!\vb A(\vb r,t)=0$, and set the scalar potential to zero.
We assume periodicity
$\vb A(\vb r,t)=\vb A(\vb r,t+2\pi/\wL)$;
for a monochromatic field we write the time dependence in terms of a
complex positive-frequency amplitude,
\begin{equation}
\vb A(\vb r,t)
\coloneqq
\mathrm{Re}
\left[
\Ap(\vb r)
e^{\iu(\omega_{\rm L}t+\varphi)}
\right].
\label{eq:Acos}
\end{equation}
Equivalently, with
\(\Ap=\vb A_c-\iu\vb A_s\),
\[
\vb A(\vb r,t)
=
\vb A_c(\vb r)\cos(\wL t+\varphi)
+
\vb A_s(\vb r)\sin(\wL t+\varphi),
\]
($\vb A_{s,c}$ are real).
Linear polarization is recovered when the two quadratures are
parallel, and in particular for \(\vb A_s=0\).
Throughout the paper we reserve $\wL$ for the laser drive
frequency, while $\omega$ without subscript denotes the generic spectral
variable appearing as the frequency argument of response and correlation
functions. The same prescribed field enters the projectile and target
Hamiltonians introduced below.

\subsection{Projectile in the light field}
\label{sec:hamiltonian-projectile}
Let $\hat{\vb{r}}_0, \hat{\vb{p}}_0$ denote the projectile's coordinate
and momentum. The projectile Hamiltonian in the light field reads
\begin{equation}
\hat H_0(t) \coloneqq \frac 1{2\me} \qty[\hat{\vb{p}}_0 - \frac{q_{\mathrm p}}{c}\, \vb A(\hat{\vb r}_0,t)]^{2}.
\label{eq:H0}
\end{equation}
The same standard minimal coupling, $\hat{\vb p}_i-Q_i\vb A/c$, is used
for all target charges, with the signed charges $q_{\mathrm p}$ and $Q_i$
defined below. In the gauge adopted here the transverse electric field
of the prescribed potential is therefore $\EEv=-c^{-1}\partial_t\vb A$.
Relative phases between projectile and target dressing
(\secref{sec:two-level}) require that both couplings use this common
convention.

\paragraph*{Leading projectile-side kinematic relativistic correction.}
At TEM/PINEM operating energies the projectile Lorentz factor
$\gamma_{\mathrm p}\equiv 1+\epsilon_{\rm k}/(m_{\mathrm e}c^{2})$ is
appreciably larger than unity --
$\gamma_{\mathrm p}\approx 1.20$, $1.39$, $1.59$ at
$\epsilon_{\rm k}=100,\ 200,\ 300\,\mathrm{keV}$, respectively. (We use
$\gamma_{\mathrm p}$ throughout to distinguish the projectile Lorentz
factor from the Drude collision rate $\gamma$ of \secref{sec:plasmon}.)
The leading kinematic
relativistic correction is confined to the projectile-side eikonal/PINEM
coupling and is effected by the substitution
\begin{equation}
m_{\mathrm e}\;\longrightarrow\; m^\ast_\perp=\gamma_\mathrm p m_\mathrm e,\
m^\ast_\parallel=\gamma_\mathrm p^3m_\mathrm e
\label{eq:rel-mass-substitution}
\end{equation}
in the projectile transverse $\perp$ and longitudinal $\parallel$
equation of motion, respectively.
Consequently, for purely transverse optical coupling one obtains the 
familiar rescalings
$\boldsymbol{\alpha}_0\to\boldsymbol{\alpha}_0/\gamma_\mathrm p$
and $U_\mathrm p\to U_\mathrm p/\gamma_\mathrm p$.
For general polarization these scalar rescalings are replaced by anisotropic expressions 
\pref{eq:alpha-NR-rel};
the Bessel--Floquet
sideband structure $J_{n}(\Qv\!\cdot\!\boldsymbol\alpha_{0})$ is
preserved, with only its argument corrected.
Moreover, the relativistic kinematic relations
$p\equiv\hbar k=\gamma_{\mathrm p}m_{\mathrm e}v$ and
$\epsilon_{\rm k}=(\gamma_{\mathrm p}-1)m_{\mathrm e}c^{2}$ translate
between acceleration voltage, projectile momentum, velocity, and
scattering angle~\footnote{The cross section
prefactor $(m_{\mathrm e}/2\pi\hbar^{2})^{2}$ inherited from the
nonrelativistic Schrödinger normalization is \emph{not} modified
within the present scope: a consistent rescaling would require a
relativistic flux and final-state density-of-states derivation, which
is beyond the leading correction adopted here.}.

The substitution \pref{eq:rel-mass-substitution} applies \emph{exclusively}
to the projectile sector and does \emph{not} affect target-side
electron masses: in particular, the $m_{\mathrm e}$ of the
target-electron $f$-sum rule \pref{eq:fsum}, of the Drude plasma
frequency $\omega_{p}^{2}=4\pi n e^{2}/m_{\mathrm e}$, of the
diffuson density of states $\nu_F$, and of the target kinetic operator
$\hat{\vb p}_i^{2}/(2m_{\mathrm e})$ remain the physical electron rest
mass throughout. Equation~\pref{eq:rel-mass-substitution} implements ingredient~(1) of the
relativistic catalog of footnote~\cite{fn:rel-catalogue}. Relativistic
kinematics are used only in the projectile coupling and in the
validity estimates of \appref{app:trajectory-bending}; the absolute
scattering normalization, spin/current coupling, retardation, and
relativistic final-state density of states -- ingredients~(2) and~(3)
of the catalog -- remain outside the present theory, consistent with
the footnote's own statement that quantitative EELS at
$100$--$300\,\mathrm{keV}$ generally requires ingredient~(3) and, in
some settings, elements of~(2).

\subsection{Hamiltonian of the target}
\label{sec:hamiltonian-target}
The target many-body Hamiltonian
\begin{equation}
\hat H_1(t) \equiv
\hat H_1\qty(\qty{\hat{\vb{r}}_i},\qty{\hat{\vb{p}}_i};\; \vb A(\hat{\vb{r}}_i,t))
\label{eq:H1}
\end{equation}
involves the coordinates and momenta of particles with indices $i=1\ldots N$.
We assume that the target's constituent particles
see the same (external) field as the projectile, namely, \epref{eq:Acos}.
The resulting time-translational invariance
$\hat H_1(t) = \hat H_1(t + \frac {2\pi}\wL)$ will be used to 
adopt a Floquet representation of the target's many-body states.

\subsection{Collision term}
\label{sec:hamiltonian-collision}
The projectile and the target interact via a Coulomb potential, as specified in \secref{sec:scope}.
Assuming that the target contains particles with charges $Q_i$, we can
write it in the form
\begin{align}
\hat V &= \sum_i Q_i v(\hat{\vb{r}}_i - \hat{\vb{r}}_0) \\
       &= \int v(\vb r - \hat{\vb r}_0) \hat \varrho_1(\vb r)\, \dd^3\vb r,
\end{align}
where the integral expression incorporates the charge density operator
\begin{equation}
\hat \varrho_1 (\vb r) \coloneqq \sum_iQ_i\,\delta(\vb r - \hat{\vb r}_i).
\label{eq:rho}
\end{equation}
The manuscript
assumes the bare Coulomb term~\footnote{The projectile--target
interaction \(\hat V\) is assumed negligible outside the collision region.},
with screening absorbed
later into the target density response $\chi_{\rho\rho}$ and the
loss function $\mathcal L(\vb q,\omega)$
(Secs.~\ref{sec:plasmon-drude}--\ref{sec:plasmon-diffusive}).

\subsection{Charge-density convention}
\label{sec:hamiltonian-charge-density}
We adopt charge-density conventions throughout this paper.
Equation~\pref{eq:rho} defines $\hat\varrho_1$ as the \emph{charge}, not number,
density of the target: $\hat\varrho_1\to -e\,\hat n$ for an all-electron
target, where $\hat n=\sum_i\delta(\vb r-\hat{\vb r}_i)$ is the
number-density operator. Consistently with this choice, $v(\vb q)$ in
\pref{eq:vq} is the Coulomb kernel \emph{per unit target charge},
$v(\vb q)=4\pi q_{\mathrm p}/q^{2}$ in Gaussian--cgs units with
$q_{\mathrm p}$ the projectile charge (we reserve $\vb q$ exclusively
for momentum throughout); for an electron projectile $q_{\mathrm p}=-e$
and $v(\vb q)=-4\pi e/q^{2}$. The projectile--target Coulomb potential
energy $\hat V=\int v\,\hat\varrho_1\,\dd^{3}\vb r$ then carries both
charges correctly. The dynamic structure factor
$S(\vb q,\omega)\propto|\bra{f}\hat\varrho_1(\vb q)\ket{i}|^{2}$
introduced in \secref{sec:floquet} is therefore the
\emph{charge-density} structure factor; it carries one factor of $e^{2}$
relative to the number-density structure factor of the EELS literature,
a difference that propagates into the prefactor of the $f$-sum rule
(\appref{app:fsum}) and the loss-function identity
\pref{eq:S-loss-function}.

\section{Scattering matrix in the first Born approximation}
\label{sec:1Born}
 To first order in $\hat V$, the $S$-matrix element reads
\begin{equation}
\label{eq:smat}
S(f\leftarrow i)\, =\, S^\text{0B}(f\leftarrow i) + S^\text{1B}(f\leftarrow i) + \order{\hat V^2}
\end{equation}
where the zero-order (identity) term takes form of an overlap
\begin{equation}
\label{eq:S0B}
S^\text{0B} \coloneqq \braket{\chi_{f}}{\chi_i} \braket{\Phi_{f}}{\Phi_i}
\end{equation}
and
the first-Born scattering matrix element~\cite{Bilitewski2015} reads
\begin{multline}
S^\text{1B}(f\leftarrow i) = -\frac{\iu}{\hbar}
\int_{-\infty}^\infty  \dd t\,\bra{\chi_{f}(t)}\bra{\Phi_f(t)} \hat V\\
\times\ket{\Phi_i(t)}\ket{\chi_{i}(t)}.
\label{eq:S1B}
\end{multline}
In the above expressions, the subscripts $i$ and $f$ label initial and
final states and the symbols $\chi$ and $\Phi$ are reserved for
projectile and target wavefunctions, respectively.
The projectile and target states satisfy the Schrödinger equations
\begin{subequations}
\label{eq:Sche}
\begin{align}
\iu \hbar \partial_t\ket{\chi(t)} &= \hat H_0(t) \ket{\chi(t)}
\label{eq:Sch-chi}
\\
\iu \hbar \partial_t\ket{\Phi(t)} &= \hat H_1(t) \ket{\Phi(t)}
\end{align}
\end{subequations}
with the Hamiltonians introduced in the preceding section.

\paragraph*{Asymptotic in/out states.}
The projectile and target states entering
Eqs.~\pref{eq:smat}--\pref{eq:S1B} are channel states of the
reference Hamiltonian
\begin{equation}
    \hat H_{\mathrm{ch}}(t)
    \coloneqq \hat H_0(t)+\hat H_1(t).
\end{equation}
Since the projectile and target degrees of freedom are decoupled in
$\hat H_{\mathrm{ch}}(t)$, these channel states factorize into a
projectile solution of $\hat H_0(t)$ and a target solution of
$\hat H_1(t)$. Because $\hat H_{\mathrm{ch}}(t)$ is periodic, its
solutions may be chosen as Floquet states. In the spatially homogeneous
model, the projectile part is given exactly by Gordon--Volkov states,
labeled by a conserved canonical wavevector and a quasienergy.

For a spatially structured but localized optical field, we use a
distorted-wave Born construction~\cite{Taylor2006,Peng2004}: the optical interaction contained in
$\hat H_0(t)$ is treated as part of the reference dynamics, while the
projectile--target interaction $\hat V$ is treated perturbatively. The
entrance-channel distorted wave $\chi_{\mathbf k_i}^{(+)}$ is normalized
to a field-free plane wave upstream of the interaction region, whereas
the exit channel is represented in the transition amplitude by the bra
$\langle\chi_{\mathbf k_f}^{(-)}|$, normalized to a field-free plane
wave downstream. Thus $\mathbf k_i$ and $\mathbf k_f$ denote the
field-free asymptotic momenta; in particular, $\hbar\mathbf k_f$ is the
momentum recorded by the detector. The superscripts $(+)$ and $(-)$
retain the standard scattering-theory boundary-condition convention
and should not be interpreted simply as the direction of propagation.
We approximate these reference-channel solutions by eikonal--Volkov
waves, as detailed below, and retain $\hat V$ as the perturbation.

In the strictly homogeneous monochromatic model there are no
field-free spatial asymptotes, and the Floquet--Volkov channel basis is
therefore the natural description throughout. For a spatially localized external
field, by contrast, the eikonal--Volkov kernel connects the
laser-dressed dynamics in the interaction region to field-free
asymptotic momentum states; its homogeneous limit recovers the usual
Volkov--Bessel kernel.

The target factors $\Phi_i(t)$ and $\Phi_f(t)$ are Floquet eigenstates
of $\hat H_1(t)$. Since the target is assumed to remain spatially
confined, they require no additional scattering boundary condition.

\paragraph*{Zero-order term.}
For inelastic target channels, \(S^\text{0B}\) vanishes by orthogonality of
the target channel states, \epref{eq:S0B}. For target-elastic channels, however,
\(S^\text{0B}\) need not vanish: it is the reference-channel scattering
amplitude between the incoming and outgoing asymptotic projectile states.
Only in the absence of reference-channel scattering does this term reduce
to the trivial identity contribution. Thus the first-Born term \(S^\text{1B}\)
is the leading contribution to the inelastic cross sections derived
below, whereas elastic observables may require the coherent combination
of \(S^\text{0B}\) and \(S^\text{1B}\). An explicit example is discussed in
\appref{app:elastic}.

\section{Scattering matrix and cross section in the Floquet--Fourier expansion}
\label{sec:floquet}
This section establishes the exact Floquet--Fourier framework that
underlies the remainder of the paper. The projectile dynamics and the
driven target are separated into an exact projectile kernel and Floquet
density components, allowing the scattering amplitude and the
corresponding cross section to be written in a fully factorized form.
This representation identifies the bi-momentum Floquet structure factor
as the fundamental target response. The following sections derive
successive reductions of the projectile kernel, culminating in the
homogeneous-field (Volkov) limit.

\subsection{Scattering matrix}
\label{sec:Smatrix-floquet}
We adopt here the Floquet--Fourier form for the solutions of
\pref{eq:Sche},
\begin{subequations}
\label{eq:FF}
\begin{align}
\ket{\chi(t)}  &= e^{-\iu \epsilon(\vb k)t/\hbar} \sum_n e^{-\iu\wL n t}\ket{\chi^{(n)}}, \\
\ket{\Phi(t)} &= e^{-\iu Et/\hbar} \sum_n e^{-\iu\wL n t}\ket{\Phi^{(n)}}.
\end{align}
\end{subequations}

The integrand in \pref{eq:S1B}
\begin{multline*}
e^{-\iu\qty(\epsilon(\vb{k}_i) - \epsilon(\vb{k}_f) + E_{i}-E_{f}) t/\hbar}\\
 \times\sum_{m_i,m_f}\sum_{n_i,n_f} e^{-\iu\wL(n_i - n_f + m_i - m_f)t}\\
\times\bra{\chi_{f}^{(n_f)}}\bra{\Phi_f^{(m_f)}} \hat V
\ket{\Phi_i^{(m_i)}}\ket{\chi_i^{(n_i)}}
\end{multline*}
is integrated straightforwardly. Introducing $n=n_f-n_i$ and
$N = m_f + n_f - m_i  - n_i$ one obtains
\begin{multline}
S^\text{1B}(f\leftarrow i) = -2\pi\iu
\sum_{N}\delta\bigl(\epsilon(\vb{k}_f)- \epsilon(\vb{k}_i)\\
+ E_{f} - E_{i} + N\hbar\wL\bigr)
\times\\
\sum_{m_i,n_i,n}\bra{\chi_{f}^{(n_i + n)}}\bra{\Phi_f^{(m_i + N -n)}}\hat V
\ket{\Phi_i^{(m_i)}}\ket{\chi_i^{(n_i)}},
\label{eq:S1B-Floquet}
\end{multline}
 or
\begin{multline}
S^\text{1B}(f\leftarrow i) = -2\pi\iu
\sum_{N}\delta\bigl(\epsilon(\vb{k}_f)- \epsilon(\vb{k}_i)\\
+ E_{f} - E_{i} + N\hbar\wL\bigr)\mathcal  M_N(f \leftarrow i),
\end{multline}
where we defined the transition amplitude
\begin{multline}
\label{eq:MN}
\mathcal M_N(f \leftarrow i) \coloneqq \sum_{m_i,n_i,n} \bra{\chi_{f}^{(n_i + n)}}\bra{\Phi_f^{(m_i + N -n)}}\\
\times\hat V
\ket{\Phi_i^{(m_i)}}\ket{\chi_i^{(n_i)}},
\end{multline}
sketched in \figref{fig:geometry} schematically.
For $N=0$ the Dirac delta-function reduces to the familiar energy
conservation of field-free inelastic scattering. More generally,
$N\hbar\wL$ is the net energy that projectile and target jointly release
into the light field during the collision, and \pref{eq:S1B-Floquet}
invites the interpretation of $N$ photons being emitted into (or, for
$N<0$, absorbed from) the field.

\subsection{Momentum-space representation and the factored amplitude}
\label{sec:planewave}
It is advantageous to adopt the plane-wave expansion
\begin{equation}
\braket{\vb r_0}{\chi^{(n)}} = \int \frac{\dd^3\vb q }{\qty(2\pi)^\frac{3}{2}}
\chi^{(n)}({\vb q}) e^{\iu \vb{q\cdot r}_0}
\label{eq:chi-q}
\end{equation}
for the initial and final states of the projectile. The matrix element of
the potential
\begin{multline}
\bra{\chi_{f}^{(n_i + n)}}\bra{\Phi_f^{(m_i + N -n)}} \hat V
\ket{\Phi_i^{(m_i)}}\ket{\chi_i^{(n_i)}} = \\
\frac 1{\qty(2\pi)^3} \int \dd^3\vb q'\,
v(\vb q')\bra{\Phi_f^{(m_i + N -n)}}\hat\varrho_1(\vb q')\ket{\Phi_i^{(m_i)}}\\
\times\int\dd^3\vb q \,\chi_f^{(n_i + n) \star}(\vb q + \vb q')
\chi_i^{(n_i)}(\vb q)
\label{eq:Mij-q}
\end{multline}
is expressed through the Fourier-transformed charge density,
\begin{equation}
\hat \varrho_1 (\vb q) \coloneqq
\int \dd^3\vb r\,e^{-\iu \vb r\cdot \vb q}\,\hat \varrho_1 (\vb r),
\label{eq:rhoq}
\end{equation}
the Fourier-transformed potential
\begin{equation}
v(\vb q) \coloneqq \int \dd^3\vb r\,e^{-\iu \vb r\cdot \vb q}\,v(\vb r),
\label{eq:vq}
\end{equation}
and a projectile-only $\vb q$-overlap. We introduce few useful definitions:

\paragraph*{Projectile kernel.}
Summing the projectile overlap in \pref{eq:Mij-q} over the internal
sideband index $n_i$ defines the \emph{exact projectile kernel}
\begin{multline}
  \mathcal P^{(n)}_{\vb k_f\vb k_i}(\vb q)
  \;\coloneqq\;
  \sum_{n_i}
  \int\!\dd^3\vb q''\,\\
  \chi_{\vb k_f}^{(-)(n_i+n)*}(\vb q''+\vb q)\,
  \chi_{\vb k_i}^{(+)(n_i)}(\vb q'').
  \label{eq:proj-kernel}
\end{multline}
It encodes the coupling of a net $n$ projectile sideband quanta to the
field, integrated over the exact quantum-mechanical
Floquet--Fourier components of both the incoming and outgoing projectile
states.
The wavefunction superscripts indicate the appropriate boundary condition
discussed in
\secref{sec:1Born}: (+) is the Floquet component of the state with 
incoming boundary condition and ($-$) is the component of the outgoing state.
For (field-free, undressed) planewaves, the kernel
resolves to $\propto \delta_{n,0}\delta(\vb q - \vb k_f +\vb k_i)$.
Explicit kernels reflecting the projectile dynamics in a laser field are given
in later sections.

\paragraph*{Target density component.}
The sum over the target Floquet index $m_i$ in \pref{eq:Mij-q} gives
the Floquet density component
\begin{equation}
  \tilde\varrho_{fi}^{(M)}(\vb q)
  \;\coloneqq\;
  \sum_{m_i}
  \bra{\Phi_f^{(m_i+M)}}\hat\varrho_1(\vb q)\ket{\Phi_i^{(m_i)}},
  \label{eq:varrho-exact}
\end{equation}
the $M$-th Floquet--Fourier component of the transition density.  The
choice \(f=i\) gives the corresponding harmonic of the coherent mean
density; \(f\ne i\) describes a target-changing transition.

\paragraph*{Factored amplitude.}
With \pref{eq:proj-kernel} and \pref{eq:varrho-exact}, the
amplitude \pref{eq:MN} takes the form
\begin{multline}
  \mathcal M_N(f\leftarrow i) = \\
  \sum_n
  \int\!\dd^3\vb q'\,
  v(\vb q')\,
  \Pbar^{(n)}_{\vb k_f\vb k_i}(\vb q')\,
  \tilde\varrho_{fi}^{(N-n)}(\vb q'),
  \label{eq:MN-factored}
\end{multline}
with the normalized kernel
$\Pbar^{(n)}_{\vb k_f\vb k_i}\coloneqq(2\pi)^{-3}\mathcal P^{(n)}_{\vb k_f\vb k_i}$.
This is a coherent sum over internal photon partitions: $n$ quanta
couple to the projectile and $N-n$ quanta to the target, for every
split consistent with the net external index $N$.

\subsection{Differential cross sections}
\label{sec:dsigma}
The differential cross section of the projectile per energy and solid
angle, with a transition of the target
$\Phi_i \to \Phi_f$, follows from \pref{eq:S1B-Floquet}
by using a standard procedure for inelastic scattering \cite{Taylor2006}; the result
reads~\cite{Dorr1994,Vucic1995}
\begin{multline}
\frac{\dd^2 \sigma \qty(\Phi_f \leftarrow \Phi_i)}{\dd \Omega \dd \epsilon(\vb k_f)} =
\qty(\frac{\me}{2\pi\hbar^2})^2
\sum_{N} \frac{k_f}{k_i}\\
\times\delta\bigl(\epsilon(\vb{k}_f)- \epsilon(\vb{k}_i)+ E_{f} - E_{i} + N\hbar\wL\bigr)\\
\times\Bigl|\mathcal M_N(f\leftarrow i)\Bigr|^2,
\label{eq:dsigma}
\end{multline}
with \pref{eq:MN}, $k_i \coloneqq |\vb{k}_i|$.

We adopt the convention that $\epsilon(\vb k)\equiv\hbar^2 k^2/(2\me)$ throughout
\pref{eq:FF} and \pref{eq:dsigma} -- \ie\ the kinetic energy
of the asymptotic free electron. The ponderomotive shift $\Up$ that
distinguishes the Volkov quasienergy from the kinetic energy of the
projectile (see \secref{sec:projdyn-volkov} below) enters $\epsilon(\vb k_i)$ and
$\epsilon(\vb k_f)$ on the same footing and therefore cancels in the
energy-conservation delta.
The prefactor $(\me/2\pi\hbar^{2})^{2}$ of
\pref{eq:dsigma} is the Born normalization for channel states
normalized to unit-amplitude plane waves, $\chi_{\vb k}(\vb r)\to
e^{\iu\vb k\cdot\vb r}$, i.e.\
$\chi_{\vb k}(\vb q)=(2\pi)^{3/2}\delta^{(3)}(\vb q-\vb k)$ in the
convention \pref{eq:chi-q}; this is the normalization of the Volkov and
eikonal--Volkov states \pref{eq:Volkov-FF} and \pref{eq:EV-FF-r}.
Expression~\eqref{eq:dsigma} is the \(\lvert S^\text{1B}\rvert^2\)
contribution for a specified target transition. For \(f=i\), a complete
target-elastic observable also contains the
\(S^\text{0B}\)--\(S^\text{1B}\) interference discussed in
\secref{sec:1Born}.

Equation~\pref{eq:dsigma} reproduces, in the
homogeneous Volkov limit of
Secs.~\ref{sec:projdyn-volkov}--\ref{sec:applications}, the
Bessel--Bethe--Floquet form \pref{eq:dsigma-SF}, and in the
field-free limit the conventional Bethe cross section \pref{eq:bethe}
of \appref{app:limits}. The atomic limit of
Joachain--D\"orr--Vu\v ci\'c--Potvliege
$(e,e')$ is exposed in \appref{app:limits}.

\paragraph*{The scattering normalization.}
Equation~\pref{eq:dsigma} uses the nonrelativistic scattering normalization
together with the dispersion
$\epsilon(\vb k)=\hbar^{2}k^{2}/(2\me)$ of the scalar projectile theory of
\secref{sec:hamiltonian-projectile}.
 When comparing with
$100$--$300\,\mathrm{keV}$ TEM data, the experimentally quoted beam
energy and scattering angle should be converted to $k_i$, $k_f$, $\Qv$,
and $v_{\rm rel}$ using the relativistic kinematic relations
$p=\hbar k=\gamma_{\mathrm p}m_{\mathrm e}v$,
$\epsilon_{\rm k}=(\gamma_{\mathrm p}-1)m_{\mathrm e}c^{2}$ of
\pref{eq:rel-mass-substitution}. This is an external kinematic
conversion: it does not modify the prefactor
$(\me/2\pi\hbar^{2})^{2}$, the energy-conservation delta, or the
Floquet sideband algebra of \pref{eq:dsigma}.

\paragraph*{Sideband-resolved cross section.}
For later use we define 
the sideband-resolved angular cross section, equivalently the angular weight of the $N$th sideband peak
\(\qty(\dd\sigma/\dd\Omega)_N\), by decomposing the
energy-differential signal into sideband peaks,
\begin{multline}
\frac{\dd^2 \sigma \qty(\Phi_f \leftarrow \Phi_i)}
{\dd \Omega\, \dd \epsilon(\vb k_f)}
\eqqcolon
\sum_N
\delta\!\bigl(
\epsilon(\vb k_f)-\epsilon(\vb k_i)+E_f-E_i+N\hbar\wL
\bigr)
\\
\times
\left(
\frac{\dd\sigma}{\dd\Omega}
\right)_N
\qty(\Phi_f \leftarrow \Phi_i).
\label{eq:sideband-resolved-angular-weight}
\end{multline}
For finite experimental resolution, the same quantity is obtained by
integrating the doubly differential signal over an energy window
\(\Delta\epsilon_N\) containing only the \(N\)-th sideband peak,
\begin{equation}
\left(
\frac{\dd\sigma}{\dd\Omega}
\right)_N
\qty(\Phi_f \leftarrow \Phi_i)
=
\int_{\Delta\epsilon_N}\!\dd \epsilon(\vb k_f)\,
\frac{\dd^2\sigma \qty(\Phi_f \leftarrow \Phi_i)}
{\dd\Omega\,\dd \epsilon(\vb k_f)} .
\label{eq:sideband-window-weight}
\end{equation}

\subsection{The bi-momentum Floquet structure factor}
\label{sec:exact-kernel}

The factorized first-Born amplitude~\pref{eq:MN-factored} separates the
projectile and target dynamics into the exact projectile kernel
\(\mathcal P^{(n)}_{\vb k_f\vb k_i}\) and the Floquet density
components \(\tilde\varrho_{fi}^{(M)}\). Squaring this amplitude,
summing over unresolved target final states, and using
\pref{eq:dsigma} yields the first-Born cross section within the
Floquet--Fourier framework. The target dependence enters exclusively
through a single spectral quantity, defined below as the bi-momentum
Floquet structure factor.

\paragraph*{Exact cross section.}

Squaring \pref{eq:MN-factored}, summing over unresolved target final
states \(f\), including \(f=i\) unless it is explicitly excluded, and
using \pref{eq:dsigma} gives the doubly differential
cross section
\begin{multline}
  \frac{\dd^2\sigma}{\dd\Omega\,\dd\epsilon}
  =
  \Bigl(\frac{\me}{2\pi\hbar^2}\Bigr)^{\!2}
  \sum_N\frac{k_f}{k_i}\,
  \sum_{n,n'}
  \int\!\dd^3\vb q_1\,\dd^3\vb q_2
  \\
  \times
  v(\vb q_1)\,v^*(\vb q_2)\,
  \Pbar^{(n)}_{\vb k_f\vb k_i}(\vb q_1)\,
  \Pbar^{(n')*}_{\vb k_f\vb k_i}(\vb q_2)
  \\
  \times
  \SF^{(N-n)(N-n')}(\vb q_1,\vb q_2;\omega_N),
  \label{eq:dsigma-bimomentum}
\end{multline}
with the normalized kernels $\Pbar=(2\pi)^{-3}\mathcal P$ of
\pref{eq:MN-factored},
where
\(
\hbar\omega_N
\coloneqq
\epsilon(\vb k_i)-\epsilon(\vb k_f)-N\hbar\wL
\)
is the energy transferred to the target in the externally resolved
sideband \(N\). Equation~\pref{eq:dsigma-bimomentum} is a
total cross section: $\SF$ is built from the extensive density
operator \pref{eq:rhoq}. In \pref{eq:dsigma-bimomentum},
\(k_f\) denotes the fixed detected outgoing wave number of the doubly
differential measurement. The sum over target final states is therefore
performed at fixed \(\epsilon(\vb k_f)\), while the corresponding energy
constraint is contained in the target response through
\(\omega_N\).

Equation~\pref{eq:dsigma-bimomentum} has a transparent bilinear
structure: two projectile--Coulomb legs,
\(v(\vb q_1)\Pbar^{(n)}(\vb q_1)\) and
\(v^*(\vb q_2)\Pbar^{(n')*}(\vb q_2)\),
are connected through a single target response function.

\paragraph*{Bi-momentum Floquet structure factor.}

The target quantity appearing in
\pref{eq:dsigma-bimomentum} is the bi-momentum Floquet structure factor,
\begin{multline}
\SF^{nn'}(\vb q_1,\vb q_2;\omega)
\coloneqq
\sum_f
\tilde\varrho_{fi}^{(n)}(\vb q_1)\,
\tilde\varrho_{fi}^{(n')*}(\vb q_2)
\\
\times
\delta(\hbar\omega+E_i-E_f).
\label{eq:SF-bimomentum}
\end{multline}

It is Hermitian in both the Floquet indices \((n,n')\) and the momentum
arguments \((\vb q_1,\vb q_2)\). On the momentum diagonal it reduces to
the homogeneous Floquet dynamic structure factor,
\[
\SF^{nn'}(\vb q,\vb q;\omega)
=
S_{\mathrm F}^{nn'}(\vb q,\omega)
\]
(extensive, at finite target volume $V$).

The bilinear projectile--response structure of the cross section is not
tied to a pure Floquet eigenstate. Equation~\pref{eq:SF-bimomentum} is
the closed-system Floquet--Lehmann realization adopted here. For a
periodically driven open steady state the same scattering structure is
obtained, with the corresponding unsymmetrized Floquet density
correlator evaluated using the stationary density matrix. The
pure-state assumption therefore affects the \emph{representation of the
target response}, not the projectile--target factorization
\pref{eq:dsigma-bimomentum} itself.
Off-diagonal Floquet contractions
\(n\neq n'\) arise from squaring the coherent internal photon-partition
sum and are retained whenever the driven target possesses off-diagonal
Floquet density correlations.

The two momentum arguments encode the translational structure of the
target. For a translationally invariant bulk,
\begin{equation}
\SF^{nn'}(\vb q_1,\vb q_2;\omega)
=(2\pi)^3
\delta^{(3)}(\vb q_1-\vb q_2)\,
s_{\mathrm F}^{nn'}(\vb q_1,\omega),
\label{eq:bimomentum-translation}
\end{equation}
with $s_{\mathrm F}\coloneqq S_{\mathrm F}/V$,
so that the response becomes diagonal in momentum; the regulated
value $(2\pi)^3\delta^{(3)}(\vb 0)\to V$ recovers the diagonal identity
above. The intensive $s_{\mathrm F}$ is the structure factor per unit
volume.
Genuine off-diagonal weight
\(\vb q_1\neq\vb q_2\) therefore requires target-side translational
symmetry breaking, for example by surfaces, interfaces, finite-size
structure, crystal symmetry, or an inhomogeneous external driving
field. Projectile-side spatial dressing does not itself create
off-diagonal target spectral weight; it only modifies the projectile
kernel sampling the target response. Recovering the full bi-momentum
structure factor therefore requires varying the projectile kernel---for
example through the beam impact parameter, incidence angle---and inverting the resulting family of
convolutions.

\paragraph*{Reduction hierarchy.}

Three successive reductions convert
\pref{eq:dsigma-bimomentum} into the working formulæ developed in the
remainder of the paper.

\emph{(i) Eikonal projectile}. In many
situations of physical interest, the laser field varies slowly in space.
In \secref{sec:projdyn-eikonal-kernel} we develop semiclassical eikonal 
expressions for the projectile kernel.

\emph{(ii) Translationally invariant target.}
Using \pref{eq:bimomentum-translation}, the double momentum integral
collapses to a single integral over the momentum diagonal.

\emph{(iii) Projectile-only limit for the connected target response}
(\secref{sec:applications}).
After the coherent mean-density sector has been split off exactly
according to \pref{eq:SF-coherent-connected} below, the reduction
\(
\SFc^{MM'}(\vb q_1,\vb q_2;\omega)
\rightarrow
\delta_{M0}\delta_{M'0}
\mathscr S_{\mathrm c}(\vb q_1,\vb q_2;\omega)
\)
forces \(n=n'=N\), collapsing the internal partition sum while leaving
the momentum convolution unchanged. Only after additionally imposing
translational invariance does the result reduce to
\(
(2\pi)^3\int\dd^3\vb q\,|v|^2|\Pbar^{(N)}|^2s_{\mathrm c}
\),
which in the homogeneous projectile limit
\(
\Pbar^{(N)}
\rightarrow
J_N\,e^{\iu N(\varphi+\theta_{\Qv}^{\rm hom})}\delta^{(3)}(\vb q-\Qv)
\)
[\epref{eq:P-hom}], with the regulated $(2\pi)^3\delta^{(3)}(\vb 0)\to V$,
recovers the Bessel-weighted field-free cross section
\(
|v(\Qv)|^2J_N^2S_{\mathrm c}(\Qv,\omega_N),\ S_{\mathrm c}=Vs_{\mathrm c}.
\)
The coherent sector is retained separately and supplies the
target-elastic PINEM channel of \secref{sec:pinem-bridge} and
\pref{eq:dsigma-loss-plus-pinem}.

Accordingly,
\pref{eq:dsigma-bimomentum}
provides the natural starting point for extensions beyond the eikonal
approximation, beyond translational invariance of the target, or beyond
negligible dressing of the connected target response.

\subsection{Coherent mean density, connected fluctuations, and PINEM}
\label{sec:pinem-bridge}

The structure factor in \pref{eq:SF-bimomentum} contains two physically
different sectors.  To display them, let
\begin{equation}
R_M(\vb q)\coloneqq\tilde\varrho_{ii}^{(M)}(\vb q)
\label{eq:R-mean-density}
\end{equation}
denote the \(M\)-th harmonic of the charge density in the prepared target
state.  Equivalently, \(R_M\) is the corresponding harmonic of the
pump-induced mean density
\(\bar\varrho(\vb q,t)=
\langle\Phi_i(t)|\hat\varrho_1(\vb q)|\Phi_i(t)\rangle\).
Writing
\(\hat\varrho_1=\bar\varrho+\delta\hat\varrho_1\) gives the exact
separation
\begin{multline}
\SF^{MM'}(\vb q_1,\vb q_2;\omega)
=R_M(\vb q_1)R_{M'}^*(\vb q_2)\,\delta(\hbar\omega)
{}\\
{}+\SFc^{MM'}(\vb q_1,\vb q_2;\omega).
\label{eq:SF-coherent-connected}
\end{multline}
The first term is target elastic: the material returns to the same
Floquet state, although the electron may gain or lose optical quanta
through a nonzero harmonic \(R_{M\ne0}\).  The connected term instead
describes target-changing density fluctuations.  Thus a sum over all
final target states is not synonymous with a purely inelastic
structure factor.

This separation also exposes how standard PINEM is embedded in the
perturbative theory. We remark that the following
anticipates the straight-trajectory eikonal construction of
Secs.~\ref{sec:projdyn-eikonal}--\ref{sec:dsigma-inhom}
and the impact-parameter representation of \appref{app:elastic}.
Inserting the first term of
\pref{eq:SF-coherent-connected} into
\pref{eq:dsigma-bimomentum} produces the modulus square of the
first-Born amplitude for scattering from the coherent, oscillating
potential generated by \(\bar\varrho\).  Its spectral delta function
sets the residual target excitation energy to zero; the resolved
electron sideband energy is carried by the Floquet harmonic.  In the
straight-trajectory, nonrecoil limit this one-photon Born amplitude
defines the weak-coupling PINEM amplitude \(g_{\rm ind}\).  More
explicitly, let the electron propagate with speed \(v_{\rm e}\) along
\(z\), at transverse impact parameter \(\vb R\), and write the
fundamental harmonic of the induced field as
\[
\EE_{{\rm ind},z}(\vb R,z,t)
=\EE_{{\rm ind},z}^{(+)}(\vb R,z)e^{-\iu\wL t}
+\mathrm{c.c.}
\]
(For induced fields, densities and potentials the superscript $(+)$
denotes the coefficient of $e^{-\iu\wL t}$, whereas $\Ap$ in
\pref{eq:Acos} multiplies $e^{+\iu\wL t}$.)
In the standard PINEM convention \cite{Park2010,ParkZewail2012,
deAbajo2010} the dimensionless target-induced
coupling is
\begin{equation}
g_{\rm ind}(\vb R)
\coloneqq
\frac{e}{\hbar\wL}
\int_{-\infty}^{\infty}\!\dd z\,
\EE_{{\rm ind},z}^{(+)}(\vb R,z)
e^{-\iu\wL z/v_{\rm e}},
\label{eq:gind-field}
\end{equation}
where \(e>0\) and the projectile charge is \(q_{\rm p}=-e\).  The
exponential is the longitudinal phase-matching factor sampled along
the trajectory.  To connect this definition directly to the density
harmonics above, write
\(\bar\varrho(\vb q,t)=
\bar\varrho_{\vb q}^{(+)}e^{-\iu\wL t}+\mathrm{c.c.}\), so that
\(\bar\varrho_{\vb q}^{(+)}=R_{-1}(\vb q)\) in the
Floquet--Fourier convention of \pref{eq:FF}.  The sign of this Floquet
index should not be compared directly with the electron-energy ladder
index often used in PINEM: in the present manuscript positive \(N\)
denotes electron energy loss, so the target-elastic \(N=+1\) channel
samples \(R_{+1}\), whereas the common PINEM ladder convention takes
positive \(N\) to denote electron energy gain and its \(+1\) amplitude
therefore samples \(R_{-1}\).  In the nonretarded scalar
limit, \pref{eq:gind-field} is equivalently
\begin{equation}
g_{\rm ind}(\vb R)
=\frac{\iu}{\hbar v_{\rm e}}
\int\!\frac{\dd^2\vb q_\perp}{(2\pi)^2}\,
e^{\iu\vb q_\perp\cdot\vb R}
v(\vb q)\,R_{-1}(\vb q)
\bigg|_{q_z=\wL/v_{\rm e}},
\label{eq:gind-density}
\end{equation}
with the Coulomb kernel \(v(\vb q)=4\pi q_{\rm p}/q^2\) defined in
\secref{sec:hamiltonian-charge-density}.  Thus \(g_{\rm ind}\) is not
an additional phenomenological parameter: it is the phase-matched
impact-parameter representation of the coherent one-photon Coulomb
Born amplitude generated by the pump-induced mean charge.

PINEM goes beyond that lowest order in one specific and controlled
way.  Repeated insertions of the same coherent \(c\)-number potential
factorize along the projectile trajectory.  The corresponding part of
the perturbation series can therefore be exponentiated,
\begin{equation}
\mathcal U_{\rm coh}
=\exp\!\left[g_{\rm ind}^{*}\hat T_{\rm L}-g_{\rm ind}\hat T_{\rm L}^{\dagger}\right],
\qquad \hat T_{\rm L}|N\rangle=|N+1\rangle ,
\label{eq:pinem-coherent-exponential}
\end{equation}
where $\hat T_{\rm L}$ raises the electron energy loss by
$\hbar\wL$, and the Fourier coefficients of this phase are the standard PINEM
amplitudes, $C_N(g)=e^{-\iu N\arg g}J_N(2|g|)$ in the loss index
($C_{+1}\simeq g^*$, consistent with \pref{eq:gind-density} and
\pref{eq:weak-pinem-born-parent}), with probabilities
\(P_N=J_N^2(2|g_{\rm ind}|)\) at the fundamental harmonic \cite{Kfir2019}.  In this
sense the first-Born coherent sector supplies the elementary PINEM
vertex, while its repeated target-elastic action supplies the Bessel
ladder.  One must not exponentiate the full density correlator:
connected target fluctuations generate target-changing processes and
in general require a different resummation, or higher connected
correlators.

If a prescribed localized projectile field already produces a
coupling \(g_0\) introduced in \epref{eq:g0}, with its phase fixed
there in the same convention as \pref{eq:gind-field}, the coherent fields add at amplitude level,
\(g_{\rm tot}=g_0+g_{\rm ind}\).  Incorporating \(g_{\rm ind}\) into
the projectile channel Hamiltonian then requires the residual target
response to be built from
\(\delta\hat\varrho_1=\hat\varrho_1-\bar\varrho\), so that the mean
field is not counted twice.  The leading impact-parameter-to-angle
transform of the resummed coherent channel, \pref{eq:pinem-angular},
is given at the end of \appref{app:elastic}. The general amplitude-level matching,
including  higher coherent
harmonics and the distinct resummations of connected fluctuations,
will be presented in a forthcoming publication.
This all-orders channel reorganization is not implemented by the
reference Hamiltonian used below, which contains only the prescribed
external field.  Its explicit realization requires adding the induced
longitudinal scalar potential, or in a retarded formulation the
corresponding electromagnetic four-potential, to the projectile
channel.  Here we establish the first-Born matching and state the
no-double-counting prescription without claiming that induced-field
dressing has already been incorporated at Hamiltonian level.

\section{Projectile dressing}
\label{sec:projdyn}
We give explicit expressions for the projectile states in this section:
first the homogeneous laser field is treated, showing the familiar
Gordon--Volkov solutions. Second, the case of
a weakly-varying field is treated in a semiclassical eikonal
approximation.

\subsection{Classical variables}
We now specialize to a spatially homogeneous monochromatic field (or
equivalently, in the
dipole approximation); \epref{eq:Acos} reduces to
\begin{equation}
\label{eq:A-homog}
\vb A(\vb r,t)
\to
\mathrm{Re}
\left[
\vb A_0^{(+)} e^{\iu(\omega_{\rm L}t+\varphi)}
\right]
\end{equation}
with the complex $\vb A_0^{(+)}$ constant.

\label{sec:projdyn-classical}
The classical equation of motion of a free electron in the field
\pref{eq:A-homog} admits a uniform drift superimposed with a quivering
motion of (complex) amplitude
\begin{equation}
\alphav \coloneqq -\frac{q_{\mathrm p}\,\vb A_0^{(+)}}{\me\wL c},
\label{eq:quiver}
\end{equation}
the \emph{quiver vector}: the classical quiver displacement is
$\mathrm{Re}[-\iu\,\alphav e^{\iu(\wL t+\varphi)}]$. The cycle-averaged kinetic energy of the
quivering motion is the \emph{ponderomotive energy}
\begin{equation}
\Up \coloneqq \frac{q_{\mathrm p}^{2}\, |\vb{A}_0^{(+)}|^{2}}{4\me c^{2}}.
\label{eq:Up}
\end{equation}
Both $|\alphav|$ and $\Up$ set the strength of non-perturbative dressing of
the projectile.

\paragraph*{Nonrelativistic vs.\ relativistic quiver amplitude.}
Equations~\pref{eq:quiver} and \pref{eq:Up} are the nonrelativistic scalar-core
expressions used throughout the formal development. The
leading kinematic relativistic correction of
\pref{eq:rel-mass-substitution} gives
\begin{align}
\boldsymbol\alpha_{0,\mathrm{rel}}^{(+)}
&\coloneqq
-\frac{q_{\rm p}}{\omega_{\rm L}c}
\left(
\frac{\vb A_{0\perp}^{(+)}}{\gamma_{\rm p}m_{\rm e}}
+
\frac{\vb A_{0\parallel}^{(+)}}{\gamma_{\rm p}^{3}m_{\rm e}}
\right),
\\
U_{\rm p, \rm rel}
&\coloneqq
\frac{q_{\rm p}^{2}}{4m_{\rm e}c^{2}}
\left(
\frac{|\vb A_{0\perp}^{(+)}|^2}{\gamma_{\rm p}}
+
\frac{|\vb A_{0\parallel}^{(+)}|^2}{\gamma_{\rm p}^{3}}
\right).
\label{eq:alpha-NR-rel}
\end{align}
where $\Ap_{0\perp}$ and $\Ap_{0\parallel}$
are understood as the components transverse and 
parallel to the mean projectile velocity.
The body of the paper writes $\alphav$ without `rel' in
formal-theory expressions.
In numerical TEM/PINEM estimates and in \appref{app:trajectory-bending}
$\alphav$ is to be read as $\boldsymbol\alpha_{0,\mathrm{rel}}^{(+)}$.

\subsection{Dipole approximation: Gordon--Volkov states}
\label{sec:projdyn-volkov}
For the  field \pref{eq:A-homog},
Equation~\pref{eq:Sch-chi} is solved by the Gordon--Volkov
wavefunction~\cite{Volkov1935,Joachain2012}.
Its building block is the homogeneous Volkov coupling
\[
\zeta_{\vb k}^{\rm hom}
=
-\iu\,\vb k\cdot\alphav
\]
with the complex quiver amplitude \pref{eq:quiver}.
Writing
\[
\zeta_{\vb k}^{\rm hom}
=
|\zeta_{\vb k}^{\rm hom}|
e^{\iu(\theta_{\vb k}^{\rm hom}-\pi/2)},
\]
the linear Volkov phase~\footnote{The lower bound
of the time integral $\protect\int^t \cos(\wL t'+\varphi)\dd t'$ is here chosen
so as to eliminate a global, $\vb k$-dependent constant phase
$\propto\sin\varphi$; this is conventional and irrelevant for the
cross section. Different choices appear in the literature without affecting
any observable.} is
\begin{align}
\Theta_{\vb k}^{\rm hom}(t)
&=
\mathrm{Re}
\left[
\zeta_{\vb k}^{\rm hom}
e^{\iu(\omega_{\rm L}t+\varphi)}
\right]
= \\ &=
|\zeta_{\vb k}^{\rm hom}|
\sin(\omega_{\rm L}t+\varphi+\theta_{\vb k}^{\rm hom}).
\end{align}
The Schr\"odinger equation is solved by
\begin{equation}
\chi_{\vb k}(\vb r,t)
=
e^{\iu\vb k\cdot\vb r}
\exp
\left[
-\frac{\iu}{\hbar}\epsilon^{\rm V}(\vb k)t
-\iu\Theta_{\vb k}^{\rm hom}(t)
-\iu\Theta_{\vb k}^{(2)}(t)
\right].
\label{eq:GV}
\end{equation}
Here
\[
\epsilon^{\rm V}(\vb k)
=
\frac{\hbar^2k^2}{2m_{\rm e}}
+
U_{\rm p}
\]
is the Volkov quasienergy. The last term in \pref{eq:GV} is the
oscillatory part generated by \(A^2\). For a general elliptically
polarized monochromatic field,
\begin{align}
\Theta_{\vb k}^{(2)}(t)
&=
\mathrm{Re}
\left[
\zeta_{2}^{\rm hom}
e^{2\iu(\omega_{\rm L}t+\varphi)}
\right],\\
\zeta_{2}^{\rm hom}
&=
-\frac{\iu q_{\rm p}^2}{8m_{\rm e}c^2\hbar\omega_{\rm L}}\,
\vb A_0^{(+)}\cdot\vb A_0^{(+)} .
\end{align}
For linear polarization this reduces to
\[
\Theta_{\vb k}^{(2)}(t)
=
\frac{U_{\rm p}}{2\hbar\omega_{\rm L}}
\sin(2\omega_{\rm L}t+2\varphi),
\]
whereas for circular polarization
\(\vb A_0^{(+)}\cdot\vb A_0^{(+)}=0\), so the second-harmonic
oscillation vanishes and \(A^2\) contributes only to \(U_{\rm p}\).

The Floquet--Fourier components \pref{eq:chi-q}
 follow from \pref{eq:GV} by the Jacobi--Anger expansion. In the
\(U_{\rm p}\ll\hbar\omega_{\rm L}\) regime, the second-harmonic dressing generated by
\(\Theta_{\vb k}^{(2)}\) is quantitatively small. Retaining only the
linear Volkov phase gives
\begin{equation}
\chi_{\vb k}^{(n)}(\vb q)
=
(2\pi)^{3/2}
\delta^{(3)}(\vb q-\vb k)\,
J_n(|\zeta_{\vb k}^{\rm hom}|)
e^{-\iu n(\varphi+\theta_{\vb k}^{\rm hom})}.
\label{eq:Volkov-FF}
\end{equation}
The projectile Floquet states are therefore plane waves dressed by a
comb of Bessel-function sidebands. For linearly polarized light,
the complex quiver vector may be chosen real, \(\alphav \to \vb*\alpha_0\) , and
\(\zeta_{\vb k}^{\rm hom}=-\iu\,\vb k\cdot\boldsymbol{\alpha}_0\).
With the branch convention above,
\begin{equation}
\label{eq:elliptic-to-linear}
J_n(|\zeta_{\vb k}^{\rm hom}|)
e^{-\iu n\theta_{\vb k}^{\rm hom}}
=
J_n(\vb k\cdot\boldsymbol{\alpha}_0),
\end{equation}
so that \pref{eq:Volkov-FF} reduces to the familiar linear-polarization
form
\[
\chi_{\vb k}^{(n)}(\vb q)
=
(2\pi)^{3/2}
\delta^{(3)}(\vb q-\vb k)\,
J_n(\vb k\cdot\boldsymbol{\alpha}_0)
e^{-\iu n\varphi}.
\]

\subsection{Slowly varying inhomogeneous field: eikonal--Volkov states}
\label{sec:projdyn-eikonal}
The Gordon--Volkov construction of \secref{sec:projdyn-volkov} assumes a
strictly homogeneous vector potential. A useful intermediate regime
relaxes this to a spatially modulated envelope $\Ap(\vb r)$ in
\pref{eq:Acos} that varies on a length scale $L$ much larger than the
projectile de Broglie wavelength,
\begin{equation}
k L \gg 1.
\label{eq:eikonal-validity}
\end{equation}
Here field inhomogeneity refers to any spatial dependence of
\(\vb A^{(+)}(\vb r)\), including the real-space optical phase of a
propagating wave and spatial beam profiles.
This is the standard semi-classical (eikonal) regime: the projectile
follows essentially classical free-electron trajectories
$\vb r(s) = \vb r - \vb v_{\vb k}(t-s)$ with velocity
$\vb v_{\vb k} = \hbar\vb k/\me$, dressed by a slowly varying phase that
integrates the field along the trajectory.
Trajectory bending by $\nabla\vb A$ is neglected at this level.
\appref{app:trajectory-bending} shows that for typical UTEM/PINEM
parameters the trajectory-geometry inequalities are easily satisfied.
Bending also causes eikonal phase mismatch leading to a
phase-fidelity condition $\eta_{\rm def}^2\,kL\ll 1$. The
straight-line eikonal--Volkov kernel below should accordingly be
read as a controlled approximation, adequate for sideband intensities and
loss-spectrum peak positions, and for phase-coherent homodyne
interference between projectile and target.

We adopt the eikonal ansatz
\begin{equation}
\chi_{\vb k}(\vb r,t) = e^{\iu\vb k\cdot\vb r}\,
e^{-\iu \epsilon(\vb k) t/\hbar}\,e^{-\iu\Theta_{\vb k}(\vb r,t)},
\label{eq:eikonal-ansatz}
\end{equation}
with $\Theta(\vb r,t)$ slowly varying in $\vb r$ on the scale $1/k$.
Substituting into \pref{eq:Sch-chi} and dropping (i)~the
eikonal-subleading \( |\nabla\Theta|^2 \) correction to the kinetic
energy and (ii)~the \( \Up\ll\hbar\wL \)-suppressed \( \vb A^2 \) term,
one obtains the linear eikonal equation
\begin{equation}
\partial_t \Theta_{\vb k}(\vb r,t)
+
\vb v_{\vb k}\cdot\nabla\Theta_{\vb k}(\vb r,t)
=
-\frac{q_{\mathrm p}}{\me c}\,
\vb k\cdot\vb A(\vb r,t),
\label{eq:eikonal-eq}
\end{equation}
where \(\vb v_{\vb k}=\hbar\vb k/\me\). 
The solution with the adiabatic boundary condition
\(\Theta_{\vb k}\to0\) on the incoming asymptote can be written as
\begin{equation}
\Theta_{\vb k}(\vb r,t)
\coloneqq
\mathrm{Re}
\left[
\zeta_{\vb k}(\vb r)
e^{\iu(\wL t+\varphi)}
\right],
\label{eq:eikonal-Theta}
\end{equation}
with the complex eikonal coupling parameter
\begin{equation}
\zeta_{\vb k}(\vb r)
\coloneqq
-\frac{q_{\mathrm p}}{\me c}\,
\vb k\cdot
\int_0^\infty
\dd s\,
\Ap(\vb r-\vb v_{\vb k}s)
e^{-\iu\wL s},
\label{eq:eikonal-zeta}
\end{equation}
which employs the complex slowly-varying field amplitude $\Ap(\vb r)$
defined in \pref{eq:Acos}.
The integral in \pref{eq:eikonal-zeta} is the
trajectory-resolved positive-frequency Fourier amplitude of the optical
vector potential sampled by the projectile. Thus
\(|\zeta_{\vb k}(\vb r)|\) generalizes the homogeneous Volkov coupling
\(|\zeta_{\vb k}^{\rm hom}|\) to a position-dependent complex amplitude.
For a homogeneous field,
\(\Ap(\vb r)\to\vb A_0^{(+)}\), one obtains
\[
\zeta_{\vb k}(\vb r)
\to
-\iu\,
\vb k\cdot\alphav,
\]
in agreement with the homogeneous coupling
\(\zeta_{\vb k}^{\rm hom}\) introduced in \secref{sec:projdyn-volkov}.

\paragraph*{Outgoing eikonal--Volkov state.}
For a localized external laser field the scattering matrix also requires
an \emph{outgoing} channel state that reduces to a field-free plane wave
\emph{downstream} of the interaction region. Its eikonal phase satisfies
the same transport equation~\pref{eq:eikonal-eq} but with the boundary
condition imposed at $+\infty$ instead of $-\infty$.
The outgoing eikonal coupling parameter is
\begin{equation}
\zeta_{\vb k}^{(-)}(\vb r)
\;\coloneqq\;
\frac{q_{\mathrm p}}{\me c}\,
\vb k\cdot
\int_0^\infty
\dd s\,
\Ap(\vb r+\vb v_{\vb k}s)\,
e^{+\iu\wL s},
\label{eq:eikonal-zeta-out}
\end{equation}
which integrates the field along the \emph{future} straight-line
trajectory from $\vb r$, with the conjugate frequency $e^{+\iu\wL s}$
and an overall sign opposite to that of \pref{eq:eikonal-zeta},
inherited from the outgoing boundary condition.
Indeed, inserting
\[
\Theta_{\vb k}^{(-)}(\vb r,t)
=
\mathrm{Re}\!\left[
\zeta_{\vb k}^{(-)}(\vb r)e^{\iu(\wL t+\varphi)}
\right]
\]
into \(\pref{eq:eikonal-eq}\) gives
\[
\left(\iu\wL+\vb v_{\vb k}\cdot\nabla\right)
\zeta_{\vb k}^{(-)}(\vb r)
=
-\frac{q_{\mathrm p}}{\me c}\,
\vb k\cdot\Ap(\vb r).
\]
Hence the relative sign in \(\pref{eq:eikonal-zeta-out}\) follows from
differentiating the upper-boundary solution along the future trajectory.
The downstream boundary condition
\(\zeta_{\vb k}^{(-)}\to0\) is therefore satisfied for a localized field.
If the envelope varies little over the optical length
\(v_{\vb k}/\wL\) (SVE limit), then
\begin{align*}
\zeta_{\vb k}^{(-)}(\vb r)
&\simeq
\frac{q_{\mathrm p}}{\me c}\,
\vb k\cdot\Ap(\vb r)
\int_0^\infty\!\dd s\,e^{+\iu\wL s-0^+s}\\
&=
-\iu\,\vb k\cdot\alphav(\vb r),
\end{align*}
the same local Volkov value as obtained from
\(\zeta_{\vb k}^{(+)}\) of \pref{eq:eikonal-zeta}.

The relativistic straight-trajectory form corresponding to
\pref{eq:eikonal-zeta} is
\begin{subequations}
\begin{align}
\zeta_{\vb k}^{\rm rel}(\vb r)
&=
-\frac{q_{\rm p}}{\hbar c}\,
\vb v_{\vb k}\cdot
\int_0^\infty\!\dd s\,
\Ap(\vb r-\vb v_{\vb k}s)e^{-\iu\wL s},\\
\vb v_{\vb k}
&\coloneqq
\frac{\hbar\vb k}{\gamma_{\vb k}m_{\rm e}},
\end{align}\label{eq:eikonal-zeta-rel}%
\end{subequations}%
which is the relativistic-velocity envelope prescription established
for PINEM in Ref.~\cite{ParkZewail2012}. An analogous correction
applies to the outgoing solution.

A position-dependent Jacobi--Anger expansion delivers the
Floquet--Fourier components of the eikonal--Volkov state. Writing
$\zeta_{\vb k}(\vb r) = |\zeta_{\vb k}(\vb r)|\,
e^{\iu[\theta_{\vb k}(\vb r) - \pi/2]}$, so that
$\Theta(\vb r,t) = |\zeta_{\vb k}(\vb r)|\sin\!\bigl(\wL t + \varphi +
\theta_{\vb k}(\vb r)\bigr)$, one obtains
\begin{equation}
\chi_{\vb k}^{(n)}(\vb r) \coloneqq e^{\iu\vb k\cdot\vb r}\,
J_n\!\bigl(|\zeta_{\vb k}(\vb r)|\bigr)\,
e^{-\iu n[\varphi+\theta_{\vb k}(\vb r)]},
\label{eq:EV-FF-r}
\end{equation}
and in plane-wave variables \pref{eq:chi-q}
\begin{multline}
\chi_{\vb k}^{(n)}(\vb q) \coloneqq \!\int\!\frac{\dd^3\vb r}{(2\pi)^{3/2}}\,
e^{-\iu(\vb q-\vb k)\cdot\vb r}\\
\times J_n\!\bigl(|\zeta_{\vb k}(\vb r)|\bigr)\,
e^{-\iu n[\varphi+\theta_{\vb k}(\vb r)]}.
\label{eq:EV-FF-q}
\end{multline}
Equation~\pref{eq:EV-FF-q} is no longer sharply localized on $\vb q=\vb k$. Its momentum support is determined by
the trajectory phase and the specified beam or interaction window; a width
of order $1/L$ is the local-SVE estimate at fixed moderate dressing
(\appref{app:validity}); the homogeneous result \pref{eq:Volkov-FF}
is recovered as $L\to\infty$. The Floquet coefficients are derived equally
 for the incoming and outgoing states; the $\zeta_{\vb k}(\vb r), \Theta(\vb r,t)$ and
$\theta_{\vb k}(\vb r)$ inherit superscripts (+) and (--), respectively.

\section{Eikonal--Volkov scattering in spatially structured fields}
\label{sec:dsigma-inhom}
We apply the incoming and outgoing eikonal--Volkov states to
calculate the scattering amplitude in the first Born approximation,
implementing the eikonal reduction \pref{eq:proj-kernel-eikonal}
of \secref{sec:exact-kernel}.
The incoming projectile state \(\chi_{\vb k_i}^{(+)}\) is chosen to be
field-free on the incoming asymptote, while the outgoing projectile
state \(\chi_{\vb k_f}^{(-)}\) is chosen to be field-free on the
outgoing asymptote.

\subsection{Eikonal--Volkov projectile kernel}
\label{sec:projdyn-eikonal-kernel}

This subsection evaluates the exact projectile kernel
\pref{eq:proj-kernel} for the eikonal--Volkov states of
\secref{sec:projdyn-eikonal}.
We work in real space because the eikonal--Volkov states are most
naturally expressed there [Eqs.~\pref{eq:EV-FF-r}--\pref{eq:EV-FF-q}].
The two-sided construction uses both the incoming state
\(\chi_{\vb k_i}^{(+)}\) (field-free upstream, coupling
\(\zeta_{\vb k_i}^{(+)}\) of \pref{eq:eikonal-zeta}) and the outgoing
state \(\chi_{\vb k_f}^{(-)}\) (field-free downstream, coupling
\(\zeta_{\vb k_f}^{(-)}\) of \pref{eq:eikonal-zeta-out}),
consistently with the DWBA~\cite{Taylor2006} with the optical
interaction as the reference dynamics.

\paragraph*{Reduction of \pref{eq:proj-kernel} to a real-space product.}
Starting from the definition \pref{eq:proj-kernel},
\[
\mathcal P^{(n)}_{\vb k_f\vb k_i}(\vb q)
=
\sum_{n_i}\int\!\dd^3\vb q''\,
\chi_{\vb k_f}^{(n_i+n)*}(\vb q''+\vb q)\,
\chi_{\vb k_i}^{(n_i)}(\vb q''),
\]
we substitute the Fourier representations \pref{eq:EV-FF-q} for both
states and exchange the order of integration.
The \(\vb q''\) integral is
\(
\int\!\dd^3\vb q''\,
e^{\iu\vb q''\cdot(\vb r-\vb r')}
=
(2\pi)^3\delta^{(3)}(\vb r-\vb r'),
\)
which collapses the double position integral to a single one:
\begin{equation}
\mathcal P^{(n)}_{\vb k_f\vb k_i}(\vb q)
=
\sum_{n_i}
\int\!\dd^3\vb r\,
e^{\iu\vb q\cdot\vb r}\,
\chi_{\vb k_f}^{(n_i+n)*}(\vb r)\,
\chi_{\vb k_i}^{(n_i)}(\vb r).
\label{eq:proj-kernel-realspace}
\end{equation}
This form is the natural starting point because the eikonal–Volkov 
states are known explicitly in real space.

\paragraph*{Eikonal--Volkov kernel: definition and identification.}
The Floquet components of the incoming and outgoing states follow from
\pref{eq:EV-FF-r} applied with the respective couplings
\(\zeta^{(+)}\) and \(\zeta^{(-)}\):
\begin{equation}
\chi_{\vb k_{i/f}}^{(\pm)(\ell)}(\vb r)
=
e^{\iu\vb k_{i/f}\cdot\vb r}\,
J_{\ell}\!\left(|\zeta_{\vb k_{i/f}}^{(\pm)}(\vb r)|\right)
e^{-\iu\ell[\varphi+\theta_{\vb k_{i/f}}^{(\pm)}(\vb r)]},
\label{eq:EV-FF}
\end{equation}
where \(\zeta_{\vb k}^{(\pm)} = |\zeta_{\vb k}^{(\pm)}|
e^{\iu[\theta_{\vb k}^{(\pm)}-\pi/2]}\).
Substituting \pref{eq:EV-FF} into \pref{eq:proj-kernel-realspace}, the
plane-wave factors combine as \(e^{-\iu\vb k_f\cdot\vb r}
e^{\iu\vb k_i\cdot\vb r} = e^{-\iu\Qv\cdot\vb r}\)
with \(\Qv = \vb k_f - \vb k_i\).
It is convenient to absorb the remaining \(e^{-\iu\Qv\cdot\vb r}\)
into a local real-space kernel:
\begin{equation}
\mathcal F_{\vb k_f\vb k_i}^{(n)}(\vb r)
\;\coloneqq\;
e^{\iu\Qv\cdot\vb r}
\sum_{\ell=-\infty}^{\infty}
\chi_{\vb k_f}^{(-)({\ell+n})\,*}(\vb r)\,
\chi_{\vb k_i}^{(+)(\ell)}(\vb r).
\label{eq:DWBA-kernel-components}
\end{equation}
The sum over \(\ell\) is the eikonal--Volkov counterpart of the
internal sideband sum \(n_i\) in \pref{eq:proj-kernel}, and the total
Floquet index partitions as \(N=n+(N-n)\) between the projectile
kernel and the target density component.
With \pref{eq:DWBA-kernel-components}, Eq.~\pref{eq:proj-kernel-realspace}
becomes
\begin{equation}
\mathcal P^{(n)}_{\vb k_f\vb k_i}(\vb q)
\, \rightarrow\,
\int\!\dd^3\vb r\,
e^{-\iu(\Qv-\vb q)\cdot\vb r}\,
\mathcal F^{(n)}_{\vb k_f\vb k_i}(\vb r).
\label{eq:proj-kernel-Fintegral}
\end{equation}
The right-hand side is a Fourier transform.
Defining the momentum-space kernel
\begin{equation}
\tilde{\mathcal F}^{(n)}_{\vb k_f\vb k_i}(\vb K)
\;\coloneqq\;
\int\!\frac{\dd^3\vb r}{(2\pi)^3}\,
e^{-\iu\vb K\cdot\vb r}\,
\mathcal F^{(n)}_{\vb k_f\vb k_i}(\vb r),
\label{eq:Fkernel}
\end{equation}
Equation~\pref{eq:proj-kernel-Fintegral} at \(\vb K=\Qv-\vb q\) gives the
eikonal--Volkov realization of \pref{eq:proj-kernel}:
\begin{equation}
\Pbar^{(n)}_{\vb k_f\vb k_i}(\vb q)\equiv
\frac{1}{(2\pi)^3}\,
\mathcal P^{(n)}_{\vb k_f\vb k_i}(\vb q)
\, \rightarrow\,
\tilde{\mathcal F}^{(n)}_{\vb k_f\vb k_i}(\Qv-\vb q).
\label{eq:proj-kernel-eikonal}
\end{equation}
This establishes step~(i) of the reduction hierarchy of
\secref{sec:exact-kernel}: the argument \(\Qv-\vb q\) is the momentum
mismatch between the kinematic transfer \(\Qv\) and the momentum
\(\vb q\) received by the target.
Instead of an identification we write $\rightarrow$ to emphasize the
difference between incoming and outgoing asymptotics.
We now use \pref{eq:DWBA-kernel-components} to reduce the \(\ell\)-sum
to a local single-Bessel form.

\paragraph*{Where Graf's theorem enters.}
Substituting \pref{eq:EV-FF} into
\pref{eq:DWBA-kernel-components} gives
\begin{multline}
\mathcal F_{\vb k_f\vb k_i}^{(n)}(\vb r)
=
e^{\iu n[\varphi+\theta_{\vb k_f}^{(-)}(\vb r)]}
\sum_{\ell=-\infty}^{\infty}
J_{\ell}\!\left(
|\zeta_{\vb k_i}^{(+)}(\vb r)|
\right)
\\
\times
J_{\ell+n}\!\left(
|\zeta_{\vb k_f}^{(-)}(\vb r)|
\right)
e^{\iu\ell[
\theta_{\vb k_f}^{(-)}(\vb r)
-
\theta_{\vb k_i}^{(+)}(\vb r)]}.
\label{eq:DWBA-two-Bessel-sum}
\end{multline}
This is the local projectile-sideband convolution. Graf's addition
theorem~\cite{DLMF,Watson1944},
\begin{equation}
\sum_{\ell=-\infty}^{\infty}
J_{\ell}(a)\,
J_{\ell+n}(b)\,
e^{\iu\ell\phi}
=
J_n(c)\,e^{\iu n\psi},
\label{eq:Graf}
\end{equation}
with
\begin{equation}
c\coloneqq
\bigl|b-ae^{-\iu\phi}\bigr|,
\qquad
\psi\coloneqq
\arg\!\bigl(b-ae^{-\iu\phi}\bigr),
\label{eq:Graf-c-psi}
\end{equation}
evaluates the sum in \pref{eq:DWBA-two-Bessel-sum}.

For the present kernel,
\[
a(\vb r)
=
|\zeta_{\vb k_i}^{(+)}(\vb r)|,
\qquad
b(\vb r)
=
|\zeta_{\vb k_f}^{(-)}(\vb r)|,
\]
and
\[
\Delta\theta(\vb r)
=
\theta_{\vb k_f}^{(-)}(\vb r)
-
\theta_{\vb k_i}^{(+)}(\vb r).
\]
Using
\[
\zeta_{\vb k}^{(\pm)}(\vb r)
=
|\zeta_{\vb k}^{(\pm)}(\vb r)|
e^{\iu[\theta_{\vb k}^{(\pm)}(\vb r)-\pi/2]},
\]
the complex number on the right-hand side of Graf's theorem becomes
\begin{align*}
b(\vb r)-a(\vb r)e^{-\iu\Delta\theta(\vb r)}
&=
|\zeta_{\vb k_f}^{(-)}(\vb r)|
\\
&\quad-
|\zeta_{\vb k_i}^{(+)}(\vb r)|
e^{\iu[
\theta_{\vb k_i}^{(+)}(\vb r)
-
\theta_{\vb k_f}^{(-)}(\vb r)]}
\\
&=
e^{-\iu[
\theta_{\vb k_f}^{(-)}(\vb r)-\pi/2]}
\left[
\zeta_{\vb k_f}^{(-)}(\vb r)
-
\zeta_{\vb k_i}^{(+)}(\vb r)
\right].
\end{align*}
Consequently,
\[
c(\vb r)
=
\left|
b(\vb r)-a(\vb r)e^{-\iu\Delta\theta(\vb r)}
\right|
=
\left|
\zeta_{\vb k_f}^{(-)}(\vb r)
-
\zeta_{\vb k_i}^{(+)}(\vb r)
\right|,
\]
while
\[
\psi(\vb r)
=
\arg\!\left[
\zeta_{\vb k_f}^{(-)}(\vb r)
-
\zeta_{\vb k_i}^{(+)}(\vb r)
\right]
-
\left[
\theta_{\vb k_f}^{(-)}(\vb r)-\frac{\pi}{2}
\right].
\]

It is therefore convenient to define
\[
\delta\zeta_{fi}^{\rm DWBA}(\vb r)
\coloneqq
\zeta_{\vb k_f}^{(-)}(\vb r)
-
\zeta_{\vb k_i}^{(+)}(\vb r).
\]
In the branch convention used here, Graf's theorem reduces the
projectile kernel to
\begin{equation}
\mathcal F_{\vb k_f\vb k_i}^{(n)}(\vb r)
=
J_n\!\left(
|\delta\zeta_{fi}^{\rm DWBA}(\vb r)|
\right)
e^{\iu n[
\varphi+\theta_{fi}^{\rm DWBA}(\vb r)]},
\label{eq:Graf-DWBA}
\end{equation}
where
\begin{equation}
\theta_{fi}^{\rm DWBA}(\vb r)
\coloneqq
\arg\!\left[
\delta\zeta_{fi}^{\rm DWBA}(\vb r)
\right]
+\frac{\pi}{2}.
\end{equation}
Thus the two-sided projectile dressing is controlled by the difference
between the outgoing and incoming eikonal couplings, rather than by
either coupling separately.

The branch of the argument is immaterial up to integer multiples of
\(2\pi\). In the linearly polarized limit, where the real quantity
\(\Qv\cdot\boldsymbol{\alpha}_0(\vb r)\) may cross zero, the sign change
of the Bessel argument is equivalently represented by the phase jump
\(e^{\iu n\pi}=(-1)^n\). The product
\[
J_n\!\left(
|\delta\zeta_{fi}^{\rm DWBA}|
\right)
e^{\iu n\theta_{fi}^{\rm DWBA}}
\]
therefore remains the continuous signed Bessel factor.

In the slowly varying envelope (SVE) limit, both eikonal couplings
reduce locally to the Volkov value,
\[
\zeta_{\vb k}^{(\pm)}(\vb r)
\longrightarrow
-\iu\,\vb k\cdot\alphav(\vb r),
\]
and hence
\begin{multline}
\delta\zeta_{fi}^{\rm DWBA}(\vb r)
\xrightarrow{\mathrm{SVE}}
\\
-\iu(\vb k_f-\vb k_i)\cdot\alphav(\vb r)
=
-\iu\Qv\cdot\alphav(\vb r).
\label{eq:dzeta-SVE}
\end{multline}
For a strictly homogeneous field this gives
\begin{equation}
\mathcal F_{\vb k_f\vb k_i}^{(n)}(\vb r)
\longrightarrow
J_n\!\left(
|\delta\zeta_{\Qv}^{\rm hom}|
\right)
e^{\iu n(
\varphi+\theta_{\Qv}^{\rm hom})},
\label{eq:Bessel-inhom-homlimit}
\end{equation}
where $\delta\zeta_{\Qv}^{\rm hom}=-\iu\,\Qv\cdot\alphav$ is the homogeneous
coupling, defined with the momentum transfer $\Qv$ in
\pref{eq:Q}--\pref{eq:dzetahom}.
In the linearly polarized limit, \(\alphav\) may be chosen real, and
the branch of \(\theta_{\Qv}^{\rm hom}\) supplies the sign of
\(\Qv\cdot\boldsymbol{\alpha}_0\):
\begin{equation}
J_n\!\left(
|\delta\zeta_{\Qv}^{\rm hom}|
\right)
e^{\iu n\theta_{\Qv}^{\rm hom}}
=
J_n(\Qv\cdot\boldsymbol{\alpha}_0).
\label{eq:Bessel-linear-limit}
\end{equation}
Thus the familiar linearly polarized factor
\(e^{\iu n\varphi}
J_n(\Qv\cdot\boldsymbol{\alpha}_0)\)
is recovered as a special case.
In the SVE and homogeneous limits, the two-sided DWBA coupling
\(\delta\zeta_{fi}^{\rm DWBA}(\vb r)\) reduces to the single Volkov
coupling \(-\iu\Qv\cdot\alphav\) determined entirely by the
transferred momentum~\(\Qv\).
This is the direct bridge to \secref{sec:application-Bessel}, where the
homogeneous specialization of the cross section is derived without
further appeal to Graf's theorem.

\subsection{Eikonal cross section from the bi-momentum response}
\label{sec:bimomentum-sff}

The previous subsection supplies the eikonal--Volkov realization of the
general projectile kernel.
Inserting it into the bi-momentum master formula
\pref{eq:dsigma-bimomentum} gives formally
\begin{multline}
\frac{\dd^2\sigma}{\dd\Omega\dd\epsilon} =
\qty(\frac{\me}{2\pi\hbar^2})^{2}\!\sum_N\frac{k_f}{k_i}
\sum_{n,n'}\!\int\!\dd^3\vb q'_1\,\dd^3\vb q'_2 \\
v(\vb q'_1)\,v^{*}(\vb q'_2)
\,\tilde{\mathcal F}^{(n)}_{\vb k_f\vb k_i}(\Qv-\vb q'_1)\,
\tilde{\mathcal F}^{(n')\,*}_{\vb k_f\vb k_i}(\Qv-\vb q'_2) \\
\times\SF^{(N-n)(N-n')}\!(\vb q'_1,\vb q'_2;\omega_N).
\label{eq:dsigma-inhom}
\end{multline}
Here $N$ is the net number of photons exchanged between projectile and
field, and
\begin{equation}
\hbar\omega_N \;\coloneqq\; \epsilon(\vb k_i)-\epsilon(\vb k_f)-N\hbar\wL
\label{eq:omegaN-def}
\end{equation}
is the target energy argument sampled in the $N$th sideband channel.
Equation~\pref{eq:dsigma-inhom} is the eikonal--Volkov
specialization of the exact first-Born formula \pref{eq:dsigma-bimomentum}.
All target-side information remains contained in the bi-momentum
structure factor defined in \pref{eq:SF-bimomentum}.

\paragraph*{Two limiting regimes.}
Equation~\pref{eq:dsigma-inhom} encompasses two physically distinct regimes,
labeled here  by the ratio between the field scale $L$ and the inverse kinematic
momentum transfer $1/Q$; this ratio is a conditional scaling shorthand
whose meaning requires the actual kernel width (\appref{app:validity}).

\emph{(i) Slowly varying envelope, $QL\gg 1$.}
For a specified sample and beam window, evaluating the Coulomb factors
and the target response at the kinematic momentum $\Qv$ is controlled
when the actual projectile-kernel displacements are small compared with
$Q$ and with the variation scales of the complete target factors. For a
translationally invariant target this leaves the harmonic-overlap
weights of \appref{app:validity}; the homogeneous Bessel products are
recovered in the regulated homogeneous limit.
The local Volkov form $\zeta^{(\pm)}_{\vb k}\simeq-\iu\vb k\cdot
\alphav(\vb r)$ of this regime additionally requires slow variation over
the optical travel length, $v/(\wL L_\parallel)\ll1$; $QL$ is a useful
comparison only after the actual kernel width has been established,
which grows with the dressing strength.
Quantitative error estimates are given in \appref{app:validity}.

\emph{(ii) Localized-beam regime, $QL\lesssim 1$.}
For the optical field localized on length $L$ comparable to or smaller
than the inverse kinematic momentum transfer, the kernel
$\tilde{\mathcal F}^{(n)}_{\vb k_f\vb k_i}(\vb K)$ has transverse
support
$\sim 1/L\gtrsim Q$, overlapping with the kinematic point $\vb K=0$ and
neighboring sectors alike. Along the trajectory the two-sided kernel
is not localized: for straight paths of equal velocity
$\delta\zeta_{fi}(z)=-(C/v)\,e^{-\iu\wL z/v}\int\!\dd z'\,A(z')e^{\iu\wL z'/v}$
(with $C$ the coupling prefactor of \pref{eq:eikonal-zeta} and $A$ the
projected field profile), so that $\mathcal F^{(n)}\propto
e^{-\iu n\wL z/v}$ and $\tilde{\mathcal F}^{(n)}$ is centered at
$K_z=-n\wL/v$, i.e.\ $q_z=Q_z+n\wL/v$: the localized field supplies the
longitudinal momentum of the $n$ exchanged quanta. The projectile then samples a broad range
of momentum transfers, and the cross section acquires an explicit
dependence on the spatial profile of the laser field through
$\zeta_{\vb k}(\vb r)$. For a translationally invariant target this broad
projectile sampling still collapses according to
\pref{eq:bimomentum-translation}; genuine off-diagonal target response
requires the target-side symmetry breaking discussed in
\secref{sec:exact-kernel}.

Figure~\ref{fig:limits} summarizes the structural reductions of the
general formula; the field-profile assumptions of the eikonal branch are
discussed above and in \appref{app:validity}.

\input{figures/fig_limits}

\section{Homogeneous-field cross section and the Floquet dynamic structure factor}
\label{sec:applications}
We now specialize the general inhomogeneous-field cross section
to the strictly homogeneous-field dipole limit of
\secref{sec:projdyn-volkov}, in which momentum integrals drop out and
a transparent Bethe-theoretic interpretation is
available.

\subsection{Bessel-comb form of the projectile kernel}
\label{sec:application-Bessel}

For a homogeneous field, $\alphav(\vb r)\to\alphav$ (constant), so the
real-space eikonal--Volkov kernel \pref{eq:Graf-DWBA} becomes
translationally invariant. Introducing the notation
\begin{equation}
\Qv \;\coloneqq\; \vb k_f - \vb k_i
\label{eq:Q}
\end{equation}
and using the homogeneous limit of the two-sided coupling from
\pref{eq:dzeta-SVE},
\begin{equation}
\delta\zeta_{\Qv}^{\rm hom}
\;\coloneqq\;
-\iu\,\Qv\cdot\alphav,
\label{eq:dzetahom}
\end{equation}
the kernel \pref{eq:Graf-DWBA} reduces to the spatially constant
\begin{equation}
\mathcal F^{(n)}_{\vb k_f\vb k_i}(\vb r)
\;\longrightarrow\;
J_n\!\left(|\delta\zeta_{\Qv}^{\rm hom}|\right)\,
e^{\iu n(\varphi+\theta_{\Qv}^{\rm hom})}.
\label{eq:F-hom}
\end{equation}
Since \pref{eq:F-hom} is independent of $\vb r$, its Fourier
transform \pref{eq:Fkernel} is proportional to $\delta^{(3)}(\vb K)$:
\begin{equation}
\tilde{\mathcal F}^{(n)}_{\vb k_f\vb k_i}(\vb K)
\;=\;
J_n\!\left(|\delta\zeta_{\Qv}^{\rm hom}|\right)\,
e^{\iu n(\varphi+\theta_{\Qv}^{\rm hom})}\,
\delta^{(3)}(\vb K).
\label{eq:F-hom-FT}
\end{equation}
Evaluating the eikonal identification \pref{eq:proj-kernel-eikonal}
at $\vb K = \Qv-\vb q$ gives
\begin{equation}
\Pbar^{(n)}_{\vb k_f\vb k_i}(\vb q)
\;=\;
J_n\!\left(|\delta\zeta_{\Qv}^{\rm hom}|\right)\,
e^{\iu n(\varphi+\theta_{\Qv}^{\rm hom})}\,
\delta^{(3)}(\vb q-\Qv).
\label{eq:P-hom}
\end{equation}
Substituting \pref{eq:P-hom} into the factored amplitude
\pref{eq:MN-factored} and performing the $\vb q'$-integration
collapses the momentum convolution to the single transfer $\Qv$:
\begin{equation}
\mathcal M_N =
v(\Qv)\sum_n
J_n\!\left(|\delta\zeta_{\Qv}^{\rm hom}|\right)\,
e^{\iu n(\varphi+\theta_{\Qv}^{\rm hom})}\,
\tilde\varrho_{fi}^{(N-n)}(\Qv).
\label{eq:M-hom}
\end{equation}
Inserting \pref{eq:M-hom} into the cross section formula \pref{eq:dsigma}
gives the homogeneous-field result
\begin{multline}
\frac{\dd^2 \sigma}{\dd \Omega \dd\epsilon}(\Phi_f \leftarrow \Phi_i)
=
\Bigl(\frac{\me}{2\pi\hbar^2}\Bigr)^{\!2}
\sum_N \frac{k_f}{k_i}\,
|v(\Qv)|^{2}\,\delta(\Delta E_N)
\\
\times
\Bigl|
\sum_n
J_n\!\left(|\delta\zeta_{\Qv}^{\rm hom}|\right)
e^{\iu n(\varphi+\theta_{\Qv}^{\rm hom})}
\tilde\varrho_{fi}^{(N-n)}(\Qv)
\Bigr|^{2},
\label{eq:dsigma-Volkov}
\end{multline}
with
\[
\Delta E_N
\coloneqq
\epsilon(\vb{k}_f)-\epsilon(\vb{k}_i)+E_f-E_i+N\hbar\wL.
\]
For linearly polarized light, $\alphav$ may be chosen real;
Equation~\pref{eq:Bessel-linear-limit} then gives
$J_n(|\delta\zeta_{\Qv}^{\rm hom}|)e^{\iu n\theta_{\Qv}^{\rm hom}}
= J_n(\Qv\cdot\boldsymbol\alpha_0)$,
recovering the familiar factor $e^{\iu n\varphi}J_n(\Qv\cdot\boldsymbol\alpha_0)$.
Equation~\pref{eq:dsigma-Volkov} gives the
\(\lvert S^\text{1B}\rvert^2\) contribution.  For \(f=i\), a complete
target-elastic measurement additionally contains the
\(S^\text{0B}\)--\(S^\text{1B}\) interference of
\secref{sec:1Born}; beyond weak coupling, the repeated coherent
mean-field insertions are resummed as described in
\secref{sec:pinem-bridge}.

\subsection{The Floquet dynamic structure factor}
\label{sec:application-SF}
The homogeneous-field cross section is controlled by the single-momentum
Floquet dynamic structure factor. It is the momentum-diagonal counterpart
of the bi-momentum Floquet structure factor of \pref{eq:SF-bimomentum}:
\begin{multline}
S_{\mathrm F}^{nn'}(\Qv,\omega) \coloneqq \sum_f
\tilde\varrho_{fi}^{(n)}(\Qv)\,\tilde\varrho_{fi}^{(n')\,*}(\Qv)\\
\times\delta(\hbar\omega + E_i - E_f).
\label{eq:SF}
\end{multline}
Equation~\pref{eq:SF} defines the \emph{full} structure factor.  To
make its relation to the fluctuation spectrum explicit, use
\(R_n(\Qv)=\tilde\varrho_{ii}^{(n)}(\Qv)\) from
\pref{eq:R-mean-density} and introduce
\(\delta\tilde\varrho_{fi}^{(n)}
\coloneqq\tilde\varrho_{fi}^{(n)}-\delta_{fi}R_n\).  For the pure
initial Floquet state considered here,
\[
S_{\mathrm F}^{nn'}(\Qv,\omega)
=R_n(\Qv)R_{n'}^*(\Qv)\delta(\hbar\omega)
+S_{\mathrm{F,c}}^{nn'}(\Qv,\omega),
\]
where
\begin{multline*}
S_{\mathrm{F,c}}^{nn'}(\Qv,\omega)
\coloneqq
\sum_f
\delta\tilde\varrho_{fi}^{(n)}(\Qv)
\delta\tilde\varrho_{fi}^{(n')\,*}(\Qv)\\
\times
\delta(\hbar\omega+E_i-E_f)
\end{multline*}
is equivalently the sum over \(f\ne i\).  Thus
\(S_{\mathrm{F,c}}\) is the connected fluctuation correlator and
contains only target-changing transitions; the omitted \(f=i\) term
is the coherent target-elastic outer product.  Both are Hermitian
matrices in the Floquet indices \((n,n')\).  Their diagonal components
are sums of squared \(n\)-photon matrix elements, while their
off-diagonal components encode interference between Floquet sectors.
For the full \(S_{\mathrm F}\), those off-diagonal components include
the coherent rank-one outer product when the mean density has several
harmonics.

Summing \pref{eq:dsigma-Volkov} over unresolved final states \(f\) of
the target gives the homogeneous Bethe--Floquet first-Born
cross section
\begin{multline}
\frac{\dd^2\sigma}{\dd\Omega\dd\epsilon}
=
\left(\frac{\me}{2\pi\hbar^2}\right)^2
|v(\Qv)|^2
\sum_N
\frac{k_f}{k_i}
\sum_{n,n'}
\\
\times
J_n\!\left(|\delta\zeta_{\Qv}^{\rm hom}|\right)
J_{n'}\!\left(|\delta\zeta_{\Qv}^{\rm hom}|\right)
e^{\iu(n-n')(\varphi+\theta_{\Qv}^{\rm hom})}
\\
\times
S_{\mathrm F}^{(N-n)(N-n')}
(\Qv,\omega_N).
\label{eq:dsigma-SF}
\end{multline}
Equation~\pref{eq:dsigma-SF} is correspondingly the \emph{full}
\(\lvert S^{\mathrm{1B}}\rvert^2\) cross section.  The separation
between its two target sectors is exact, not an additional
approximation.  In the compact notation
\begin{align*}
\mathcal C_{\Qv}
&\coloneqq
\left(\frac{\me}{2\pi\hbar^2}\right)^2
\frac{k_f}{k_i}|v(\Qv)|^2,\\
B_n(\Qv)
&\coloneqq
J_n\!\left(|\delta\zeta_{\Qv}^{\rm hom}|\right)
e^{\iu n(\varphi+\theta_{\Qv}^{\rm hom})},
\end{align*}
the \(N\)-th term of \pref{eq:dsigma-SF} reads
\[
\left.
\frac{\dd^2\sigma^{\mathrm{1B}}}{\dd\Omega\,\dd\epsilon}
\right|_N
=
\left.
\frac{\dd^2\sigma_{\mathrm{coh}}^{\mathrm{1B}}}{\dd\Omega\,\dd\epsilon}
\right|_N
+
\left.
\frac{\dd^2\sigma_{\mathrm c}}{\dd\Omega\,\dd\epsilon}
\right|_N ,
\]
with
\begin{multline*}
\left.
\frac{\dd^2\sigma_{\mathrm{coh}}^{\mathrm{1B}}}{\dd\Omega\,\dd\epsilon}
\right|_N
=
\mathcal C_{\Qv}\\
\times
\left|\sum_n B_n(\Qv)R_{N-n}(\Qv)\right|^2
\delta(\hbar\omega_N),
\end{multline*}
\begin{multline*}
\left.
\frac{\dd^2\sigma_{\mathrm c}}{\dd\Omega\,\dd\epsilon}
\right|_N
=
\mathcal C_{\Qv}
\sum_{n,n'}B_n(\Qv)B_{n'}^*(\Qv)\\
\times
S_{\mathrm{F,c}}^{(N-n)(N-n')}(\Qv,\omega_N).
\end{multline*}
The first line is target elastic and contains the square of the
coherent first-Born amplitude; the second is the connected
target-changing signal.  The two lines do not interfere because their
final target sectors are orthogonal.  A complete target-elastic
observable may nevertheless contain interference of the first line
with \(S^{\mathrm{0B}}\), as explained in \secref{sec:1Born} and
\secref{sec:pinem-bridge}.
The off-diagonal Floquet components produce interferences between
processes in which the projectile and the target exchange different
photon numbers but arrive at the same total sideband index $N$. For
elliptically polarized light, the projectile-side interference phase is
shifted by the polarization-dependent phase \(\theta_{\Qv}^{\rm hom}\),
so that the relative phase is \((n-n')(\varphi+\theta_{\Qv}^{\rm hom})\).
For linearly polarized light this reduces to the familiar phase
\((n-n')\varphi\), with the sign of \(\Qv\cdot\boldsymbol{\alpha}_0\)
carried by the Bessel functions. The closed-system status of
$S_{\mathrm F}^{nn'}$ in \pref{eq:SF}--\pref{eq:dsigma-SF} is the one
set in \secref{sec:scope}, item (v); the open-system extension required
for absorptive driven targets is treated in \secref{sec:plasmon-scope}
and \secref{sec:outlook}.

\paragraph*{Projectile-only reduction of the connected target response.}
When the off-equilibrium Floquet components of the
\emph{connected} target response are small, its Floquet dynamic
structure factor may be approximated by the conventional field-free
connected dynamic structure factor,
\begin{equation}
S_{\mathrm{F,c}}^{nn'}(\Qv,\omega) \simeq
\delta_{n,0}\,\delta_{n',0}\,S_{\mathrm c}(\Qv,\omega).
\label{eq:SF-undressed}
\end{equation}
This is a target-response approximation made while the projectile
Bessel dressing is retained; it is therefore clearer than denoting it
as a limit \(\vb A_0\to0\), which would simultaneously remove that
projectile dressing.  Crucially, \pref{eq:SF-undressed} is inserted
only into the connected second line above, not in place of the full
\(S_{\mathrm F}\) in \pref{eq:dsigma-SF}.  The coherent first line
remains separate.  Equation~\pref{eq:SF-undressed} then forces
\(N-n=0\) and \(N-n'=0\), hence \(n=n'=N\). The doubly differential
connected signal in the \(N\)-th Floquet channel then factorizes into
a projectile Bessel factor and the ordinary connected target dynamic
structure factor,
\begin{multline}
\left.
\frac{\dd^2\sigma_{\mathrm c}}{\dd\Omega\,\dd\epsilon}
\right|_N
=
\left(\frac{\me}{2\pi\hbar^2}\right)^2
\frac{k_f}{k_i}
|v(\Qv)|^2\\
\times
J_N^2\!\left(|\delta\zeta_{\Qv}^{\rm hom}|\right)
S_{\mathrm c}(\Qv,\omega_N),
\label{eq:dsigma-factorized}
\end{multline}
with $\omega_N$ the sideband loss frequency \pref{eq:omegaN-def}.
The target spectral delta in \(S_{\mathrm c}(\Qv,\omega_N)\) retains
the energy-density character of this expression.  Integrating
\pref{eq:dsigma-factorized} over a resolved sideband window
\(\Delta\epsilon_N\) gives the angular sideband weight defined in
\pref{eq:sideband-window-weight}.
The target sees an effective comb of probes of momentum
$\Qv$ and energy $\hbar\omega_N$, with relative weights controlled by
the projectile-only Bessel factor
\(J_N^2\!\left(|\delta\zeta_{\Qv}^{\rm hom}|\right) =
J_N^2\!\left(|\Qv\cdot\alphav|\right)\). This is the natural
fast-electron analogue of the comb of sidebands observed in
PINEM~\cite{Barwick2009,deAbajo2010}, but it dresses a
target-changing connected loss and is not by itself the
target-elastic PINEM term identified in
\secref{sec:pinem-bridge}. The relation of the homogeneous-field
Bessel-comb factorization \pref{eq:dsigma-factorized} to the
Kroll--Watson low-frequency result for laser-assisted elastic
scattering~\cite{KrollWatson1973} is developed in \appref{app:limits}.

\subsection{Inelastic Kroll--Watson sideband closure}
\label{sec:KW-closure}
A useful sideband-summed closure follows from the homogeneous
Bethe--Floquet cross section \pref{eq:dsigma-SF}: when all projectile
sidebands associated with one target loss are summed, the Volkov
Bessel factors close to unity and the off-diagonal Floquet coherences
of the target cancel from the summed signal. This is the inelastic
analogue of the Kroll--Watson sideband closure~\cite{KrollWatson1973}, \epref{eq:elastic-KW-closure}.
We use this closure for a target-changing loss
\(\omega\ne0\), where the coherent
\(\delta(\hbar\omega)\) sector is absent and
\(S_{\mathrm F}=S_{\mathrm{F,c}}\). Three conditions make it precise.

\paragraph*{Conditions.}
(i)~\emph{Fixed target loss, not fixed detector energy.} At fixed
outgoing projectile energy the implied target loss $\hbar\omega_N$
[\epref{eq:omegaN-def}] depends on $N$, so summing over the formal
sideband index would mix different target excitations and no Bessel
closure could apply. The closure instead fixes a target excitation
frequency $\omega$ and sums over \emph{its} comb of sideband
replicas, located at the projectile energies
\begin{equation}
\epsilon^{(N)}(k_f)
\coloneqq
\epsilon(k_i)-\hbar\omega-N\hbar\wL,
\qquad N\in\mathbb Z,
\label{eq:fixed-loss-sideband-energy}
\end{equation}
so that $\omega_N=\omega$ identically in every term of the sum.
(ii)~\emph{Resolved or consistently coarse-grained sidebands.}
The sideband energy windows \(\Delta\epsilon_N\), in the sense of
\pref{eq:sideband-window-weight}, are centered at the energies
\(\epsilon^{(N)}(k_f)\) of \(\pref{eq:fixed-loss-sideband-energy}\) and must
isolate the same target excitation.
 Alternatively, after the
prescribed coarse graining (\appref{app:lehmann-coarse-graining}), the
windows should not mix spectral weight from unrelated target excitations
or spectral features displaced by integer multiples of $\hbar\wL$.
(iii)~\emph{Soft-sideband kinematics.} As in the elastic
Kroll--Watson closure~\cite{KrollWatson1973}, the relevant range of
$N$ satisfies
\begin{align*}
k_N&\simeq k,\qquad
\Qv_N\simeq\Qv,\\
v(\Qv_N)&\simeq v(\Qv),\qquad
x_{\Qv_N}\simeq x_{\Qv},
\end{align*}
with $x_{\Qv}\coloneqq|\Qv\cdot\alphav|$, so that the projectile
kinematic prefactors and Bessel arguments take the common value
$x_{\Qv}$ across the whole comb.

\paragraph*{Derivation.}
Under conditions~(i) and (ii), but before invoking the
soft-sideband approximation~(iii), \epref{eq:dsigma-SF} gives the exact
fixed-loss comb sum
\begin{multline*}
\sum_N
\left(\frac{\dd\sigma}{\dd\Omega}\right)_N^{(\omega)} =
\left(\frac{\me}{2\pi\hbar^2}\right)^2
\sum_N\frac{k_N}{k_i}\,
\abs{v(\Qv_N)}^2 \\ \times
\sum_{n,n'}
J_n(x_{\Qv_N})J_{n'}(x_{\Qv_N})\\ \times
e^{\iu(n-n')\varphi'}
S_{\mathrm{F,c}}^{(N-n)(N-n')}(\Qv_N,\omega).
\end{multline*}
Here
\(\Qv_N\coloneqq\vb k_N-\vb k_i\),
\(x_{\Qv_N}\coloneqq\abs{\Qv_N\cdot\alphav}\), and
\(\varphi'\coloneqq\varphi+\theta_{\Qv}^{\rm hom}\).
This equality retains the full sideband dependence and is exact
within the homogeneous Bethe--Floquet expression.

Condition (iii) allows moving the $\sum_N$ to the right.
Setting $m=n-n'$, $M=N-n$, the resulting sideband sum is
\begin{multline*}
\sum_N\sum_{n,n'}
J_n(x_{\Qv})J_{n'}(x_{\Qv})e^{\iu(n-n')\varphi'}\\
\times S_{\mathrm{F,c}}^{N-n,N-n'}(\Qv,\omega)
\\
=
\sum_m e^{\iu m \varphi'}
\sum_n J_n(x_{\Qv})J_{n-m}(x_{\Qv})\\
\times\sum_M S_{\mathrm{F,c}}^{M,M+m}(\Qv,\omega).
\end{multline*}
The Bessel completeness relation
\begin{equation}
\sum_n J_n(x)J_{n-m}(x)
=
\delta_{m0}
\label{eq:Bessel-closure}
\end{equation}
leaves only $m=0$, collapsing the double Floquet sum to the diagonal
trace $\sum_M S_{\mathrm{F,c}}^{MM}(\Qv,\omega)$.

\paragraph*{Result.}
Substituting back into \pref{eq:dsigma-SF}, the closure reads
\begin{multline}
\sum_N
\qty(\frac{\dd\sigma}{\dd\Omega})_N^{(\omega)}
\, \simeq \\
\left(\frac{\me}{2\pi\hbar^2}\right)^2
\frac{k}{k_i}
|v(\Qv)|^2
\sum_M S_{\mathrm{F,c}}^{MM}(\Qv,\omega).
\label{eq:inelastic-KW-sum}
\end{multline}
Here \(\left(\dd\sigma/\dd\Omega\right)_{N}^{(\omega)}\) denotes the
sideband-resolved angular cross section obtained from the window
\(\Delta\epsilon_N\) centered at \(\epsilon^{(N)}(\vb k_f)\) in
\(\pref{eq:fixed-loss-sideband-energy}\).
The sideband-summed measurement is blind to the off-diagonal Floquet
coherences $S_{\mathrm{F,c}}^{MM'}$ with $M\neq M'$, and measures only
the incoherent Floquet-sector trace of the driven target at the fixed
target loss $\omega$. In view of Parseval's theorem, the sum $\sum_M$
is equivalent to cycle-averaged transition probability (with $\omega$ fixed).

The target-elastic comb at \(\omega=0\) is not governed by
\pref{eq:inelastic-KW-sum}.  It additionally contains the coherent
mean-density sector of \pref{eq:SF-coherent-connected} and the
reference-channel interference of \secref{sec:1Born}.  For a purely
coherent all-orders channel its sideband-summed statement is instead
the unitarity of the PINEM ladder,
\(\sum_NJ_N^2(2|g_{\rm tot}|)=1\), with
\(g_{\rm tot}=g_0+g_{\rm ind}\).


\paragraph*{Discussion.}
Experimentally, the sum in \pref{eq:inelastic-KW-sum} corresponds to
integrating or reading off a comb of windows centered at the
$N$-dependent electron energies $\epsilon^{(N)}(k_f)$, all associated with
the same target loss $\omega$ -- it is not a sum over sideband labels
at one fixed spectrometer channel. The closure is most useful for
targets with a discrete excitation $\hbar\omega$ selected by energy
conservation; we discuss such an example in \secref{sec:two-level}.
In the homogeneous-field limit it is exact up to approximation~(iii).
In an inhomogeneous eikonal near-field geometry the scalar Bessel
weights are replaced by momentum-space kernels, so the same statement
survives only in the narrow-kernel limit; otherwise the
sideband-summed signal becomes a kernel-weighted momentum average
rather than the single-$\Qv$ trace in \pref{eq:inelastic-KW-sum}.
An application of the closure is shown in \secref{sec:two-level}
for a discrete Rabi-type two-level system.

\section{Floquet response functions: exact properties and Floquet-zone representative covariance}
\label{sec:floquet-response}
The structure factors $S_{\mathrm F}^{nn'}(\Qv,\omega)$ and
$\SF^{nn'}(\vb q_1,\vb q_2;\omega)$, introduced
in \pref{eq:SF} and \pref{eq:SF-bimomentum}, are Lehmann spectral
correlators for a chosen initial Floquet state.
This section collects their general properties: Hermiticity,
positivity, Cauchy--Schwarz bounds, the $f$-sum rule,
polarization selection rules, translational selection rules, and
Floquet-zone covariance.
A relation to the physical
retarded response is not used; \appref{app:lehmann-coarse-graining}
distinguishes that response from a scattering-adapted auxiliary spectral
matrix, and \appref{app:fsum} gives the multi-species $f$-sum
generalization.

\subsection{Initial state and Floquet spectral correlator}
\label{sec:floquet-response-FDT}
We take the target density matrix to be concentrated
on a single Floquet eigenstate $\ket{\Phi_0}$ of $\hat H_1(t)$, namely
the Floquet eigenstate adiabatically continued from the equilibrium
ground state of the undriven target as $\vb A_0$ is switched on. The
quantity \pref{eq:SF} is then the corresponding (zero-temperature)
spectral correlator,
\begin{multline}
S_{\mathrm F}^{nn'}(\Qv,\omega) \coloneqq \sum_a
\tilde\varrho_{a0}^{(n)}(\Qv)\,\tilde\varrho_{a0}^{(n')\,*}(\Qv)\\
\times\delta(\hbar\omega + E_0 - E_a),
\label{eq:SF-pure}
\end{multline}
with $\ket{\Phi_a}$ ranging over Floquet eigenstates of $\hat H_1(t)$
and $\hbar\omega_{a0}\coloneqq E_a-E_0$ the quasienergy difference.
The relation of this spectral correlator
to a retarded
density-density susceptibility, together with the continuum and
coarse-grained interpretation needed for macroscopic targets, is
collected in \appref{app:lehmann-coarse-graining}. Here we keep only the
structural identities that are needed in the main text.

\paragraph*{Connected fluctuation sector.}
For the same pure-state preparation, the connected correlator is
obtained either by replacing
\(\hat\varrho\) with
\(\delta\hat\varrho=\hat\varrho-\langle\hat\varrho\rangle\), or
equivalently by restricting the Lehmann sum in
\pref{eq:SF-pure} to \(a\ne0\):
\begin{multline*}
S_{\mathrm{F,c}}^{nn'}(\Qv,\omega)
=
\sum_{a\ne0}
\tilde\varrho_{a0}^{(n)}(\Qv)
\tilde\varrho_{a0}^{(n')\,*}(\Qv)\\
\times
\delta(\hbar\omega+E_0-E_a).
\end{multline*}
The full correlator is therefore
\(S_{\mathrm F}^{nn'}=
R_nR_{n'}^*\delta(\hbar\omega)+S_{\mathrm{F,c}}^{nn'}\).
The connected matrix separately inherits Hermiticity, positive
semidefiniteness, and the Cauchy--Schwarz bounds discussed below.
Moreover, subtracting the \(c\)-number mean density does not change a
retarded commutator:
\([\delta\hat\varrho(t),\delta\hat\varrho(t')]
=[\hat\varrho(t),\hat\varrho(t')]\).
The auxiliary spectral matrix $\chiaux^{nn'}$ of
\appref{app:lehmann-coarse-graining}, built from the connected Lehmann
weights, satisfies, with the Hermitian spectral part
$\ImH\chi\coloneqq(\chi-\chi^\dagger)/2\iu$,
$(\chi^\dagger)_{nn'}\coloneqq\chi_{n'n}^*$,
\[
-\frac{1}{\pi}\ImH\chiaux(\Qv,\omega)
=S_{\mathrm{F,c}}(\Qv,\omega)-S_{\mathrm{F,c}}(\Qv,-\omega),
\]
which reduces to $S_{\mathrm{F,c}}$ at $\omega>0$ when all quasienergy
gaps $\omega_{a0}$ of the chosen representative are positive. An
entrywise imaginary part would be wrong for the complex off-diagonal
entries. $\chiaux$ is not the Floquet transform of the physical retarded
commutator, whose antiresonant branch carries different weights; the
cross section is computed directly from the unsymmetrized correlator.
The coherent \(R_nR_{n'}^*\delta(\hbar\omega)\) line is absent from the
retarded response and must be restored separately when the
\emph{full} scattering correlator is required.  Thus the structural
statements below apply separately to \(S_{\mathrm{F,c}}\); the
coherent sector drops out of the symmetrized $f$-sum rule
\pref{eq:fsum}, whereas one-sided moments require subtraction of its
contribution.

\subsection{Hermiticity, positivity, Cauchy--Schwarz, and the
$f$-sum rule}
\label{sec:floquet-identities}
Viewed as a matrix in the Floquet-index pair $(n,n')$ at fixed
$(\Qv,\omega)$, the spectral correlator \pref{eq:SF-pure} -- and its
thermal generalization -- inherits from the underlying Lehmann form
\begin{equation}
\bigl[S_{\mathrm F}^{nn'}\bigr]^{*} = S_{\mathrm F}^{n'n},\qquad
S_{\mathrm F} \succeq 0,
\label{eq:SF-Hermitian}
\end{equation}
\ie\ Hermiticity and positive semi-definiteness, and consequently the
Cauchy--Schwarz bound on Floquet coherences
\begin{equation}
\bigl|S_{\mathrm F}^{nn'}(\Qv,\omega)\bigr|^{2} \;\leq\;
S_{\mathrm F}^{nn}(\Qv,\omega)\,S_{\mathrm F}^{n'n'}(\Qv,\omega).
\label{eq:CS}
\end{equation}
Equation~\pref{eq:CS} is a model-independent constraint: any approximate
calculation of $S_{\mathrm F}^{nn'}$ violating it is necessarily
unphysical.
The connected matrix obeys the same constraints separately,
\[
\bigl[S_{\mathrm{F,c}}^{nn'}\bigr]^*
=S_{\mathrm{F,c}}^{n'n},
\qquad
S_{\mathrm{F,c}}\succeq0,
\]
and its off-diagonal entries satisfy \pref{eq:CS} with every
\(S_{\mathrm F}\) replaced by \(S_{\mathrm{F,c}}\).  Thus removing the
coherent rank-one outer product preserves, rather than weakens, the
positivity tests.

The standard collection of assumptions: density--density interactions,
local minimal coupling, and a single electronic species of charge $-e$ and
mass $\me$, leads to the Floquet $f$-sum rule
(for any prescribed $\vb A(\vb r,t)$, which commutes with the density),
\begin{multline}
\frac12\sum_{n} \int_{-\infty}^{\infty}\!\dd\omega'\,(\omega' + n\wL)\,
\bigl[s_{\mathrm F}^{nn}(\Qv,\omega')+s_{\mathrm F}^{nn}(-\Qv,\omega')\bigr]
\\
=\frac{e^{2}\,Q^{2}\,\bar n_{\mathrm e}}{2\me},
\label{eq:fsum}
\end{multline}
with the per-volume $s_{\mathrm F}=S_{\mathrm F}/V$ of
\pref{eq:bimomentum-translation}. The one-sided form (without the
$-\Qv$ term and the factor $\tfrac12$) follows when the prepared state
respects a generalized inversion or generalized time-reversal symmetry
of the driven Hamiltonian that equates the moments at $\Qv$ and
$-\Qv$; it fails, e.g., for Floquet states carrying a cycle-averaged
current (\appref{app:fsum}).
Here $\omega'$ is the spectral integration variable, $\bar n_{\mathrm e}$ 
denotes the cycle-averaged electron
number density of the driven target, and the factor of $e^{2}$ on the
right-hand side is the consequence of the charge-density convention
adopted in \secref{sec:hamilton}. The shift $\omega'+n\wL$ on the
left-hand side accounts for the photon energy absorbed or emitted by
the target in the $n$th Floquet sector. 
See \appref{app:fsum} for
the derivation and for the appropriate generalization to
multi-species targets.
Equation~\pref{eq:fsum} is a moment of the \emph{full}
correlator. Its coherent sector contributes
$\Delta_{\rm coh}(\Qv)=(\wL/\hbar V)\sum_n n|R_n(\Qv)|^2$ to a one-sided
moment; since $R_n(-\Qv)=R_{-n}(\Qv)^*$, $\Delta_{\rm
coh}(-\Qv)=-\Delta_{\rm coh}(\Qv)$, so the coherent sector cancels in
\pref{eq:fsum} for any illumination, and the connected correlator alone
has the same symmetrized moment (\appref{app:fsum}).

Under the
Floquet--Fourier convention of \pref{eq:FF}, the physical frequency
carried by the $n$th sector of $S_{\mathrm F}^{nn}(\Qv,\omega')$ is
$\omega'+n\wL$, so the integrand of \pref{eq:fsum} weights each
sector by its physical energy; this is the source of the $n\wL$ shift
and fixes the sign convention used in \appref{app:fsum}. Equation~\pref{eq:fsum} is the Floquet
analogue of the equilibrium $f$-sum rule, $\int\dd\omega'\,\omega'\,
s(\Qv,\omega') = e^{2}Q^{2}n_{\mathrm e}/(2\me)$ (for equilibrium
states with equal $\pm\Qv$ moments), and is a
powerful consistency check that holds independently of the
quasi-equilibrium assumption.

\subsection{Polarization-induced selection rules}
\label{sec:floquet-selection}
The polarization of the drive imposes selection rules on
$S_{\mathrm F}^{nn'}$. Consider
a \emph{circularly polarized} field
$\vb A(t) = A_0[\hat{\vb x}\cos\wL t + \hat{\vb y}\sin\wL t]$,
or $\Ap=A_0
\left(\hat{\vb x} -\iu\hat{\vb y} \right)$.
The joint
operation of spatial rotation by $\delta\phi$ about $\hat z$ and
time-shift by $\delta\phi/\wL$ leaves $\hat H_1(t)$ invariant whenever
the equilibrium target is rotationally symmetric about $\hat z$ \cite{Neufeld2019}. The
Floquet eigenstates $\ket{\Phi_a}$ are then simultaneous eigenstates of
this combined operation with conserved \emph{co-rotating} angular
momentum $J_z(a) = m_a\hbar$; correspondingly the $n$th Fourier
component $\ket{\Phi_a^{(n)}}$ carries lab-frame angular momentum
$L_z=(m_a+n)\hbar$ (the absorbing part $\propto e^{-\iu\wL t}$ of this
field couples through $D_x+\iu D_y$ and raises $L_z$). The density operator $\hat\varrho_1(\Qv\!\parallel\!\hat z)$
is rotationally invariant ($\Delta L_z=0$), so the matrix elements in
$\tilde\varrho_{a0}^{(n)}(\Qv\!\parallel\!\hat z)$ require
$m_a+n = m_0$ on every term of the sum; with the equilibrium ground
state $m_0=0$, this fixes $m_a=-n$. Each Floquet sector is therefore fed
by Floquet final states of a single co-rotating angular momentum
$-n\hbar$, and the Floquet structure factor is diagonal:
\begin{equation}
S_{\mathrm F}^{nn'}(\Qv\!\parallel\!\hat z,\omega) = 0 \quad \text{for}\ n\neq n'.
\label{eq:circ-selection}
\end{equation}
The connected diagonal entries
$S_{\mathrm{F,c}}^{nn}(\Qv\!\parallel\!\hat z,\omega)$ remain nonzero:
they probe target excitations of co-rotating angular momentum $-n\hbar$ and yield a
photon-resolved loss spectrum in each Floquet sector.  The full
\(n=0\) entry may additionally contain the coherent target-elastic
line \(R_0R_0^*\delta(\hbar\omega)\); under the stated circular--axial
symmetry, nonzero-harmonic mean-density terms are forbidden by the
same selection rule.  Equation~\pref{eq:circ-selection} forbids only
off-diagonal \emph{coherences} between sectors, not the connected
diagonal sidebands themselves. The coherences are restored either by
\emph{linear} or elliptical polarization, which breaks the joint
rotation+time-shift symmetry, or by non-axial $\Qv$, which makes
$\hat\varrho_1(\Qv)$ carry nonzero angular momentum. The off-diagonal
spectral weight predicted in \secref{sec:application-SF} is therefore
an experimental signature of broken circular--axial symmetry: linearly
polarized drives, transverse momentum transfer, or both.

In crystalline targets the continuous rotational symmetry weakens to a
discrete subgroup; \epref{eq:circ-selection} is then replaced by
group-theoretic statements that allow a restricted set of off-diagonal
Floquet coherences -- a sharper experimental fingerprint of the
underlying crystal symmetry under the drive.

\subsection{Floquet-zone convention and representative covariance}
\label{sec:floquet-gauge}
The covariance discussed here is the relabeling of
Floquet-zone representatives under shifts of the quasienergy by
$\hbar\wL$, not the electromagnetic gauge covariance of the underlying
$\vb A(\vb r,t)$ -- which is already fixed to Coulomb gauge in
\secref{sec:hamiltonian-light-field}.
The Floquet eigenstates $\ket{\Phi_a(t)}$ of a periodic Hamiltonian are
defined only up to a shift of their quasienergies by integer multiples
of $\hbar\wL$, accompanied by a corresponding relabeling of the
Fourier components:
\begin{equation}
E_a\;\to\;E_a + \ell_a\hbar\wL,\qquad
\ket{\Phi_a^{(m)}}\;\to\;\ket{\Phi_a^{(m+\ell_a)}},
\label{eq:floquet-gauge}
\end{equation}
with $\ell_a\in\mathbb Z$ chosen independently for each Floquet state.
The full Floquet state $\ket{\Phi_a(t)} = e^{-\iu E_a t/\hbar}\sum_m
e^{-\iu m\wL t}\ket{\Phi_a^{(m)}}$ is invariant under
\pref{eq:floquet-gauge}, but the labels $E_a$ and the index $m$
individually are not.

Under \pref{eq:floquet-gauge} the Floquet density components
transform as $\tilde\varrho_{a0}^{\prime\,(n)}=\tilde\varrho_{a0}^{(n+\ell_a-\ell_0)}$,
i.e.\ the same physical contribution moves from label $n$ to
$n-(\ell_a-\ell_0)$, a relabeling that depends on \emph{both} initial- and
final-state representatives. The individual quantities
$\tilde\varrho_{a0}^{(n)}$ and $S_{\mathrm F}^{nn'}$ therefore depend
on the chosen representatives of the Floquet quasienergies: a single
global covariance law for the entire summed object
$S_{\mathrm F}^{nn'}$ would require, in general, a state-by-state
relabeling, which cannot be absorbed into a single transformation of
$(n,n')$. The same caveat applies to
$\SF^{nn'}$ and to $\chiaux^{nn'}$.

The physical cross section
\pref{eq:dsigma-Volkov}--\pref{eq:dsigma-SF} is invariant under a
change of Floquet-zone representatives provided that all quasienergies,
target Fourier indices, projectile Fourier indices, and external
sideband labels are reindexed consistently. Although the individual
auxiliary labels depend on the representative convention, the physical
energy-conservation combination
\[
E_f-E_i+N\hbar\wL
\]
and the complete summed cross section are unchanged. Concretely,
with $\delta\ell\coloneqq\ell_f-\ell_i$ the same contribution
acquires the labels $M'=M-\delta\ell$ and $N'=N-\delta\ell$, so that
$(E_f'-E_i')+N'\hbar\wL=(E_f-E_i)+N\hbar\wL$ and the projectile order
$n=N-M$ is unchanged. The covariance is therefore
a discrete index-relabeling covariance rather than an electromagnetic
gauge transformation or a universal cancellation of
phases~\cite{Bilitewski2015}. Fixing the adiabatic-continuation
convention, as we do below, removes this ambiguity in the intermediate
quantities.

For definiteness we use the
\emph{adiabatic-continuation convention}: each Floquet state $\ket{\Phi_a}$
and its quasienergy $E_a$ are continued continuously from the
field-free target eigenstate of energy $\lim_{\vb A_0\to
0}E_a=E_a^{(0)}$. In this gauge the equilibrium ground state $\ket{\Phi_0}$
has quasienergy $E_0$ continuously connected to the undriven ground
state, and the Floquet sectors $n,n'$ retain their natural meaning as
net photon-number indices relative to that reference. Other gauges --
e.g.\ restriction of all quasienergies to a fundamental Floquet
zone $[0,\hbar\wL)$ -- are equally valid and produce the same
cross section once labels are fixed consistently throughout the
calculation.

\section{Worked example: a metal in the Drude and diffusive limits}
\label{sec:plasmon}
The formalism of Secs.~\ref{sec:applications}--\ref{sec:floquet-response}
admits an analytical evaluation whenever the equilibrium target has a
closed-form linear-response function. The simplest non-trivial example
is a metal, for which two complementary regimes are accessible
analytically: a local Drude description that captures the
high-frequency bulk plasmon (Secs.~\ref{sec:plasmon-drude}--\ref{sec:plasmon-cross-section}),
and a diffusive description that captures the low-frequency
particle--hole continuum of a disordered metal
(\secref{sec:plasmon-diffusive}). We carry out both calculations in
full and compare with the corresponding limits of the literature.

\subsection{Status of the metallic examples}
\label{sec:plasmon-scope}
The metallic response separates according to
\pref{eq:SF-coherent-connected}.  A useful approximation for a
linearly driven metal is
\begin{multline}
\SF^{MM'}(\vb q_1,\vb q_2;\omega)
\simeq
R_M(\vb q_1)R_{M'}^*(\vb q_2)\,\delta(\hbar\omega)
{}\\
{}+\delta_{M,0}\delta_{M',0}\,
\mathscr S_{\mathrm{eq,c}}(\vb q_1,\vb q_2;\omega).
\label{eq:metal-coherent-connected}
\end{multline}
The first term retains the laser-induced mean charge and generates the
target-elastic PINEM channel.  The second assumes that, to the accuracy
of the worked example, the connected density fluctuations retain their
equilibrium form.  This is the appropriate refinement of applying the
projectile-only approximation to the \emph{full} correlator:
dropping \(R_MR_{M'}^*\) would remove the coherent optical response
rather than approximate it.

The analytical Drude and diffusive spectra plotted below evaluate only
the target-changing, connected part of
\pref{eq:metal-coherent-connected}.  On the momentum diagonal their
working reduction is
\begin{equation}
S_{\mathrm{F,c}}^{nn'}(\Qv,\omega)\;\longrightarrow\;
\delta_{n,0}\,\delta_{n',0}\,S_{\mathrm c}(\Qv,\omega),
\label{eq:weak-target-dressing}
\end{equation}
so the prescribed optical field dresses the fast projectile through the
Volkov phase, while the residual target response is supplied as its
equilibrium (or quasi-equilibrium) connected density structure factor.
The missing coherent term is restored explicitly in
\secref{sec:plasmon-cross-section}.

The Drude
collision rate $\gamma$ in $\varepsilon^{\mathrm D}(\omega)$
\pref{eq:drude} and the diffusion coefficient $D$ -- which together
with the screening wavevector $\kappa$ fixes the screened relaxation
scale $D(q^{2}+\kappa^{2})$ of \pref{eq:S-diff-screened} -- enter as
response-function inputs to \pref{eq:dsigma-loss} and
\pref{eq:dsigma-diff-inhom} -- they are \emph{not} non-Hermitian
modifications of $\hat H_1(t)$ -- and are phenomenological parameters
of the equilibrium dissipative response supplied to the scattering
formula. Inserting an equilibrium $\gamma$ or $\tau$ does not amount to
constructing an open driven Floquet state; a genuinely driven
absorptive target requires a Floquet--Keldysh or kinetic treatment.
This restricted status is consistent with assumption~(v) of
\secref{sec:scope}; the genuinely driven-dissipative regime is
collected for the metallic case in \secref{sec:plasmon-discussion} and
discussed more broadly in \secref{sec:outlook}.
Moreover, the bulk response functions alone do not determine
\(R_M(\vb q)\): the illumination amplitude, surface geometry, and
electromagnetic boundary conditions are needed to calculate the
coherent induced near field.  Consequently the formulas below give a
complete bulk benchmark for the connected loss sector, while the
PINEM sector becomes quantitative only after that optical
boundary-value problem has been supplied.

\subsection{Loss function and Drude dielectric response}
\label{sec:plasmon-drude}
The conventional EELS identity~\cite{Pines1999,Egerton2011} relates the
connected target structure factor defined by the second line of
\pref{eq:SF-coherent-connected} -- in the reduction
\pref{eq:weak-target-dressing} -- to the loss
function $\mathcal L(\vb q,\omega)\coloneqq\mathrm{Im}[-1/\varepsilon(\vb q,\omega)]$
of the longitudinal dielectric function $\varepsilon(\vb q,\omega)$,
\begin{equation}
S_{\mathrm c}(\vb q,\omega) =
\frac{q^{2}}{4\pi^{2}}\,\mathcal L(\vb q,\omega),
\label{eq:S-loss-function}
\end{equation}
in Gaussian--cgs units. The relation follows from the standard density
fluctuation--dissipation theorem together with the RPA-like identity
$\varepsilon^{-1}(\vb q,\omega) = 1 + V_{\mathrm C}(\vb q)\chi_{nn}(\vb q,\omega)$
with $V_{\mathrm C}(\vb q) = 4\pi e^{2}/q^{2}$. Here $S_{\mathrm c}$ denotes the
connected \emph{charge-density} structure factor per unit volume
[the intensive $s_{\mathrm c}$ of \pref{eq:bimomentum-translation};
correspondingly, the connected bulk loss formulas below are per unit target
volume, $\dd^2\Sigma_{\mathrm c}/\dd\Omega\,\dd\epsilon\coloneqq
V^{-1}\dd^2\sigma_{\mathrm c}/\dd\Omega\,\dd\epsilon$], consistent
with the convention adopted in \secref{sec:hamilton}; with the
number-density convention used in much of the EELS literature the
right-hand side of \pref{eq:S-loss-function} differs by a factor of
$e^{2}$ and the relation reads
$S_{n n,\mathrm c}(\vb q,\omega)=
[q^{2}/(4\pi^{2}e^{2})]\mathcal L(\vb q,\omega)$.

In the Drude approximation, $\varepsilon$ is local and given by
\begin{equation}
\varepsilon^{\mathrm D}(\omega) = 1 - \frac{\omega_{p}^{2}}{\omega(\omega+\iu\gamma)},
\qquad \omega_{p}^{2} \coloneqq \frac{4\pi n_{\mathrm e}\,e^{2}}{\me},
\label{eq:drude}
\end{equation}
with $\omega_{p}$ the bulk plasma frequency, $n_{\mathrm e}$ the
conduction-electron number density, and $\gamma$ a phenomenological
collision rate entering the equilibrium Drude response (see
\secref{sec:plasmon-scope}).

The same dielectric function also controls the coherent mean response,
but through its full complex value rather than through the loss
function.  For illustration, in a longitudinal scalar geometry define
\(\bar\varrho^{(+)}(\vb q)\) by
\(\bar\varrho(\vb q,t)=
\bar\varrho^{(+)}(\vb q)e^{-\iu\wL t}+\mathrm{c.c.}\).
Then
\begin{multline}
v(\vb q)\bar\varrho^{(+)}(\vb q)
=q_{\rm p}\phi_{\rm ind}^{(+)}(\vb q)
\\
=q_{\rm p}
\left[\frac{1}{\varepsilon^{\mathrm D}(\wL)}-1\right]
\phi_{\rm ext}^{(+)}(\vb q),
\label{eq:drude-induced-mean}
\end{multline}
where
\begin{equation}
\frac{1}{\varepsilon^{\mathrm D}(\wL)}-1
=
\frac{\omega_p^2}
{\wL(\wL+\iu\gamma)-\omega_p^2}.
\label{eq:drude-coherent-response}
\end{equation}
Thus the connected EELS sector samples
\(\operatorname{Im}[-1/\varepsilon^{\mathrm D}(\omega_N)]\) at the
residual target-loss frequency, whereas the coherent PINEM sector
samples the amplitude and phase of
\(\varepsilon^{\mathrm D}(\wL)^{-1}-1\) at the laser frequency.
Equation~\pref{eq:drude-induced-mean} is only a constitutive relation: a
localized PINEM field still requires the external illumination and the
surface boundary conditions.  Once those are supplied,
\(\bar\varrho^{(+)}=R_{-1}\) enters the trajectory coupling
\pref{eq:gind-density}.

Direct algebra yields the loss function in closed form,
\begin{equation}
\mathcal L^{\mathrm D}(\omega) = \frac{\gamma\,\omega\,\omega_{p}^{2}}
{(\omega^{2}-\omega_{p}^{2})^{2}+\gamma^{2}\omega^{2}},
\label{eq:drude-loss}
\end{equation}
a Lorentzian-like peak centered on the plasmon frequency with full width
$\sim\gamma$ (in frequency units) controlled by the collision rate.
In the clean limit $\gamma\to 0^{+}$ the representation
$\eta/(x^{2}+\eta^{2})\to\pi\delta(x)$ gives
\begin{equation}
\mathcal L^{\mathrm D}(\omega)\xrightarrow{\gamma\to0^{+}}
\frac{\pi\omega_{p}}{2}\,\delta(\omega-\omega_{p})
\quad(\omega>0),
\label{eq:drude-clean}
\end{equation}
and the connected structure factor \pref{eq:S-loss-function} then collapses
to the delta-function plasmon line
$S_{\mathrm c}^{\mathrm D}\propto q^{2}\delta(\omega-\omega_{p})$,
saturating the $f$-sum rule restricted to the plasmon mode.

\subsection{Plasmon-loss cross section under a homogeneous laser drive}
\label{sec:plasmon-cross-section}
For the connected sector in the reduction
\pref{eq:weak-target-dressing}, substituting
\pref{eq:S-loss-function} into \pref{eq:dsigma-factorized} with
$|v(\Qv)|^{2} = 16\pi^{2}e^{2}/Q^{4}$ yields
\begin{multline}
\left.
\frac{\dd^2\Sigma_{\mathrm c}}{\dd\Omega\,\dd\epsilon}
\right|_N
=
\frac{m_e^2 e^2}{\pi^2\hbar^4 Q^2}
\frac{k_f}{k_i}\\
\times
J_N^2\!\left(|\Qv\cdot\alphav|\right)
\mathcal L(\Qv,\omega_N),
\label{eq:dsigma-loss}
\end{multline}
with $\hbar\omega_{N} = \epsilon(\vb k_i)-\epsilon(\vb k_f)-N\hbar\wL$ the energy
transferred to the target in the $N$-photon channel and the Bessel
argument \pref{eq:dzetahom}.
Equation~\pref{eq:dsigma-loss} is the Bethe--Floquet generalization of the
standard EELS loss-function formula~\cite{Egerton2011}; it is the
target-changing term, not the entire signal of the driven metal.

The target-elastic coherent term follows from the first term of
\pref{eq:metal-coherent-connected}.  In the homogeneous or
single-momentum reduction, with $\mathcal C_{\Qv}$ and $B_n(\Qv)$
as defined below \pref{eq:dsigma-SF}, and to first Born order in the target-induced potential, the improved
cross section is
\begin{multline}
\left.
\frac{\dd^2\sigma}{\dd\Omega\,\dd\epsilon}
\right|_N
\simeq
V\left.
\frac{\dd^2\Sigma_{\mathrm c}}{\dd\Omega\,\dd\epsilon}
\right|_N
{}\\
+\mathcal C_{\Qv}
\left|
\sum_n B_n(\Qv)R_{N-n}(\Qv)
\right|^2
\delta(\hbar\omega_N).
\label{eq:dsigma-loss-plus-pinem}
\end{multline}
All photon partitions leading to the same \(N\) interfere inside the
modulus square.  They do not interfere with the connected loss term,
because the latter leaves the target in an orthogonal final state.
The delta function in the second term fixes only the residual target
excitation to zero; the electron can still appear in a nonzero optical
sideband. Both terms are total, target-integrated cross sections for the
same sample volume $V$: the connected term is $V$ times the per-volume
bulk result, whereas the scaling of the coherent term with $V$ depends
on the geometry of the induced mean density.

If the prescribed projectile dressing is switched off,
\(B_n\to\delta_{n,0}\), the fundamental components give the one-photon
coherent Born amplitudes
\begin{equation}
f_N^{(1)}(\Qv)=
-\frac{\me}{2\pi\hbar^2}\,
v(\Qv)R_N(\Qv),
\qquad N=\pm1.
\label{eq:weak-pinem-born-parent}
\end{equation}
After the standard nonrecoil impact-parameter transform,
implemented explicitly by \pref{eq:gind-density}, this is the
weak-PINEM result
\(P_{\pm1}=|g_{\rm ind}|^2+\order{|g_{\rm ind}|^4}\), the leading
term of \(J_{\pm1}^2(2|g_{\rm ind}|)\).  The full Bessel law is obtained
by moving the coherent mean potential into the projectile channel
Hamiltonian and exponentiating its repeated target-elastic action, as
outlined in \secref{sec:pinem-bridge}.  If an independently prescribed
localized coupling \(g_0\) is already present, the elastic-target
amplitude must first be combined with the zero-Born projectile
amplitude.  The coherent result then resums to
\(J_N^2(2|g_0+g_{\rm ind}|)\) at the fundamental harmonic, while the
residual structure factor is constructed from
\(\delta\hat\varrho_1\) to prevent double counting.


Substituting the Drude loss function \pref{eq:drude-loss}
gives the Bessel-weighted plasmon-loss comb,
\begin{equation}
\frac{\dd^{2}\Sigma_{\mathrm c}}{\dd\Omega\,\dd\epsilon}\bigg|_{N} =
\frac{\me^{2}\,e^{2}}{\pi^{2}\hbar^{4}\,Q^{2}}\,\frac{k_{f}}{k_{i}}\,
J_{N}^{2}(|\Qv\cdot\alphav|)\,\mathcal L^{\mathrm D}(\omega_{N}),
\label{eq:dsigma-Drude}
\end{equation}
with the $N$th peak centered at the projectile energy loss
\begin{equation}
\epsilon(\vb k_i)-\epsilon(\vb k_f) \;=\; \hbar\omega_{p} + N\hbar\wL,
\qquad N\in\mathbb Z,
\label{eq:plasmon-comb-energies}
\end{equation}
when $\gamma\ll\wL$, so that the Lorentzian $\mathcal
L^{\mathrm D}(\omega_{N})$ is sharp near $\omega_{N}=\omega_{p}$.
The Bessel factor $J_{N}^{2}(|\Qv\cdot\alphav|)$ controls the relative
weight of each peak; see \figref{fig:metal-spectra} for illustrative examples.
The Lorentzian width $\sim\gamma$ is the
only spectral broadening; in real metallic films optical screening,
surfaces, and plasmon dispersion replace $\mathcal L^{\mathrm D}$
by an appropriate geometry-dependent loss function
in \pref{eq:dsigma-loss}. In the clean limit $\gamma\to0^{+}$,
$\mathcal L^{\mathrm D}\to\frac{\pi\omega_{p}}{2}\delta(\omega_{N}-\omega_{p})$
and \pref{eq:dsigma-Drude} reduces to the idealized delta-function comb.

\subsection{Diffusive metal: non-local density susceptibility}
\label{sec:plasmon-diffusive}
In a disordered metal the low-frequency density response is controlled
by diffusion. The standard diffuson result for the irreducible
density--density correlator at low momentum and frequency
reads~\cite{AltshulerAronov1985,Akkermans2007}
\begin{equation}
\chi_{0}^{\mathrm{diff}}(\vb q,\omega) = -\nu_F\,\frac{D q^{2}}{D q^{2} - \iu\omega},
\label{eq:chi-diff}
\end{equation}
with $\nu_F$ the density of states per unit volume at the Fermi level
and $D = v_F^{2}\tau/d$ the diffusion coefficient ($d$ the spatial
dimension). The validity of \pref{eq:chi-diff} is restricted to the
hydrodynamic diffusive window
\begin{equation}
q\,\ell \ll 1,\qquad |\omega|\,\tau \ll 1,
\label{eq:diffusive-validity}
\end{equation}
with $\ell=v_F\tau$ the elastic mean free path. The diffuson enters
the cross sections below through the projectile-only reduction of the
\emph{connected} target response,
\pref{eq:weak-target-dressing}, as an equilibrium-target benchmark. Its
pole reflects diffusive density relaxation generated by elastic
disorder scattering in the thermodynamic, disorder-averaged response,
and does not by itself require an external bath; a bath or kinetic
description is needed only to prepare and stabilize a genuinely driven
absorptive Floquet steady state (see \secref{sec:plasmon-scope}). The
screened correlator is
\begin{multline}
\chi^{\mathrm{diff}}(\vb q,\omega) = \frac{\chi_{0}^{\mathrm{diff}}}
{1 - V_{\mathrm C}(\vb q)\,\chi_{0}^{\mathrm{diff}}}\\
= -\nu_F\,\frac{D q^{2}}{D(q^{2}+\kappa^{2})-\iu\omega},
\label{eq:chi-diff-screened}
\end{multline}
with $\kappa^{2}\coloneqq 4\pi e^{2}\nu_F$ the Thomas--Fermi
wavevector. In a genuine three-dimensional charged metal, long-range
Coulomb screening is not optional: it changes the diffusive pole from
$Dq^{2}$ to $D(q^{2}+\kappa^{2})$, and the screened form
\pref{eq:chi-diff-screened} -- not the bare \pref{eq:chi-diff} -- is
the physical prediction for the bulk metallic response. We use it as
the principal result below; the unscreened \pref{eq:chi-diff} is
retained only as the simpler intermediate correlator from which
\pref{eq:chi-diff-screened} is obtained by RPA, and as the
$\kappa\to0$ comparison limit.

\paragraph*{Number- versus charge-density convention.}
The correlator $\chi_{0}^{\mathrm{diff}}$ of \pref{eq:chi-diff} is the
\emph{number-density} response. Since the structure factor $S$ used in
the body of the manuscript (and in the cross section) is the
charge-density structure factor, the corresponding structure function
carries an additional factor $e^{2}$,\ie{}
$$
S^{\mathrm{diff}} = -(e^{2}/\pi)\,
\mathrm{Im}\,\chi_{0}^{\mathrm{diff}}(\vb q,\omega)\,\Theta(\omega)
$$ at $T=0$, where
the Heaviside step encodes the positive-frequency support of the $T=0$
spectral function (only absorptive transitions out of the ground state
contribute; the Lehmann form has $E_{f}-E_{0}\ge 0$)
\footnote{At finite temperature $\Theta(\omega)$ is replaced by the appropriate
detailed-balance factor; the worked example below stays at $T=0$.}.
With these conventions the unscreened structure factor reads
\begin{multline}
S^{\mathrm{diff}}(\vb q,\omega) = \frac{e^{2}\,\nu_F}{\pi}\,
\frac{D q^{2}\,\omega}{(D q^{2})^{2} + \omega^{2}}\,\Theta(\omega)
\qquad
(T=0),
\label{eq:S-diff}
\end{multline}
and the screened structure factor -- obtained by replacing the
denominator relaxation rate $Dq^{2}\to D(q^{2}+\kappa^{2})$ while
retaining the numerator factor $Dq^{2}$ unchanged, as required by
charge conservation -- is
\begin{equation}
S^{\mathrm{diff}}_{\mathrm{scr}}(\vb q,\omega) =
\frac{e^{2}\,\nu_F}{\pi}
\frac{D q^{2}\,\omega\,\Theta(\omega)}{\bigl[D(q^{2}+\kappa^{2})\bigr]^{2} + \omega^{2}}
\qquad
(T=0).
\label{eq:S-diff-screened}
\end{equation}
The screened form \pref{eq:S-diff-screened} is used as the principal
metallic prediction below.

Interpreting the maximum of \pref{eq:S-diff-screened} at
$\omega=D(q^{2}+\kappa^{2})$ as a diffusive relaxation peak requires
the additional consistency condition
\[
D(q^{2}+\kappa^{2})\,\tau\ll1,
\quad\text{equivalently}\quad
(q^{2}+\kappa^{2})\,\ell^{2}/d\ll1,
\]
which follows from $D=v_F^{2}\tau/d$ and $\ell=v_F\tau$ and sharpens
the hydrodynamic window \pref{eq:diffusive-validity}. When this
condition is violated -- as may occur in a well-screened good metal
with $\kappa\ell\gg1$ -- the screened diffuson expression remains
applicable only on its low-frequency side; the response near the
nominal maximum then requires a dynamical theory beyond the diffusive
approximation.

Both \pref{eq:S-diff} and \pref{eq:S-diff-screened} are connected
charge-density structure factors: they describe target-changing
fluctuations and contain no pump-induced mean density.  The unscreened
spectral function \pref{eq:S-diff} is positive-frequency,
rises linearly at $\omega\to 0^{+}$, peaks at $\omega=D q^{2}$, and
decays as $\omega^{-1}$ within the hydrodynamic window
\pref{eq:diffusive-validity}; the screened
\pref{eq:S-diff-screened} has its peak shifted to
$\omega = D(q^{2}+\kappa^{2})$.
Substituting \pref{eq:S-diff} into the connected cross section
\pref{eq:dsigma-factorized}, using $|v(\Qv)|^{2}=16\pi^{2}e^{2}/Q^{4}$ and
the identity $4\pi e^{2}\nu_F = \kappa^{2}$, one obtains
\begin{multline}
\frac{\dd^{2}\Sigma_{\mathrm c}}{\dd\Omega\,\dd\epsilon}\bigg|_{N}^{\mathrm{diff}} =
\frac{\me^{2}\,e^{2}\,\kappa^{2}\,D}{\pi^{2}\hbar^{4}\,Q^{2}}\,\frac{k_{f}}{k_{i}} \\
\times\,J_{N}^{2}(|\Qv\cdot\alphav|)\,
\frac{\omega_{N}}{(D Q^{2})^{2} + \omega_{N}^{2}}\,\Theta(\omega_{N}).
\label{eq:dsigma-diff}
\end{multline}
The corresponding screened result, from \pref{eq:S-diff-screened} by
the same substitution, is the screened low-frequency bulk-metal
expression within the joint diffusive validity window,
\begin{multline}
\frac{\dd^{2}\Sigma_{\mathrm c}}{\dd\Omega\,\dd\epsilon}\bigg|_{N}^{\mathrm{diff},\mathrm{scr}} =
\frac{\me^{2}\,e^{2}\,\kappa^{2}\,D}{\pi^{2}\hbar^{4}\,Q^{2}}\,\frac{k_{f}}{k_{i}} \\
\times\,J_{N}^{2}(|\Qv\cdot\alphav|)\,
\frac{\omega_{N}}{\bigl[D(Q^{2}+\kappa^{2})\bigr]^{2} + \omega_{N}^{2}}\,\Theta(\omega_{N}).
\label{eq:dsigma-diff-screened}
\end{multline}
All displayed predictions in the remainder of this section, and in the
inhomogeneous-field extension below, use the screened result
\pref{eq:dsigma-diff-screened}. The unscreened \pref{eq:dsigma-diff}
is retained only as the intermediate object from which it is obtained,
and as the $\kappa\to0$ benchmark; the two differ by the single
replacement $DQ^{2}\to D(Q^{2}+\kappa^{2})$ in the denominator alone,
the numerator $DQ^{2}$ being fixed by charge conservation.
The Heaviside $\Theta(\omega_{N})$ inherits the positive-frequency
support of the $T=0$ diffusive spectral function and selects, at a
given projectile energy loss
$\hbar\omega_{\mathrm{loss}}=\epsilon(\vb k_i)-\epsilon(\vb k_f)$, only the
sidebands with $\omega_{N}\equiv\omega_{\mathrm{loss}}-N\wL>0$.
Summing over Floquet sectors at fixed $\omega_{\mathrm{loss}}$ gives
the connected diffusive spectrum
\begin{multline}
\frac{\dd^{2}\Sigma_{\mathrm c}}{\dd\Omega\,\dd\epsilon} =
\frac{\me^{2}\,e^{2}\,\kappa^{2}\,D}{\pi^{2}\hbar^{4}\,Q^{2}}\,\frac{k_{f}}{k_{i}}\\
\times\sum_{N} J_{N}^{2}(|\Qv\cdot\alphav|)\,
\frac{\omega_{N}}{\bigl[D(Q^{2}+\kappa^{2})\bigr]^{2} + \omega_{N}^{2}}\,
\Theta(\omega_{N}),
\label{eq:dsigma-diff-sum}
\end{multline}
with $\omega_{N} = \omega_{\mathrm{loss}} - N\wL$.

\paragraph*{Coherent channel in the diffusive regime.}
Equations~\pref{eq:dsigma-diff}--\pref{eq:dsigma-diff-sum} do not by
themselves constitute the complete driven-metal signal.  They describe
target-changing diffusive losses with \(\omega_N>0\).  A pump-induced
mean charge instead produces \(R_{-1}=\bar\varrho^{(+)}\), hence the
target-elastic coupling \(g_{\rm ind}\) of
\pref{eq:gind-density}.  If the laser frequency also lies within the
diffusive window \(|\wL|\tau\ll1\), this mean charge may be calculated
from the \emph{complex} screened susceptibility
\(\chi^{\mathrm{diff}}(\vb q,\wL)\) together with the external
potential and the sample boundary conditions.  Outside that window,
the appropriate finite-frequency complex response must be used.
In neither case can \(g_{\rm ind}\) be reconstructed from the
positive-frequency structure factor alone.

The coherent contribution has residual target-energy support
\(\delta(\hbar\omega_N)\), rather than the
\(\Theta(\omega_N)\) diffusive continuum.  At first Born order it is
therefore added through the second term of
\pref{eq:dsigma-loss-plus-pinem}; at all orders it is incorporated by
replacing \(g_0\) with \(g_0+g_{\rm ind}\) in the projectile channel
and retaining only the connected diffusive response above.  The
diffusion constant and screening wavevector can affect the complex
induced field, but they do not turn the target-elastic PINEM line into
a diffusive loss band.

Unlike the sharp connected plasmon comb
\pref{eq:dsigma-Drude}, the diffusion cross section per energy
consists of a
Bessel-weighted superposition of asymmetric, positive-frequency
Lorentzian-like bands. At $T=0$ each sideband is referenced to the
photon-shifted threshold
$\omega_{\mathrm{loss}}=N\wL$ (below which $\omega_{N}<0$ and the
sideband is forbidden by the $\Theta(\omega_{N})$ factor), with its
maximum displaced by the screened diffusive relaxation rate to
$\omega_{\mathrm{loss}}=N\wL+D(Q^{2}+\kappa^{2})$; the unscreened
position $N\wL+DQ^{2}$ is recovered only in the $\kappa\to0$
benchmark. The momentum transfer $\Qv$ sets the Bessel argument
through $|\delta\zeta_{\Qv}^{\rm hom}|$ and thereby controls the
Floquet-sideband strengths, while the momentum transfer $Q$ and the
screening wavevector $\kappa$ jointly set the screened relaxation scale
$D(Q^{2}+\kappa^{2})$, which controls the linewidth of each diffusive
band. The non-locality of
\pref{eq:chi-diff-screened} -- absent in the Drude form \pref{eq:drude} --
therefore directly shapes the spectral profile of the Floquet sidebands.

\subsection{Inhomogeneous laser field: eikonal--Volkov extension}
\label{sec:plasmon-inhom}
The Drude (\secref{sec:plasmon-cross-section}) and diffusive
(\secref{sec:plasmon-diffusive}) cross sections assume a strictly
homogeneous monochromatic drive. The eikonal--Volkov construction of
Secs.~\ref{sec:projdyn-eikonal}--\ref{sec:dsigma-inhom}
extends the cross section to slowly varying envelopes $\Ap(\vb r)$
via the bi-momentum master formula \pref{eq:dsigma-inhom}. We carry out
this extension first for the connected metallic loss sector and then
state its coherent PINEM completion.

For a translationally invariant equilibrium target -- a Drude metal or
a homogeneous disordered metal -- the bi-momentum Floquet structure
factor reduces, via \pref{eq:bimomentum-translation}, to its
momentum-diagonal counterpart. In the projectile-only reduction of the
connected target response relevant to
\secref{sec:plasmon-cross-section} and
\secref{sec:plasmon-diffusive}, the Floquet sectors collapse to
$n=n'=0$ and the bi-momentum factor reduces to
\begin{equation}
\SFc^{nn'}(\vb q_1,\vb q_2;\omega) \to
\delta_{n,0}\,\delta_{n',0}\,(2\pi)^3\delta^{(3)}(\vb q_1-\vb q_2)\,
S_{\mathrm c}(\vb q_1,\omega).
\label{eq:SF-weak-trans}
\end{equation}
Substituting \pref{eq:SF-weak-trans} into the inhomogeneous master
formula \pref{eq:dsigma-inhom} collapses the Floquet sums onto
$n=n'=N$ and the momentum integration onto its diagonal $\vb q'_1 = \vb q'_2$,
yielding the natural inhomogeneous-field analog of the loss-function
cross section \pref{eq:dsigma-loss}:
\begin{multline}
\frac{\dd^{2}\Sigma_{\mathrm c}}{\dd\Omega\,\dd\epsilon}\bigg|_{N} =
\qty(\frac{\me}{2\pi\hbar^{2}})^{2}\frac{k_{f}}{k_{i}}\,\frac{(2\pi)^3}{V}
\!\int\!\!\dd^{3}\vb q'\,|v(\vb q')|^{2}\\
\times\abs{\tilde{\mathcal F}^{(N)}_{\vb k_f\vb k_i}(\Qv-\vb q')}^{2}\,
S_{\mathrm c}(\vb q',\omega_{N}).
\label{eq:dsigma-inhom-metal}
\end{multline}
The kinematic momentum transfer $\Qv$ has been replaced by a
momentum-space convolution between the projectile Bessel kernel of
\secref{sec:dsigma-inhom} and the target structure
factor. The homogeneous limit must be taken at finite volume to avoid
a formal $\delta^{(3)}(\vb K)^{2}$: we replace $\Ap(\vb r)\to\Ap_{0}$
inside a sample of volume $V$, so that
\begin{equation}
\tilde{\mathcal F}_{\vb k_f\vb k_i}^{(N)}(\vb K)
\to
J_N\!\left(|\Qv\cdot\alphav|\right)
e^{\iu N(\varphi+\theta_{\Qv}^{\rm hom})}
\frac{V}{(2\pi)^3}\,g_V(\vb K).
\label{eq:F-finite-V}
\end{equation}
Here
\(
g_V(\vb K)\equiv V^{-1}\int_V\dd^3\vb r\,
e^{-\iu\vb K\cdot\vb r}
\)
is peak normalized, \(g_V(\vb 0)=1\), and has spectral width
of order \(V^{-1/3}\); this is the same finite-volume regularization used in
\secref{sec:dsigma-inhom}. $V$ here is the sample-volume
regulator and is distinct from the field-inhomogeneity scale $L$ of
\secref{sec:dsigma-inhom}. Squaring at finite $V$ and
using the distributional identity
\(
\bigl[V/(2\pi)^{3}\bigr]\,|g_{V}(\vb K)|^{2}
\xrightarrow{V\to\infty}\delta^{(3)}(\vb K),
\)
the weighted kernel $(2\pi)^3V^{-1}|\tilde{\mathcal F}^{(N)}|^2$ of \pref{eq:dsigma-inhom-metal} reduces to
$J_{N}^{2}(|\delta\zeta_{\Qv}^{\rm hom}|)\,\delta^{(3)}(\vb K)$, and
the $\vb q'$-integration collapses on $\vb q'=\Qv$, recovering the
per-volume result \pref{eq:dsigma-loss}.

\paragraph*{Drude metal.}
The Drude loss function \pref{eq:drude-loss} is local in $\vb q'$
and therefore factors out of the convolution integral in
\pref{eq:dsigma-inhom-metal}.  Using $|v(\vb q')|^{2}=16\pi^{2}e^{2}/q'^{4}$
and \pref{eq:S-loss-function} gives
\begin{equation}
\frac{\dd^{2}\Sigma_{\mathrm c}}{\dd\Omega\,\dd\epsilon}\bigg|_{N}^{\mathrm D,\,\mathrm{inh}} =
\frac{\me^{2}\,e^{2}}{\pi^{2}\hbar^{4}}\,\frac{k_{f}}{k_{i}}\,
\mathcal L^{\mathrm D}(\omega_{N})\,\mathcal K_{N}(\Qv),
\label{eq:dsigma-Drude-inhom}
\end{equation}
with the \emph{inhomogeneous Bessel--Coulomb weight}
\begin{equation}
\mathcal K_{N}(\Qv) \coloneqq \frac{(2\pi)^3}{V}\int\!\frac{\dd^{3}\vb q'}{q'^{2}}\,
\abs{\tilde{\mathcal F}^{(N)}_{\vb k_f\vb k_i}(\Qv-\vb q')}^{2}.
\label{eq:K_N}
\end{equation}
The spectral shape $\mathcal L^{\mathrm D}(\omega_{N})$ and the
angular weight $\mathcal K_{N}(\Qv)$ carry entirely separate
physics and factorize exactly because the Drude plasmon is
dispersionless (independent of $\vb q'$).
The plasmon-loss spectrum is therefore a comb of Lorentzian peaks
at energies $\hbar\omega_{p}+N\hbar\wL$, with Lorentzian width
$\sim\hbar\gamma$; only the sideband intensities are modified by the
field inhomogeneity, from the homogeneous form
$J_{N}^{2}(|\Qv\cdot\alphav|)/Q^{2}$ to the momentum-convolved
$\mathcal K_{N}(\Qv)$. In the strict homogeneous limit
$\mathcal K_{N}(\Qv)\to J_{N}^{2}(|\Qv\cdot\alphav|)/Q^{2}$,
recovering \pref{eq:dsigma-Drude} in the homogeneous limit; in the localized-field
limit, $\mathcal K_{N}(\Qv)$ acquires the angular and spatial profile
of the eikonal coupling parameter $\zeta_{\vb k}(\vb r)$ integrated
along the projectile trajectory.

\paragraph*{Diffusive metal.}
Using the screened diffuson \pref{eq:S-diff-screened} yields
\begin{multline}
\frac{\dd^{2}\Sigma_{\mathrm c}}{\dd\Omega\,\dd\epsilon}\bigg|_{N}^{\mathrm{diff},\,\mathrm{inh}} =
\frac{\me^{2}\,e^{2}\,\kappa^{2}\,D\,\omega_{N}}{\pi^{2}\hbar^{4}}\,\frac{k_{f}}{k_{i}}\,
\Theta(\omega_{N})\\
\times\frac{(2\pi)^3}{V}\!\int\!\frac{\dd^{3}\vb q'}{q'^{2}}\,
\abs{\tilde{\mathcal F}^{(N)}_{\vb k_f\vb k_i}(\Qv-\vb q')}^{2}\,
\frac{1}{\bigl[D(q'^{2}+\kappa^{2})\bigr]^{2} + \omega_{N}^{2}}.
\label{eq:dsigma-diff-inhom}
\end{multline}
Two physically distinct effects of the field inhomogeneity appear here:
(i)~the projectile momentum transfer is smeared over the
width of the actual projectile kernel (of order $1/L$ for smooth
local-SVE profiles at fixed moderate dressing)
by the kernel $|\tilde{\mathcal F}^{(N)}_{\vb k_f\vb k_i}|^{2}$, exactly as in the
Drude case; and (ii)~the diffusive linewidth $D(q'^{2}+\kappa^{2})$ is itself
sampled by the kernel and therefore inherits a width from the spatial
gradients of $\Ap(\vb r)$. Each diffusive sideband of the spectrum is
broadened in two qualitatively distinct ways simultaneously -- by the
projectile kinematic smearing through the kernel, and by the
target dissipative lineshape through the same $\vb q'$-integration.
The non-locality of $\chi^{\mathrm{diff}}$ ensures that these two
effects are entangled: in the homogeneous limit $L\to\infty$ both
collapse onto the sharp kinematic delta $\delta^{(3)}(\Qv-\vb q')$ and
the spectrum reduces to the Lorentzian-band result
\pref{eq:dsigma-diff-sum}; for localized near-fields the two combine
into a single broadened lineshape with width inherited from both the
kernel and the target.

\paragraph*{Coherent PINEM completion.}
 Equations~\pref{eq:dsigma-inhom-metal}--\pref{eq:dsigma-diff-inhom}
are connected loss contributions.  The coherent part omitted from
\pref{eq:SF-weak-trans} is the generally non-diagonal outer product
\(R_M(\vb q_1)R_{M'}^*(\vb q_2)\delta(\hbar\omega)\) of
\pref{eq:SF-coherent-connected}.  Its explicit first-Born
contribution is
\begin{multline}
\left.
\frac{\dd^2\sigma_{\rm coh}^{\rm 1B}}
{\dd\Omega\,\dd\epsilon}
\right|_N
=
\left(\frac{\me}{2\pi\hbar^2}\right)^2\frac{k_f}{k_i}
\delta(\hbar\omega_N)
{}\\
\times
\left|
\sum_n\int\!\dd^3\vb q'\,
v(\vb q')\,
\tilde{\mathcal F}^{(n)}_{\vb k_f\vb k_i}(\Qv-\vb q')\,
R_{N-n}(\vb q')
\right|^2 .
\label{eq:dsigma-inhom-coherent}
\end{multline}
Like the coherent term of \pref{eq:dsigma-loss-plus-pinem}, this is a
total cross section. Thus the off-diagonal momenta and all internal photon partitions
interfere inside one modulus square; they must not be replaced by the
momentum-diagonal convolution of
\pref{eq:dsigma-inhom-metal}.  For the same sample volume $V$,
$V$ times \pref{eq:dsigma-inhom-metal} plus \pref{eq:dsigma-inhom-coherent} gives the connected and
coherent pieces of \(|S^{\rm 1B}|^2\), respectively.  A
target-elastic measurement can additionally contain interference of
the latter amplitude with the reference-channel \(S^{\rm 0B}\)
amplitude, as discussed in \secref{sec:1Born}.

Quantitatively completing the example therefore requires the induced
density of the actual film, surface, or nanostructure.  It determines
\(g_{\rm ind}(\vb R)\) through \pref{eq:gind-density}.  For an
all-orders PINEM treatment, the preferable organization is to replace
the prescribed coupling by \(g_{\rm tot}=g_0+g_{\rm ind}\) inside the
eikonal projectile kernel and to retain only
\(\SFc\) for residual target excitations.  In that
organization \pref{eq:dsigma-inhom-coherent} is not added again: its
repeated coherent vertices are already resummed into the structured
PINEM Bessel amplitude.  This keeps that amplitude inside the
transverse Fourier integral and avoids counting the mean field twice.

\subsection{Physical interpretation of the metallic spectra}
\label{sec:plasmon-discussion}
The paragraphs below interpret the cross sections of
Secs.~\ref{sec:plasmon-cross-section}--\ref{sec:plasmon-inhom} through their
field-free reductions, coupling requirements, and the physical distinction
between the Drude-plasmon and diffusive regimes.
Unless stated otherwise, these cross sections are the connected,
target-changing contributions.  The coherent target-elastic channel is
the separate term \pref{eq:dsigma-loss-plus-pinem} in the homogeneous
case and \pref{eq:dsigma-inhom-coherent} in the inhomogeneous case.
The comparison with the Pines--Ritchie, PINEM sidebands, and
Altshuler--Aronov limits, with remarks on what is new in the
combined results, is collected in \appref{app:plasmon-literature}.

\input{figures/fig_metal_spectra}


\paragraph{Field-free reduction.} In the absence of the drive,
\(\alphav\to0\), \(R_{M\ne0}\to0\), and
\(J_N\to\delta_{N,0}\).  The optically induced coherent sidebands
therefore disappear, while the connected loss sector retains only
\(N=0\).  Equation~\pref{eq:dsigma-Drude} then reduces to the standard bulk-plasmon
EELS loss-function formula,
\begin{equation}
\frac{\dd^{2}\Sigma_{\mathrm c}}{\dd\Omega\,\dd\epsilon} =
\frac{\me^{2}\,e^{2}}{\pi^{2}\hbar^{4}\,Q^{2}}\,\frac{k_{f}}{k_{i}}\,
\mathcal L^{\mathrm D}(\omega_{0}),
\label{eq:dsigma-plasmon-bulk}
\end{equation}
in agreement with the standard dielectric-formulation
result~\cite{Pines1999,Egerton2011,RitchieRitchie1957}.
In the clean limit $\gamma\to0^{+}$ the Lorentzian $\mathcal
L^{\mathrm D}(\omega_{0})\to\frac{\pi\omega_{p}}{2}\delta(\omega_{0}-\omega_{p})$
recovers the idealized bulk-plasmon delta function.

\paragraph{Comb structure and coupling strength.}
The Bessel factor in the connected spectra is generated by the
prescribed projectile dressing.  Its dimensionless sideband argument is
\[
|\delta\zeta_{\Qv,\mathrm{rel}}^{\rm hom}|
=
\left|
\Qv\cdot\boldsymbol{\alpha}_{0,\mathrm{rel}}^{(+)}
\right|,
\]
with \(\boldsymbol{\alpha}_{0,\mathrm{rel}}^{(+)}\) the projectile-side
relativistically corrected complex quiver amplitude of
\pref{eq:alpha-NR-rel}. It reaches order unity when the projection of the
complex quiver amplitude along the momentum transfer becomes comparable
to the inverse momentum transfer scale,
\[
|\delta\zeta_{\Qv,\mathrm{rel}}^{\rm hom}|
\sim 1.
\]
For linearly polarized light this reduces to the usual projection
\[
|\delta\zeta_{\Qv,\mathrm{rel}}^{\rm hom}|
=
|\Qv\cdot\boldsymbol{\alpha}_{0,\mathrm{rel}}|
=
Q|\boldsymbol{\alpha}_{0,\mathrm{rel}}|\,
|\cos\beta_{\Qv}|,
\]
where \(\beta_{\Qv}\) is the angle between \(\Qv\) and the real quiver direction.

This argument must not be identified with the target-induced PINEM
argument \(2|g_{\rm ind}|\).  The former redistributes a
target-changing Drude or diffusive loss among optical sidebands; the
latter describes a target-elastic electron phase modulation by the
induced near field.  If both coherent fields are retained to all
orders, the target-elastic sideband weights are
\(J_N^2(2|g_0+g_{\rm ind}|)\), while the connected loss formulas use
the same total coherent field in their projectile kernel and the
connected target response only.

For a $1\,\mathrm{eV}$ drive and $Q\sim 1\,\mathrm{nm}^{-1}$,
the nonrelativistic value
$|\alphav|\approx 0.21\,\mathrm{nm}$ at peak intensity
$I=10^{19}\,\mathrm{erg\,s^{-1}cm^{-2}}$ follows in Gaussian--cgs
from \(I=c\EE_0^2/(8\pi)\), giving
\(\EE_0\simeq9.2\times10^4\,\mathrm{statV\,cm^{-1}}\).  It is
reduced at $200\,\mathrm{keV}$ to
$|\alphav^{\mathrm{rel}}|\approx |\alphav|/\gamma_{\mathrm p}\approx
0.15\,\mathrm{nm}$. Since $|\alphav^{\mathrm{rel}}|\propto\sqrt{I}$,
the nonrelativistic estimate for
$|\Qv\cdot\alphav|\sim1$ is
\(I\simeq2.3\times10^{20}\,\mathrm{erg\,s^{-1}cm^{-2}}\), whereas the
relativistically corrected value is
\(I\simeq4.4\times10^{20}\,\mathrm{erg\,s^{-1}cm^{-2}}\).
At the latter intensity
\(\Up^{\mathrm{rel}}\equiv\Up^{\mathrm{NR}}/\gamma_{\mathrm p}
\simeq4.6\,\mathrm{eV}\), well above
\(\hbar\wL=1\,\mathrm{eV}\); thus the unity-argument estimate at
\(Q\sim1\,\mathrm{nm}^{-1}\) lies outside the controlled
linear-Volkov window of \secref{sec:projdyn-volkov}.  In fact,
\(\Up^{\mathrm{rel}}\simeq\hbar\wL\) already near
\(I\simeq9.7\times10^{19}\,\mathrm{erg\,s^{-1}cm^{-2}}\), where
\(|\Qv\cdot\alphav^{\mathrm{rel}}|\simeq0.47\) for this \(Q\).
Accordingly, first- and second-order sidebands and comb fine structure
can appear for arguments of order \(0.3\)--\(0.5\), but the upper end
already approaches the boundary of the linear-Volkov regime; reaching
order-unity arguments while retaining control requires, for example, a
larger momentum transfer or a shorter-wavelength drive.

\paragraph{Plasmon dispersion.}
The local Drude form \pref{eq:drude} misses plasmon dispersion; the
leading Lindhard correction
$\omega_{p}^{2}(\vb q)\simeq\omega_{p}^{2} + 3v_{\mathrm F}^{2}q^{2}/5$
shifts each plasmon line to $\hbar\omega_{p}(\Qv) + N\hbar\wL$, leaving
the comb structure intact but the peak positions $\Qv$-dependent~\footnote{Sum-rule, selection-rule, and driven-target consistency follow respectively from \pref{eq:fsum}, \pref{eq:circ-selection}, and the open-system extension of \secref{sec:scope}; here we only note the Bessel-completeness redistribution $\protect\sum_N J_N^2=1$ and the $\Qv\cdot\alphav=0$ collapse to $N=0$ for circular polarization and axial $\Qv$.}.

\paragraph{Plasmon vs.\ diffusive limits.} The connected Drude result
\pref{eq:dsigma-Drude} and the diffusive result
\pref{eq:dsigma-diff-sum} are complementary descriptions of the same
metallic target in two non-overlapping regimes of energy loss. The
Drude calculation produces a Bessel-weighted comb of Lorentzian
plasmon-loss peaks of width $\hbar\gamma$ centered at
$\hbar\omega_{p}+N\hbar\wL$ -- the high-frequency response
of the collective plasmon mode; the comb is sharp and well-resolved
when $\gamma\ll\wL$. The diffusive calculation produces a
Bessel-weighted superposition of positive-frequency Lorentzian-like
bands with onset at $\hbar\omega_{\mathrm{loss}}=N\hbar\wL$ and peak
displaced by the screened diffusive relaxation rate to
$\hbar\omega_{\mathrm{loss}}=N\hbar\wL+\hbar D(Q^{2}+\kappa^{2})$ -- the
low-frequency particle--hole continuum. The dimensionless ratio
\begin{equation}
\xi_{\rm scr}(\Qv) \coloneqq \frac{D (Q^{2}+\kappa^{2})}{\wL}
\label{eq:xi}
\end{equation}
controls whether the Floquet comb is resolved in the diffusive regime:
for $\xi_{\rm scr}\ll 1$ the diffusive sidebands are narrow and the
Bessel-comb structure is clearly visible; for
$\xi_{\rm scr}\gtrsim 1$ the bands overlap and the comb washes out
into a Bessel-weighted smooth continuum. Because $\xi_{\rm scr}$
depends on the Thomas--Fermi wavevector $\kappa$ in addition to $D$ and
$Q$, and is dominated by $D\kappa^{2}$ whenever $\kappa\gg Q$, we quote
it parametrically rather than attaching a numerical range to a
particular metal: the resolved-comb regime requires
$D(Q^{2}+\kappa^{2})\ll\wL$, which is demanding for a well-screened
bulk metal and is most readily approached in weakly screened or
reduced-carrier-density settings.
Figure~\ref{fig:metal-spectra}(b) illustrates both cases
at $|\Qv\cdot\alphav|=1.5$; panel~(a) shows the corresponding Bessel-weight distributions at three representative couplings.
Neither panel contains the target-elastic PINEM sidebands.  Those lie
at \(\omega_N=0\) before instrumental broadening and are controlled by
the complex induced mean field, rather than by the connected linewidth
\(\gamma\) or \(D(Q^2+\kappa^2)\).  Their absence from the figure is
therefore a matter of missing optical geometry, not a prediction that
they vanish.

\paragraph{Inhomogeneous-field extension.} For the connected sector, the eikonal--Volkov
generalization of \secref{sec:plasmon-inhom} delivers the convolution
forms \pref{eq:dsigma-Drude-inhom} and \pref{eq:dsigma-diff-inhom} for
the Drude and diffusive cases respectively. For the Drude metal the
plasmon-loss line position is preserved (the plasmon is dispersionless,
hence insensitive to the momentum-transfer smearing) but the
Bessel sideband weight $J_{N}^{2}(|\delta\zeta_{\Qv}^{\rm hom}|)/Q^{2}$ is replaced
by the inhomogeneous Bessel--Coulomb weight $\mathcal K_{N}(\Qv)$ of
\pref{eq:K_N}, which inherits the spatial profile of $\Ap(\vb r)$.
For the diffusive metal the same spatial profile additionally broadens
each diffusive sideband through the convolution of the lineshape
$1/\bigl\{[D(q'^{2}+\kappa^{2})]^{2}+\omega_{N}^{2}\bigr\}$ with the kernel: the non-locality of
$\chi^{\mathrm{diff}}$ -- which makes \pref{eq:dsigma-diff-inhom}
qualitatively richer than \pref{eq:dsigma-Drude-inhom} -- produces a
feature absent from the Drude limit, in which the spatial profile of
the laser drive controls not only sideband intensities but also
sideband \emph{widths}.  The coherent sector has a different
mathematical structure: \pref{eq:dsigma-inhom-coherent} is a modulus
square of a momentum amplitude, not a diagonal convolution of a loss
function.  Its all-orders form is obtained from the trajectory
coupling \pref{eq:gind-density}, including its transverse phase, before
performing the far-field transform.

\paragraph{Consistent combination.}
There are two internally consistent levels of description.  At strict
first Born order one adds the coherent outer-product contribution to
$V$ times the connected per-volume Drude or diffusive loss formula, keeping any
target-elastic \(S^{\rm 0B}\)--\(S^{\rm 1B}\) interference at amplitude
level.  In the PINEM-resummed organization one instead puts
\(g_0+g_{\rm ind}\) into the projectile channel Hamiltonian and
constructs every residual metallic response from
\(\delta\hat\varrho_1=\hat\varrho_1-\bar\varrho\).  Adding the coherent
outer product once more in the latter organization would double count
the induced mean field.

\section{Off-diagonal Floquet coherences: a driven two-level worked example}
\label{sec:two-level}

The metallic examples of \secref{sec:plasmon} all operate in the
projectile-only limit for the connected target response,
\pref{eq:weak-target-dressing}, where the
off-diagonal $n\neq n'$ content of $S_{\mathrm{F,c}}^{nn'}$ vanishes
identically and the connected cross section reduces to a
Bessel-weighted sampling of the equilibrium structure factor.  The
full \(S_{\mathrm F}^{nn'}\) may nevertheless retain the separate
coherent outer product \(R_nR_{n'}^*\delta(\hbar\omega)\).  Away from
that target-elastic line, the connected sector of the Floquet matrix
developed in \secref{sec:floquet-response} contains, in addition to
its diagonal photon-resolved spectra, off-diagonal sectors that encode
coherences between Floquet components. To exhibit these coherences in closed form we treat
here the simplest analytically tractable example: a target
consisting of a single two-level dipole transition driven near
resonance by the optical field. Free-electron probing of driven two- and
three-level targets without projectile dressing was analyzed in
Ref.~\cite{ArqueLopez2022}; here the same field also dresses the
projectile, which produces the two-path Bessel interference derived below.

\subsection{Floquet eigenstates}
\label{sec:two-level-eigenstates}
The target has two levels $\ket{g}$ (energy $0$) and $\ket{e}$
(energy $\hbar\omega_0$), coupled to the optical drive of frequency
$\wL$ through a dipole interaction with bare Rabi amplitude
$\Omega\propto|\Ap_0|$. For the charge-density operator we use a
leading transition-density two-level truncation. Writing
$\varrho_{ab}(\vb q)\coloneqq
\bra{a}\hat\varrho_1(\vb q)\ket{b}$, we assume
\begin{equation}
\varrho_{ee}(\vb q)=\varrho_{gg}(\vb q),
\qquad
\varrho_{eg}(\vb q)=\varrho_{ge}(\vb q)
\eqqcolon q_{\rm dip}F(\vb q),
\label{eq:two-level-density-model}
\end{equation}
with $q_{\rm dip}$ a real scale; \epref{eq:two-level-density-model}
is the phase and form-factor convention adopted by this minimal model.
The common transition form factor
$F(\vb q)$ need not itself be real.
Indeed, $\hat\varrho_1(\vb q)^{\dagger}=\hat\varrho_1(-\vb q)$
constrains $F(-\vb q)=F(\vb q)^{*}$.
A relative phase between $\varrho_{eg}$ and $\varrho_{ge}$ would rotate
the off-diagonal Floquet coherence below, together with the phase of
the optical coupling. A nonzero diagonal contrast
$\varrho_{ee}-\varrho_{gg}$ would additionally generate an $n=-1$
Fourier component and enlarge the displayed $2\times2$ Floquet block
to the sector set $\{0,-1,-2\}$.

The drive couples through the length-gauge interaction
$-\hat{\vb D}\cdot\EEv=+\hat{\vb D}\cdot\dot{\vb A}/c$,
$\hat{\vb D}=\sum_iQ_i\hat{\vb r}_i$.
The co-rotating component of the transition matrix element is
\[
\left.\bra{e}\hat H_{\rm int}\ket{g}\right|_{\rm co}
=
\frac{\hbar\Omega_c}{2}\,
e^{-\iu(\wL t+\varphi)},
\]
with
\begin{equation}
\hbar\Omega_c=-\frac{\iu\wL}{c}\,\vb D_{eg}\cdot\vb A_0^{(+)*},
\quad \Omega\coloneqq|\Omega_c|,\quad \chi\coloneqq\arg\Omega_c .
\label{eq:two-level-coupling}
\end{equation}
Equation~\pref{eq:two-level-density-model} makes the transition density
real in real space, hence $\vb D_{eg}$ real; for linear polarization therefore
$e^{\iu\chi}=-\iu\,\mathrm{sgn}(\vb D_{eg}\cdot\vb A_0)$ and
$e^{2\iu\chi}=-1$.
We work at positive detuning $\Delta\coloneqq\wL-\omega_0>0$ in the
adiabatic-continuation convention of \secref{sec:floquet-gauge}, so
that the rotating-wave approximation (RWA)
eigenstates of $\hat H_1(t)$ connect to $\ket{g}$ and
$\ket{e}$ as $\Omega\to 0$ and read
\begin{subequations}
\begin{align}
\ket{\Phi_+(t)} &= e^{-\iu E_+t/\hbar}
\bigl[\cos\tfrac{\theta}{2}\ket{g}\nonumber\\
&\quad+\sin\tfrac{\theta}{2}\,e^{\iu\chi}e^{-\iu(\wL t+\varphi)}\ket{e}\bigr],\\
\ket{\Phi_-(t)} &= e^{-\iu E_-t/\hbar}
\bigl[-\sin\tfrac{\theta}{2}\,e^{-\iu\chi}e^{+\iu(\wL t+\varphi)}\ket{g}\nonumber\\
&\quad+\cos\tfrac{\theta}{2}\ket{e}\bigr],
\end{align}
\label{eq:two-level-states}
\end{subequations}
with quasienergies
$E_{+}=\hbar(\WR-\Delta)/2$,
$E_{-}=\hbar\omega_0 + \hbar(\Delta-\WR)/2$,
generalized Rabi frequency
$\WR\coloneqq\sqrt{\Delta^{2}+\Omega^{2}}$, and
$\tan\theta = \Omega/\Delta$. The angle $\theta$ mixes the bare
ground and excited components of the Floquet states: in $\Phi_+$, in
which the target is prepared, their weights are $\cos^2(\theta/2)$ and
$\sin^2(\theta/2)$. We further restrict to the branch $\WR<\wL$,
which is automatically satisfied in the near-resonant RWA regime
$|\Delta|,\Omega\ll\wL$ assumed at the outset. Under this restriction the
target-side loss frequency $\wL-\WR$ that shows up in the
results below [Eqs.~\pref{eq:two-level-SF} and \pref{eq:dsigma-2lev}] is
positive in the chosen Floquet representative. The laser phase $\varphi$ of
\pref{eq:Acos}, together with the coupling phase $\chi$, is carried explicitly by the $\ket{e}$-component in
$\ket{\Phi_+}$ and by the $\ket{g}$-component in $\ket{\Phi_-}$
(after the $\ell_-=+1$ Floquet-zone shift implemented by the
adiabatic-continuation convention), reflecting that the projectile and the target are
driven by the \emph{same} coherent field.

\subsection{Floquet density components}
\label{sec:two-level-density}
Using the Floquet--Fourier convention implicit in \pref{eq:varrho-exact},
the matrix element
$\bra{\Phi_-(t)}\hat\varrho_1(\vb q)\ket{\Phi_+(t)}$ obtained from
the Floquet states \pref{eq:two-level-states} has two nonzero Fourier
components:
\begin{align}
\tilde\varrho_{-+}^{(0)}(\vb q) &= q_{\rm dip}F(\vb q)\,
\cos^{2}\!\tfrac{\theta}{2},\nonumber\\
\tilde\varrho_{-+}^{(-2)}(\vb q) &= -q_{\rm dip}F(\vb q)\,
e^{2\iu\chi}e^{-2\iu\varphi}\sin^{2}\!\tfrac{\theta}{2},
\label{eq:two-level-rho-n}
\end{align}
all others vanishing. The $e^{2\iu(\chi-\varphi)}$ phase carried by
$\tilde\varrho^{(-2)}$ originates in equal parts from the $\ket{e}$-component
of the incoming $\ket{\Phi_+}$ and the conjugated $\ket{g}$-component of
the outgoing $\ket{\Phi_-}$ in \pref{eq:two-level-states}; for linear
polarization $e^{2\iu\chi}=-1$; an equivalent dressed-state phase
redefinition would redistribute the $2\varphi$ phase symmetrically
between $\tilde\varrho^{(0)}$ and $\tilde\varrho^{(-2)}$ without
altering the cross section. The connected Floquet structure factor
defined below \pref{eq:SF}, with $\nu$ denoting the spectral-argument variable
(reserved here to avoid collision with the drive frequency $\wL$),
acquires a nonzero $2\times 2$ block:
\begin{multline}
S_{\mathrm{F,c}}^{nn'}(\vb q,\nu) =
q_{\rm dip}^{2}|F(\vb q)|^{2}\,s^{nn'}(\theta,\varphi)\\
\times\delta\bigl(\hbar[\nu-(\wL-\WR)]\bigr),
\label{eq:two-level-SF}
\end{multline}
with $n,n'\in\{0,-2\}$ and (writing $c\equiv\cos\tfrac{\theta}{2}$,
$s\equiv\sin\tfrac{\theta}{2}$)
\begin{align}
s^{0,0}(\theta) &\coloneqq c^{4},\nonumber\\
s^{-2,-2}(\theta) &\coloneqq s^{4},\nonumber\\
s^{0,-2}(\theta,\varphi) &\coloneqq -e^{+2\iu(\varphi-\chi)}s^{2}c^{2},\nonumber\\
s^{-2,0}(\theta,\varphi) &\coloneqq -e^{-2\iu(\varphi-\chi)}s^{2}c^{2}.
\label{eq:two-level-s}
\end{align}
The rank-one connected Lehmann structure (only one intermediate Floquet state)
saturates the Cauchy--Schwarz bound \pref{eq:CS}:
$|s^{0,-2}|^{2} = s^{0,0}\,s^{-2,-2}$.

The  connected
Floquet block in Eqs.~\pref{eq:two-level-SF}--\pref{eq:two-level-s}
describes the target-changing
channel \(\Phi_+\to\Phi_-\).  It therefore does not include the
target-elastic coherent mean-density channel \(f=i=+\).  For the same
two-level density model, that omitted channel has
\begin{align}
R_0(\vb q)&=\varrho_{gg}(\vb q),\nonumber\\
R_{\pm1}(\vb q)&=
q_{\rm dip}F(\vb q)\sin\frac{\theta}{2}\cos\frac{\theta}{2}
\nonumber\\
&\quad\times e^{\pm\iu(\varphi-\chi)}.
\label{eq:two-level-coherent-R}
\end{align}
The harmonics \(R_{\pm1}\) produce target-elastic PINEM sidebands at
\(\omega_N=0\), distinct from the Rabi-shifted target-changing line
worked out below; they return the target to $\Phi_+$. Substituting $R_{-1}$ into
\pref{eq:gind-density} gives the explicit two-level PINEM coupling
\begin{multline}
g_{\rm ind}^{\rm 2lev}(\vb R)
=\frac{\iu q_{\rm dip}\sin\theta\,e^{\iu(\chi-\varphi)}}{2\hbar v_{\rm e}}
\int\!\frac{\dd^2\vb q_\perp}{(2\pi)^2}\,
e^{\iu\vb q_\perp\cdot\vb R}\\
\times v(\vb q)F(\vb q)\Big|_{q_z=\wL/v_{\rm e}},
\label{eq:two-level-gind}
\end{multline}
with $\sin\theta=\Omega/\WR$. Its modulus grows as $\Omega/\Delta$
at weak drive and saturates at maximal mixing; its phase matters
whenever another coherent amplitude $g_0$ is present. In the
localized-field, straight-trajectory description the coherent mean
potential generates the PINEM probabilities
$P_N(\vb R)=J_N^2(2|g_0+g_{\rm ind}^{\rm 2lev}|)$; the angular
distribution follows from \pref{eq:pinem-angular} of \appref{app:elastic},
and its relation to the strict first-Born result is stated in
\appref{app:two-level}. A complete first-Born spectrum adds the coherent
term of \pref{eq:dsigma-loss-plus-pinem}, with
\pref{eq:two-level-coherent-R}, to the displayed transition; see
\pref{eq:two-level-coherent-cross-section} below.
An all-orders coherent treatment instead constructs
\(g_{\rm ind}\) from these harmonics through
\pref{eq:gind-density} and exponentiates it in the
projectile channel, while retaining the \(\Phi_+\to\Phi_-\) matrix
elements as part of the connected residual response.

\subsection{Cross section: a secondary sideband comb}
\label{sec:two-level-cross-section}
Substituting the connected block \pref{eq:two-level-SF} into the
connected part of the homogeneous-field Bethe--Floquet cross section
\pref{eq:dsigma-SF} gives the target-changing cross section
\begin{multline}
\frac{\dd^{2}\sigma_{\mathrm c}^\text{1B}}{\dd\Omega\,\dd\epsilon}\bigg|^{\text{2lev}}_{N}
=
\qty(\frac{\me}{2\pi\hbar^{2}})^{2}
|v(\Qv)|^{2}
\frac{k_f}{k_i}\,
q_{\rm dip}^{2}|F(\Qv)|^{2}\\
\times \delta\!\left(\hbar[\omega_N-(\wL-\WR)]\right)
\left|\mathcal B_N (\vb Q)\right|^{2},
\label{eq:dsigma-2lev}
\end{multline}
where we introduced 
\begin{multline}
\label{eq:BN}
\mathcal B_N (\vb Q) \coloneqq
J_N\!\left(|\Qv\cdot\alphav|\right)
\cos^{2}\frac{\theta}{2}
\\
- e^{2\iu(\chi+\theta_{\Qv}^{\rm hom})}
J_{N+2}\!\left(|\Qv\cdot\alphav|\right)
\sin^{2}\frac{\theta}{2}.
\end{multline}
The target components carry the phase $\varphi-\chi$ and the
projectile harmonics $\varphi+\theta_{\Qv}^{\rm hom}$; the absolute laser
phase $\varphi$ therefore cancels from the relative interference factor
$e^{2\iu(\chi+\theta_{\Qv}^{\rm hom})}$. This factor is convention independent
within the density convention \pref{eq:two-level-density-model}; the phase
enters doubled because the two routes differ by two harmonics. Including
the mixing minus sign, the coefficient combines the projections of the field on the momentum transfer and on the
transition dipole,
$-e^{2\iu(\chi+\theta_{\Qv}^{\rm hom})}=e^{2\iu[\arg(\Qv\cdot\vb A_0^{(+)})
-\arg(\vb D_{eg}\cdot\vb A_0^{(+)})]}$ in the nonrelativistic core. For
linear polarization this factor equals one [\epref{eq:elliptic-to-linear}
and $e^{2\iu\chi}=-1$], and the connected sideband weight is
\begin{equation}
\label{eq:bessel-diff}
\begin{aligned}
W_N^{\rm lin}&\coloneqq\bigl|\mathcal B_N^{\rm lin}(\Qv;\theta)\bigr|^2,\\
\mathcal B_N^{\rm lin}(\Qv;\theta)&\coloneqq
J_N(\Qv\cdot\boldsymbol{\alpha}_0)
\cos^{2}\frac{\theta}{2}
+
J_{N+2}(\Qv\cdot\boldsymbol{\alpha}_0)
\sin^{2}\frac{\theta}{2}.
\end{aligned}
\end{equation}
The connected, target-changing cross section is a comb
of energy losses, the
$N$th peak sitting at $\hbar\omega_{\mathrm{loss}} = N\hbar\wL +
\hbar(\wL-\WR)$. In the adiabatic Floquet branch chosen
here, $\wL-\WR$ is the quasienergy transition frequency
that connects continuously to the bare transition $\omega_0$ as
$\Omega\to 0$ at fixed $\Delta>0$ (where $\WR\to\Delta$ and
$\wL-\WR\to\wL-\Delta=\omega_0$); the displayed comb is
therefore the AC-Stark/Rabi-shifted analogue of the undriven Bethe
line at $\omega_{\mathrm{loss}}=N\wL+\omega_0$.
Equation~\pref{eq:bessel-diff} has a two-path interpretation in the
space of dressed-state components. The collision can excite the
$\ket{g}$ component of $\Phi_+$ into the $\ket{e}$ component of
$\Phi_-$, or de-excite its $\ket{e}$ component into the $\ket{g}$
component of $\Phi_-$. The first route has mixing weight
$\cos^2(\theta/2)$. The second has weight
$-\sin^2(\theta/2)\,e^{2\iu\chi}e^{-2\iu(\wL t+\varphi)}$: one optical
phase comes from the incoming component and the other from the
conjugated outgoing component, and the dipole coupling fixes
$e^{2\iu\chi}=-1$ for linear polarization, giving the plus sign in
\pref{eq:bessel-diff}. The target Fourier orders are therefore $M=0$
and $M=-2$; at fixed observed sideband $N$ the partition $n+M=N$
selects projectile orders $n=N$ and $n=N+2$. Both routes end in the
same target state and the same outgoing electron channel, so they add
coherently: $W_N^{\rm lin}$ is a perfect square, reflecting the rank-one
Lehmann structure of \pref{eq:two-level-SF}, and is not reducible to
the diagonal comb $J_N^2$ of the projectile-only limit. The potential
$M=-1$ contribution cancels because the diagonal densities in
\pref{eq:two-level-density-model} are equal; unequal diagonal densities
would add a $J_{N+1}$ term. The derivation is given in
\appref{app:two-level}, and \figref{fig:two-level} shows the
resulting sideband redistribution.

\input{figures/fig_two_level}

Since $\varphi$ cancels, an observable laser-phase dependence
requires a separately phase-controlled interaction, whose signal must
be calculated with the corresponding projectile kernel. The present two-path interference is
already visible in the sideband weights without such a phase scan.

With the same projectile coefficients $B_n$ and prefactor
$\mathcal C_{\Qv}$ as in \pref{eq:dsigma-SF}, the target-elastic
first-Born contribution of the harmonics
\pref{eq:two-level-coherent-R} is
\begin{multline}
\left.\frac{\dd^2\sigma_{\rm coh}^{\rm 1B}}{\dd\Omega\,\dd\epsilon}\right|_N
=\mathcal C_{\Qv}\,\delta(\hbar\omega_N)\,
\Bigl|B_N\varrho_{gg}(\Qv)\\
+\frac{q_{\rm dip}F(\Qv)\sin\theta}{2}
\bigl(e^{\iu(\varphi-\chi)}B_{N-1}+e^{-\iu(\varphi-\chi)}B_{N+1}\bigr)\Bigr|^2 .
\label{eq:two-level-coherent-cross-section}
\end{multline}
For linear polarization the modulus becomes
$|\varrho_{gg}J_N(x)+\iu\sigma q_{\rm dip}F\sin\theta\,J_N'(x)|$, with
$x=\Qv\cdot\boldsymbol\alpha_0$ and $\sigma=\mathrm{sgn}(\vb D_{eg}\cdot\vb A_0)$.
It adds to \pref{eq:dsigma-2lev} because the final target states are
orthogonal; a zeroth-order reference collected in the same
target-elastic channel must first be combined at amplitude level
(\appref{app:elastic}). The static term $\varrho_{gg}B_N$ describes
laser-assisted scattering from the static mean density and is not an
induced near-field contribution.

Combining the target-elastic coherent contribution with the
target-changing connected contribution, the complete first-Born
two-level spectrum in the $N$th sideband reads
\begin{multline}
\left.
\frac{\dd^{2}\sigma^{\rm 1B}}
{\dd\Omega\,\dd\epsilon}
\right|^{\rm 2lev}_{N}
\\
=
\mathcal C_{\Qv}\,
\delta(\hbar\omega_N)
\Biggl|
B_N(\Qv)\varrho_{gg}(\Qv) +
\\
\frac{q_{\rm dip}F(\Qv)\sin\theta}{2}
\left[
e^{\iu(\varphi-\chi)}B_{N-1}(\Qv)
+
e^{-\iu(\varphi-\chi)}B_{N+1}(\Qv)
\right]
\Biggr|^2
\\
+
\mathcal C_{\Qv}\,
q_{\rm dip}^{2}|F(\Qv)|^{2}
\left|\mathcal B_N(\Qv)\right|^{2}
\delta\!\left(
\hbar[\omega_N-(\wL-\WR)]
\right).
\label{eq:2lstotal}
\end{multline}
Equation~\pref{eq:2lstotal} thus contains two distinct sideband
families:
The coherent sector has peaks at electron losses $N\hbar\wL$,
whereas the connected transition has peaks at
$(N+1)\hbar\wL-\hbar\WR$. Near resonance a connected peak
therefore lies $\hbar\WR$ below the neighboring coherent peak with
$M=N+1$; the two families are separated when linewidths and energy
resolution are small compared with $\hbar\WR$. Overlap after
broadening does not create interference between their orthogonal final
target states.

Figure~\ref{fig:eels-spectrum} renders the cross section
\pref{eq:2lstotal} as an energy-loss spectrum.
For a passive target the familiar $J_N^2$ PINEM comb is recovered,
whereas target Floquet dynamics generate a second comb at
$N\hbar\wL+\hbar(\wL-\WR)$.
Equivalently, each of these peaks lies $\hbar\WR$ below the
neighboring $(N{+}1)$st target-elastic sideband.
Its non-Bessel sideband weights $W_N^{\rm lin}$ of \pref{eq:bessel-diff}
are the spectroscopic signature of the off-diagonal Floquet coherence
$s^{0,-2}$. The coherent harmonics $R_{\pm1}$ additionally reweight the
target-elastic comb through the induced-dipole term
$(\Qv\cdot\vb P_0)J_N'(x)$ (\appref{app:two-level}), while the
relative overall strength of the two combs is model dependent. The figure
uses $q_{\rm dip}^2|F|^2/|\varrho_{gg}|^2=1$ and, in the dipole limit with
real $\varrho_{gg}$, a real positive ratio
$(\Qv\cdot\vb P_0)/\varrho_{gg}=\sin\theta$; for complex $F$ or
$\varrho_{gg}$ this ratio is complex in general.

The sideband-summed closure of \secref{sec:KW-closure}
has a simple interpretation in this two-level example. At fixed target
loss and within the same soft-sideband approximation
(with $x_{\Qv} \coloneqq \left|\Qv\cdot\alphav\right|$),
\begin{align}
\sum_N |\mathcal B_N(\Qv)|^2
&=\sum_N \Bigl|J_N(x_{\Qv})c^2 \nonumber\\
&\qquad-e^{2\iu(\chi+\theta_{\Qv}^{\rm hom})} J_{N+2}(x_{\Qv})s^2 \Bigr|^2
\label{eq:sumN_BN}
\\ &= s^{0,0}(\theta) + s^{-2,-2}(\theta),
\end{align}
where \epref{eq:two-level-s} has been used.
The interference term proportional to
\(J_N(x_{\Qv})J_{N+2}(x_{\Qv})\) cancels in the
complete sideband sum due to
\pref{eq:Bessel-closure}.
Thus the sideband-integrated signal for this transition measures only
the diagonal Floquet-sector trace of the two-level block, while the
coherence responsible for the sideband-resolved interference pattern is
lost in the sum.

\paragraph*{Strong mixing.} The strong-mixing limit $\theta\to\pi/2$
-- approached by $\Omega\gg\Delta$ at fixed detuning, or by
$\Delta\to0^{+}$ at fixed $\Omega$ after the adiabatic Floquet
branch is fixed -- gives $\sin^2(\theta/2)=\cos^2(\theta/2)=\tfrac12$,
reducing \pref{eq:BN}, and for linear polarization \pref{eq:bessel-diff}
(by continuity at $x=0$), to
\begin{multline}
\mathcal B_N(\vb Q)\;\to\;
\tfrac{1}{2}\Bigl[J_N(x_{\Qv})
-e^{2\iu(\chi+\theta_{\Qv}^{\rm hom})}
J_{N+2}(x_{\Qv})\Bigr],
\\
\mathcal B_N^{\rm lin}(\Qv;\tfrac{\pi}{2})=\frac{N+1}{x}\,J_{N+1}(x).
\label{eq:BNstrong}
\end{multline}
Equal-weight interference between the $N$- and
$(N{+}2)$-channel projectile amplitudes is the minimal-model signature
of the off-diagonal coherence $s^{0,-2}$ entering the cross section.
Since $\mathcal B^{\rm lin}_{-1}=-\cos\theta\,J_1(x)$, at maximal
mixing the linearly polarized RWA model predicts a dark connected line
at $\epsilon(\vb k_i)-\epsilon(\vb k_f)=-\hbar\WR$, where the two routes cancel for every
$x$. Deleting the off-diagonal coherence instead leaves
$\cos^4(\theta/2)J_N^2+\sin^4(\theta/2)J_{N+2}^2$; a single $J_N^2$ comb
is recovered only when the second harmonic vanishes, as in the unmixed
limit $\theta\to0$.

\section{Outlook}
\label{sec:outlook}

\paragraph*{A unifying perspective.}
The formalism places three previously separate languages -- Bethe
EELS, laser-assisted inelastic Floquet scattering, and Volkov/eikonal
projectile dressing -- into a single many-body response
framework. Time-periodic driving promotes the equilibrium density correlator
$S(\vb q,\omega)$ to a matrix-valued Floquet correlator. At first Born
order the target enters through the bi-momentum object
$\SF^{nn'}(\vb q_1,\vb q_2;\omega)$ already in the
exact formula \pref{eq:dsigma-bimomentum}, independently of the
projectile approximation; a spatially inhomogeneous external field
instead broadens the projectile kernel, causing the cross section to
sample this pre-existing target response through a momentum
convolution.  Crucially, the full target correlator first separates
into a coherent mean-density outer product and connected fluctuations.
The former gives the target-elastic weak-PINEM vertex
\(g_{\rm ind}\), while the latter contains the target-changing
many-body response. The
classical limits are explicit: the projectile-only connected-response limit
\pref{eq:weak-target-dressing} recovers Bethe theory with the
Bessel-weighted projectile comb on top, and the homogeneous limit
reduces the bi-momentum kernel to the diagonal Floquet structure
factor \pref{eq:SF}. In the presence of a many-body target the
projectile Bessel sidebands are no longer merely gain/loss peaks of
the electron: they serve as phase- and photon-resolved probes of
the density--density correlations of the driven solid.

\paragraph*{Candidate next targets.}
Equation~\pref{eq:dsigma-bimomentum} reduces the general scattering
problem to the calculation of the bi-momentum Floquet density response
and its contraction with a projectile kernel. For a translationally
invariant target and a homogeneous projectile field this reduces to
\pref{eq:dsigma-Volkov}--\pref{eq:dsigma-SF} and to the single-momentum
Floquet dynamic structure factor $S_{\mathrm F}^{nn'}(\Qv,\omega)$,
whose general structure and sum rules are established in
\secref{sec:floquet-response} and whose simplest projectile-only
connected-response
application -- the bulk plasmon-loss Bessel comb of a Drude metal -- is
worked out analytically in \secref{sec:plasmon}. Beyond
this Drude warm-up, several classes of targets stand out as natural
next applications. (a)~Driven semiconductors and band insulators
near a one-photon resonance, where simplified Bloch--Volkov or Floquet
two-band models can yield analytically tractable forms of
$S_{\mathrm{F,c}}^{nn'}$ and the connected off-diagonal coherences
$S_{\mathrm{F,c}}^{n\neq n'}$ could become experimentally accessible. 
The recent free-electron homodyne technique of
Ref.~\cite{Gaida2024} uses a phase-controlled reference interaction
as a local oscillator to extract sample-induced modulations of the
free-electron wave function with attosecond temporal resolution.
It may therefore be suited to resolving such phase-sensitive
coherences.
(b)~Correlated solids subject to a sub-gap drive, where
$S_{\mathrm{F,c}}^{nn'}$
encodes Floquet-engineered band renormalizations, dressed plasmons, and
driven-superconducting features.
(c)~Finite metallic films, membranes, and interfaces, where broken
target-side translational symmetry makes the bi-momentum response
$\SF^{nn'}(\vb q_1,\vb q_2;\omega)$ of
\pref{eq:SF-bimomentum} the relevant object, and where surface
response, boundary conditions, and the dimensional crossover of the
diffusion modes modify the bulk estimates of
\secref{sec:plasmon-diffusive}; PINEM of such films and membranes is
already established experimentally~\cite{Meuret2024}. Connecting to
interference-modulated sideband combs additionally requires the
self-consistent optical field inside the film, which lies outside the
strictly external-field scope adopted here
(\secref{sec:hamiltonian-light-field}) and would have to be supplied
separately, \eg{} from a thin-film scattering
calculation~\cite{Muller2024}.
(d)~The coherent completion of the worked metallic examples:
specifying the illumination amplitude and sample geometry, solving the
electromagnetic boundary-value problem for \(R_M(\vb q)\), and
constructing \(g_{\rm ind}(\vb R)\) through
\pref{eq:gind-density}.  This would make the target-elastic terms
\pref{eq:dsigma-loss-plus-pinem} and
\pref{eq:dsigma-inhom-coherent} quantitative for realistic films and
nanostructures.  In energy space the coherent sector lies at
\(\omega_N=0\), coinciding with the zero-target-loss laser-assisted comb
and with the \(N\hbar\wL\) onset of a diffusive band before broadening.
Separating it therefore requires adequate energy resolution or, where
the induced field is laterally smooth, angular discrimination against
its near-forward weight. Strongly structured near fields can spread
that weight over transverse momenta of order the inverse structure
size, so energy discrimination remains the geometry-independent
criterion.  The amplitude-level matching to the connected hierarchy is
deferred to the companion analysis.

\paragraph*{Strongly driven absorptive regime and open-system extension.}
Beyond these specific examples, the principal open problem the present
formalism leaves on the table is the strongly driven \emph{absorptive}
regime in which assumption~(v) of \secref{sec:scope} breaks down.
The substitution $S\to S_{\mathrm F}^{nn'}$ in the cross section then
requires the nonequilibrium density correlator of the actual driven
steady state, most naturally obtained from Floquet--Keldysh or
quantum-kinetic frameworks (Floquet--RPA~\cite{Asmar2022}, Floquet--DMFT,
Floquet--Boltzmann) rather than from the pure-state Floquet--Lehmann
representation of \secref{sec:floquet-response-FDT}. The broader
landscape of free-electron quantum-nanophotonic techniques within
which these extensions sit is reviewed in the recent roadmap of
García de Abajo and coworkers~\cite{deAbajoRoadmap2025}.

\paragraph*{Closed-system identities and their open-system counterparts.}
Of the closed-system identities established in
\secref{sec:floquet-response}, Hermiticity \pref{eq:SF-Hermitian}
and positivity \pref{eq:CS} carry over to the density-correlation
matrix of any nonequilibrium steady state, while the Floquet $f$-sum
rule \pref{eq:fsum} requires only that the substitute target
correlator be obtained from a conserving response approximation. The auxiliary identity
\pref{eq:spectral-response-identity} is algebraic and does not by itself
relate the scattering correlator to the physical retarded response, in
either the closed or the open setting; any such relation requires a
specified preparation and, for a dissipative steady state, its bath model
(\appref{app:lehmann-coarse-graining}). The target-elastic coherent line is
absent from the commutator and must be supplied separately.

\paragraph*{Bottom line.}
Laser-assisted EELS in present-generation UTEM facilities now has a
two-part target: the connected Floquet density-correlation matrix
$S_{\mathrm{F,c}}^{nn'}(\Qv,\omega)$ of the driven solid, accessible
through Bessel-weighted plasmon-loss and diffusive sideband combs for
diagonal entries, and the pump-induced mean-density harmonics
\(R_M(\vb q)\), which determine the target-elastic PINEM amplitude.
The present work identifies the connected matrix as the quantity that
future Floquet--RPA, Floquet--DMFT, Floquet--Keldysh, or related
many-body treatments should supply, while the coherent harmonics
require the accompanying optical boundary-value problem. Neither is
evaluated here
for a realistically driven interacting solid, the metallic examples
being confined to the projectile-only connected-response limit in which only the
diagonal $n=n'=0$ entry survives. The plasmon-loss and screened
diffusive combs of \figref{fig:metal-spectra} provide experimentally
testable quasistatic benchmarks; quantitative comparison with thin
films, interfaces, or aloof-beam geometries requires the appropriate
screened, geometry-dependent, and where necessary retarded
electromagnetic response. The Bessel-interference signature of
\figref{fig:two-level} provides a falsifiable benchmark for few-level
or resonantly driven targets. Access to off-diagonal Floquet coherences would require a future
phase-sensitive protocol in which the projectile kernel is varied
independently while the target Floquet preparation is held fixed.
For the target-elastic sector, the corresponding all-orders
implementation must promote \(g_0+g_{\rm ind}\) into the projectile
channel and retain only connected residual target correlations, as
stated in \secref{sec:pinem-bridge}.

\section*{Data Availability}

No experimental data were generated or analyzed in this work.
The numerical data and scripts used to generate the figures are
available from the corresponding author upon reasonable request.

\begin{acknowledgments}
During manuscript preparation, the authors used Claude Opus~5
(Anthropic) and GPT-5.6-sol (OpenAI) for editorial assistance,
algebraic consistency checks, and critical examination of selected
derivations. The authors specified the questions and supplied the
relevant mathematical context, independently rederived and verified
every adopted correction, and take full responsibility for the
scientific content.

We acknowledge funding by the
Deutsche Forschungsgemeinschaft (DFG, German Research Foundation) through 
GRK 2905 (project-ID 502572516), through CRC 1277 (Project-ID 314695032, subprojects A03 and C02), 
through RTG 2905 (project number 502572516), through EV30/12-1, EV30/14-1, EV30-16-1, through the German Excellence Strategy – EXC3112/1 – 533767171 (Center for Chiral Electronics) and, in the initial phase of the project, from the Volkswagen Foundation as part of the Lichtenberg Professorship “Ultrafast nanoscale dynamics probed by time-resolved electron imaging”.
Funding by the Free State of Bavaria through
the Lighthouse project “Free-electron states as ultrafast probes for
qubit dynamics in solid-state platforms” within the Munich Quantum Valley initiative is acknowledged.

\end{acknowledgments}

\appendix

\section{Floquet density response}
\label{app:lehmann-coarse-graining}
This appendix collects material related to the full Floquet structure
function \(S_{\mathrm F}^{nn'}\) and its connected component
\(S_{\mathrm{F,c}}^{nn'}\), introduced in
\secref{sec:floquet-response}. We first distinguish the physical retarded density response
from a scattering-adapted auxiliary spectral matrix and state explicitly how the coherent
mean-density line is restored to reconstruct the full scattering
correlator. We then discuss how the connected Lehmann delta comb should
be interpreted for macroscopic targets through coarse graining. We add
a discussion of fluctuation-dissipation relations and references for
the metallic examples.

\subsection{Floquet density correlations and retarded response}
\label{sec:floquet-density-response}
We first distinguish the density correlation measured by scattering from
the retarded density response used in linear-response theory. The
scattering cross section is governed by the unsymmetrized density
correlation
\begin{equation}
C_{\rho\rho}(\Qv_1,t_1;\Qv_2,t_2)
\coloneqq
\bra{\Phi_0}
\hat\varrho(-\Qv_1,t_1)\hat\varrho(\Qv_2,t_2)
\ket{\Phi_0},
\label{eq:density-correlation}
\end{equation}
where the target is prepared in the Floquet eigenstate \(\ket{\Phi_0}\).
This object is not a retarded commutator. It is the unsymmetrized
density correlation whose positive-loss components and
target-elastic coherent line enter the scattering probability.

Inserting a complete set of Floquet states and using the Floquet--Fourier
expansion
\begin{equation}
\bra{\Phi_a}
\hat\varrho(\Qv,t)
\ket{\Phi_0}
=
e^{\iu(E_a-E_0)t/\hbar}
\sum_n
e^{\iu n\wL t}
\tilde\varrho_{a0}^{(n)}(\Qv),
\label{eq:rho-floquet-fourier-response}
\end{equation}
one obtains a double Floquet decomposition of the two-time correlation.
For \(\Qv_1=\Qv_2\equiv\Qv\), this decomposition may be written as
\begin{equation}
C_{\rho\rho}(\Qv,t_1;\Qv,t_2)
=
\sum_{n,n'}
e^{\iu n\wL t_2}
e^{-\iu n'\wL t_1}
C_{\rho\rho}^{nn'}(\Qv,t_{\rm rel}),
\label{eq:density-correlation-floquet-decomposition}
\end{equation}
where $t_{\rm rel}\coloneqq t_1-t_2$
and
\begin{equation}
C_{\rho\rho}^{nn'}(\Qv,t_{\rm rel})
\coloneqq
\sum_a
\tilde\varrho_{a0}^{(n)}(\Qv)
\tilde\varrho_{a0}^{(n')\,*}(\Qv)
e^{-\iu\omega_{a0}t_{\rm rel}},
\label{eq:density-correlation-floquet-component}
\end{equation}
with $\omega_{a0}\coloneqq {\qty(E_a-E_0)}/{\hbar}$.
The Floquet dynamic structure factor is the Fourier transform of this
relative-time Floquet component,
\begin{equation}
S_{\mathrm F}^{nn'}(\Qv,\omega)
=
\frac{1}{2\pi\hbar}
\int_{-\infty}^{\infty}
\dd t_{\rm rel}\,
e^{\iu\omega t_{\rm rel}}
C_{\rho\rho}^{nn'}(\Qv,t_{\rm rel}).
\label{eq:SF-correlation-transform}
\end{equation}
Using
\[
\frac{1}{2\pi\hbar}
\int_{-\infty}^{\infty}
\dd t_{\rm rel}\,
e^{\iu(\omega-\omega_{a0})t_{\rm rel}}
=
\delta(\hbar\omega+E_0-E_a),
\]
this gives the Lehmann form
\begin{equation}
S_{\mathrm F}^{nn'}(\Qv,\omega)
=
\sum_a
\tilde\varrho_{a0}^{(n)}(\Qv)
\tilde\varrho_{a0}^{(n')\,*}(\Qv)
\delta(\hbar\omega+E_0-E_a).
\label{eq:SF-pure-app}
\end{equation}
This is the Floquet dynamic structure factor that appears in the
homogeneous-field cross section. It is the scattering-defined structure
factor of the driven target: for a pure initial Floquet state it
collects both the \(a=0\) coherent mean-density outer product and the
positive-loss transition weights from \(\ket{\Phi_0}\) to the
target-changing Floquet states \(\ket{\Phi_{a\ne0}}\), resolved by the
pair of Floquet indices \((n,n')\).

The retarded density response is a related but distinct object~\cite{Kumar2020}.
It is defined by the density commutator,
\begin{multline}
\chi_{\rho\rho}^{\rm R}(\Qv,t;\Qv',t')
\coloneqq \\
-\frac{\iu}{\hbar}\theta(t-t')
\bra{\Phi_0}
\left[
\hat\varrho(-\Qv,t),
\hat\varrho(\Qv',t')
\right]
\ket{\Phi_0}.
\label{eq:chi-commutator}
\end{multline}
(The momentum order matches \pref{eq:density-correlation}; the
reversed order would weight the positive-frequency poles with the
$-\Qv$ structure factor.) The positive-loss Lehmann branch of this response contains the same
target-changing matrix-element products as the connected part of
\pref{eq:SF-pure}; its antiresonant branch carries the $-\Qv$
products and different harmonic shifts. It is therefore useful, for the scattering problem,
to introduce the compact scattering-adapted auxiliary spectral matrix
\begin{multline}
\chiaux^{nn'}(\Qv,\omega)
\coloneqq
\sum_a
\tilde\varrho_{a0}^{(n)}(\Qv)
\tilde\varrho_{a0}^{(n')\,*}(\Qv)
\\
\times
\left[
\frac{1}{\hbar\omega-\hbar\omega_{a0}+\iu0^+}
-
\frac{1}{\hbar\omega+\hbar\omega_{a0}+\iu0^+}
\right],
\label{eq:chiF}
\end{multline}
where \(\hbar\omega_{a0}=E_a-E_0\). It is not the Floquet
transform of \pref{eq:chi-commutator}. With the Hermitian spectral part
$\ImH\chi\coloneqq(\chi-\chi^\dagger)/2\iu$,
$(\chi^\dagger)_{nn'}=\chi_{n'n}^*$, one has exactly
\begin{multline}
-\frac{1}{\pi}\,\ImH\chiaux^{nn'}(\Qv,\omega)
\\
=S_{\mathrm{F,c}}^{nn'}(\Qv,\omega)-S_{\mathrm{F,c}}^{nn'}(\Qv,-\omega),
\label{eq:spectral-response-identity}
\end{multline}
on both
frequency signs. It is an identity for the auxiliary matrix
\pref{eq:chiF}, not a fluctuation--dissipation theorem for the physical
commutator \pref{eq:chi-commutator}. An entrywise
imaginary part would be wrong for the complex off-diagonal entries.
The target-elastic term \(a=0\) cancels identically between the two
auxiliary denominators of \pref{eq:chiF}.  Consequently the coherent
mean-density sector of \pref{eq:SF-coherent-connected} appears in the
scattering correlator as
\(R_n(\Qv)R_{n'}^*(\Qv)\delta(\hbar\omega)\), but not in  \(\chiaux^{nn'}\).  At positive residual frequency
\pref{eq:spectral-response-identity} therefore reconstructs the connected
spectrum $S_{\mathrm{F,c}}^{nn'}(\Qv,\omega)$ only when all gaps
$\omega_{a0}$ of the chosen representative are positive (a negative gap
gives negative spectral weight at positive frequency). The coherent
mean-density outer product must be supplied separately to recover the
full scattering correlator.

In a thermal or quasi-thermal driven steady state the cycle-averaged
density matrix has support on a distribution of Floquet states
$\ket{\Phi_b}$ with populations $p_b$; both the connected scattering
correlator \(S_{\mathrm{F,c}}^{nn'}\) and
$\chiaux^{nn'}$ then acquire an additional sum over $b$
weighted by $p_b$ and by quasienergy denominators referred to $E_b$
rather than to $E_0$, with the $p_b$ fixed by the system--bath coupling
outside the present framework.  The full \(S_{\mathrm F}\) is obtained
only after the corresponding coherent ensemble mean-density outer
product has been restored. ``Connected'' means subtraction of the
ensemble mean at the operator level: merely omitting the $a=b$
transitions would also discard the population variance
$\sum_bp_bR_{b,n}R_{b,n'}^*-\bar R_n\bar R_{n'}^*$, with $R_{b,n}$ the
mean-density harmonics of $\ket{\Phi_b}$ and $\bar R_n=\sum_bp_bR_{b,n}$,
which belongs to the connected weight at zero residual frequency.

\subsection{Macroscopic targets and continuum spectral density}
\label{sec:floquet-coarsegraining}

For any finite target, the connected part of
\pref{eq:SF-pure} and the spectral measure of \pref{eq:chiF} are literal sums of
delta functions over target-changing Floquet eigenstates; $\chiaux$ itself
has the corresponding poles and principal-value parts. For a
macroscopic target the many-body level spacing in the relevant energy
window is far below any physical spectral scale of interest, and this
connected delta comb acquires meaning only as a distribution against
smooth functions, or equivalently as a locally averaged continuum
spectral density, as in the standard theory of dynamic structure
factors~\cite{VanHove1954,Squires2012}.  The target-elastic outer
product \(R_nR_{n'}^*\delta(\hbar\omega)\), by contrast, is a separate
coherent line and does not dissolve into the many-body continuum.  In
an experiment it is broadened by pulse duration, dephasing of the
prepared mean field, and instrumental resolution, rather than by the
thermodynamic many-body level density. The structural identities of
\secref{sec:floquet-response} apply to the sum of these two sectors.

\paragraph*{Coarse-graining kernel.}
Introduce a normalized non-negative smooth kernel $K_\eta(x)$ of width
$\eta$, $\int\dd x\,K_\eta(x)=1$, and the coarse-grained Floquet
connected structure factor
\begin{multline}
S_{\mathrm F,c,\eta}^{nn'}(\Qv, \omega) \coloneqq
\sum_{a\ne0} \tilde\varrho_{a0}^{(n)}(\Qv)\,
\tilde\varrho_{a0}^{(n')\,*}(\Qv)\\
\times K_\eta\!\bigl(\hbar\omega + E_0 - E_a\bigr),
\label{eq:SF-coarse}
\end{multline}
obtained from the connected Lehmann form by replacing the delta
function with $K_\eta$.  For \(a\ne0\), matrix elements of
\(\delta\hat\varrho=\hat\varrho-\bar\varrho\) equal those of
\(\hat\varrho\), which justifies the form displayed above.

\paragraph*{Thermodynamic limit.}
In the thermodynamic limit $V\to\infty$, taken before $\eta\to 0$,
the Floquet sum in \pref{eq:SF-coarse} becomes an integral over a
continuous level density. We define the smooth matrix-valued spectral
density
\begin{equation}
\mathcal D_{\mathrm c,\eta}^{nn'}(\Qv, \hbar\omega)
\coloneqq \lim_{V\to\infty}
\frac1VS_{\mathrm F,c,\eta}^{nn'}(\Qv, \omega),
\label{eq:cont-density}
\end{equation}
with the thermodynamic limit taken at fixed $\eta$; the factor $1/V$
removes the extensivity of the connected bulk response.
The operational continuum representation of the \emph{full}
scattering correlator is therefore
\begin{multline}
\frac1VS_{\mathrm F}^{nn'}(\Qv,\omega)
\quad\longrightarrow\quad
\frac1VR_n(\Qv)R_{n'}^*(\Qv)\delta(\hbar\omega)
{}\\
{}+\mathcal D_{\mathrm c,\eta}^{nn'}
\bigl(\Qv,\hbar\omega\bigr),
\label{eq:SF-continuous}
\end{multline}
where only the second term is a continuum function of the spectral
frequency \(\omega\); the coherent term is kept at finite $V$, since
its scaling with volume depends on the illumination geometry.
The auxiliary matrix \pref{eq:chiF} likewise
acquires a continuum form (per unit volume, smoothed on
the scale $\eta$),
\begin{multline}
\chi_{\mathrm{aux},\eta}^{nn'}(\Qv,\omega) \;\coloneqq\;\int\!\dd E\,
\mathcal D_{\mathrm c,\eta}^{nn'}(\Qv,E)\\
\times\left[
\frac{1}{\hbar\omega-E+\iu 0^{+}} -
\frac{1}{\hbar\omega+E+\iu 0^{+}}\right],
\label{eq:chiF-continuous}
\end{multline}
in which all reference to the Floquet eigenstates $\ket{\Phi_a}$
of the target-changing sector is encoded in the smooth
\(\mathcal D_{\mathrm c,\eta}^{nn'}\); the coherent line is absent
from \(\chiaux\), as explained above. Within the hierarchy
\pref{eq:hierarchy} below, the smoothed intensive matrix
$\chi_{\mathrm{aux},\eta}^{nn'}$ approximates $V^{-1}\chiaux^{nn'}$ in the
distributional sense, when tested against energy-resolution functions
varying on scales large compared with $\eta$.

\paragraph*{Hierarchy of scales.}
The continuum spectral density \pref{eq:cont-density} is the
relevant target object whenever the hierarchy
\begin{equation}
\delta_{\mathrm{mb}}(V)\;\ll\;\eta\;\ll\;\Delta_{\mathrm{phys}}
\label{eq:hierarchy}
\end{equation}
is respected, with $\delta_{\mathrm{mb}}(V)\to 0$ in the
thermodynamic limit and $\Delta_{\mathrm{phys}}$ the smallest
physically meaningful spectral feature -- a linewidth, gap edge,
Rabi splitting, diffusion scale, or other characteristic scale of
\(\mathcal D_{\mathrm c,nn'}\). In this regime the result is independent of the
detailed shape of $K_\eta$, up to corrections controlled by powers of
$\eta/\Delta_{\mathrm{phys}}$ from smoothing of
\(\mathcal D_{\mathrm c,nn'}\) on
its physical scale and by $\delta_{\mathrm{mb}}/\eta$ from finite-size
graininess.
For the sideband-resolved Floquet regime considered in this paper one
also requires
\begin{equation}
\eta\;\ll\;\hbar\wL,
\label{eq:floquet-resolution}
\end{equation}
so that distinct photon sectors $(n,n')$ are not washed together by
the coarse graining; the opposite limit, where photon sidebands are
no longer resolved, corresponds to a strongly broadened or
sideband-unresolved driven response not treated here.

\paragraph*{Discrete-spectrum (few-level) regime.}
The opposite ordering of scales, in which $\eta$ is taken below
$\delta_{\mathrm{mb}}(V)$ or set to zero at fixed finite $V$, is
appropriate for atomic and small-molecular Floquet
targets~\cite{Shirley1965, Sambe1973, Joachain2012, Dorr1994, Vucic1995, KrollWatson1973, ArqueLopez2022, Peskin1994}
and for engineered few-level systems including the driven two-level
demonstration of \secref{sec:two-level}. The spectral object is then
a coherent target-elastic line at \(\omega=0\) plus a literal
connected delta comb on the discrete set
$\hbar\omega = E_a-E_0$ with \(a\ne0\). Each target-changing final
Floquet state and its allowed photon partitions contribute
sideband-resolved line amplitudes; the full Lehmann form
\pref{eq:SF-pure}, or equivalently the split
\pref{eq:SF-coherent-connected}, is the correct mathematical object.
The structural identities of \secref{sec:floquet-identities} hold line
by line on the discrete spectrum.

\paragraph*{Noncommuting limits.}
The connected continuum in \pref{eq:SF-continuous} and the literal
connected delta comb correspond to opposite orderings of
$V\to\infty$ and $\eta\to 0$; finite experimental resolution makes
both operationally well-defined, and hybrid connected spectra are
handled by combining the two representations.  The coherent
target-elastic line is retained separately in either ordering.

\paragraph*{Worked examples in the two regimes.}
The metallic Drude/diffusive examples of \secref{sec:plasmon} are
continuum functions of $\omega$ within the projectile-only connected-response
reduction \pref{eq:weak-target-dressing}, realizing the diagonal
$\mathcal D_{\mathrm c,00}(\Qv,E)$ of
\pref{eq:SF-continuous}. The driven two-level example of
\secref{sec:two-level} is a finite model: its target-changing sector is
a single Floquet excitation at $\nu=\wL-\WR$, for which
\pref{eq:two-level-SF} is the correct mathematical object and no
coarse-graining is needed, while its coherent harmonics
\pref{eq:two-level-coherent-R} remain a separate target-elastic line.

\subsection{Discussion of fluctuation-dissipation}

The auxiliary identity \pref{eq:spectral-response-identity} does not
determine the physical retarded response \pref{eq:chi-commutator} and is
not a fluctuation--dissipation theorem for driven systems. For a general
driven preparation, the scattering correlator and the commutator
\pref{eq:chi-commutator} must be evaluated with the actual Floquet
occupations and harmonic matrix elements~\cite{Kumar2020}; no universal
equilibrium Bose factor is assumed here. Because the coherent mean
density is absent from the commutator, any such relation concerns the
connected spectrum. An additional detailed-balance relation, as in
Floquet--Gibbs-like states, requires a specified preparation and bath
model. The cross section uses the unsymmetrized correlator directly,
with its coherent ensemble mean separated at the operator level.

The Hermiticity, positivity, and Cauchy--Schwarz identities
\pref{eq:SF-Hermitian}--\pref{eq:CS} hold separately for the full and
connected correlators, while the \(f\)-sum rule \pref{eq:fsum} is
written for the full \(S_{\mathrm F}\); in its symmetrized form the
coherent sector cancels.  The selection rules of
\secref{sec:floquet-selection} and \epref{eq:bimomentum-translation}
likewise hold in the general (non-Floquet--Gibbs) case. 

\subsection{Literature comparison for the metallic worked examples}
\label{app:plasmon-literature}
The metallic results of \secref{sec:plasmon} combine three
established limits -- the Pines--Ritchie dielectric plasmon-loss
formula~\cite{Pines1999,RitchieRitchie1957,Egerton2011} recovered by
setting $\alphav=0$ in \pref{eq:dsigma-Drude}; the
Born--Floquet $(e,e')$ Bessel comb of
D\"orr--Joachain--Potvliege--Vu\v ci\'c~\cite{Dorr1994,Vucic1995} with
the atomic generalized oscillator strength replaced by the metallic
plasmon spectral weight $\omega_p\delta(\omega_N-\omega_p)$; and the
diffuson $\chi^{\rm diff}$ as the standard Altshuler--Aronov
ingredient~\cite{AltshulerAronov1985,Akkermans2007} -- into four
combinations: the
bulk plasmon-Floquet comb \pref{eq:dsigma-Drude}, the
Floquet-diffusive cross section \pref{eq:dsigma-diff-screened}, the
inhomogeneous-field plasmon-loss comb \pref{eq:dsigma-Drude-inhom}
with $\mathcal K_N(\Qv)$-modulated sideband intensities, and the
inhomogeneous-field diffusive sidebands \pref{eq:dsigma-diff-inhom}
whose linewidths inherit the spatial profile of $\vb A(\vb r)$.

All four combinations are connected, target-changing loss spectra.
They should be distinguished from standard PINEM, in which the target
returns to the prepared state and the electron is modulated by the
coherent induced near field.  The projectile-only loss comb is governed
by \(J_N^2(|\Qv\cdot\alphav|)\) multiplying a positive-frequency
structure factor, whereas the target-elastic PINEM comb is governed,
after coherent resummation, by
\(J_N^2(2|g_0+g_{\rm ind}|)\) at \(\omega_N=0\).  The same dielectric
function can enter both sectors, but in different ways: its loss part
sets the connected EELS spectrum, while its full complex response,
together with the illumination and boundary geometry, determines
\(R_M\) and \(g_{\rm ind}\).  The bulk response functions used in the
worked examples therefore do not suffice to draw a quantitative PINEM
curve.

\section{Derivation of the Floquet \(f\)-sum rule}
\label{app:fsum}

We derive the Floquet \(f\)-sum rule \pref{eq:fsum} using the
charge-density and Floquet conventions adopted in the main text,
tracking explicitly the factors of \(\hbar\), \(e\), and
\(\bar n_{\mathrm e}\). The familiar equilibrium sum rule will follow
at the end as the undriven reduction of the Floquet result.

\paragraph*{Working assumptions.}
We write the target Hamiltonian as
\[
\hat H_1(t)
=
\hat T_1(t)+\hat V_{\mathrm{int}}+\hat V_1
\]
and make three standard assumptions:
(a)~\(\hat V_{\mathrm{int}}\) is a density--density interaction and
\(\hat V_1\) is a static potential linear in the density;
(b)~the target particles are coupled minimally to the drive through
\[
\hat{\vb p}_i
\longrightarrow
\hat{\vb p}_i-\frac{Q_i}{c}\vb A(\hat{\vb r}_i,t);
\]
and
(c)~the prescribed $\vb A(\vb r,t)$
is a function of position (Coulomb gauge, no scalar potential), so that
it commutes with the density operator; spatial homogeneity is not
required. These assumptions
ensure that the double commutator entering the spectral moment is
unchanged by the drive.

\paragraph*{Drive-independent double commutator.}
Let
\[
\hat\varrho(\vb q)
=
\sum_i Q_i e^{-\iu\vb q\cdot\hat{\vb r}_i}
\]
be the charge-density operator \pref{eq:rhoq}. For mobile electrons of
charge \(-e\), mass \(\me\), and kinetic energy
\[
\hat T_1(t)
=
\sum_i
\frac{
\left[
\hat{\vb p}_i+(e/c)\vb A(\hat{\vb r}_i,t)
\right]^2
}{2\me},
\]
one finds
\begin{equation}
\begin{split}
\bigl[
\hat\varrho(\vb q),
\bigl[
\hat H_1(t),
\hat\varrho^\dagger(\vb q)
\bigr]
\bigr]
&=
\bigl[
\hat\varrho(\vb q),
\bigl[
\hat T_1,
\hat\varrho^\dagger(\vb q)
\bigr]
\bigr]
\\
&=
\frac{e^2\hbar^2q^2}{\me}\,
\hat N_{\mathrm e},
\end{split}
\label{eq:doublecomm}
\end{equation}
where \(\hat N_{\mathrm e}=\sum_i1\).

The interaction and static-potential terms do not contribute under
assumption~(a). The term in \(\hat T_1(t)\) linear in \(\vb A\)
produces an inner commutator proportional to
$\sum_i e^{\iu\vb q\cdot\hat{\vb r}_i}\,\vb q\cdot\vb A(\hat{\vb r}_i,t)$, a
function of the positions only; the outer commutator therefore vanishes
because
functions of the positions commute.
The term proportional to $\vb A^2(\hat{\vb r}_i,t)$
is a function of the positions, commutes with the density, and likewise
does not contribute. Thus only the bare kinetic operator
\(\sum_i\hat{\vb p}_i^{\,2}/(2\me)\) survives in
\pref{eq:doublecomm}, at every instant of time.

For a target containing several mobile species, the right-hand side
generalizes to
\[
\hbar^2q^2
\sum_s
\frac{Q_s^2}{m_s}\hat N_s .
\]
The result used below is the electronic sum rule relevant to the
applications considered in the main text.

\paragraph*{Spectral-moment convention.}
With the energy-delta convention used throughout the paper,
\begin{equation}
S(\vb q,\omega')
\coloneqq
\sum_f
\left|
\bra{f}
\hat\varrho(\vb q)
\ket{0}
\right|^2
\delta(\hbar\omega'+E_0-E_f),
\label{eq:S-energydelta}
\end{equation}
the structure function carries one inverse power of energy. Changing
variables from \(\omega'\) to \(u=\hbar\omega'\) gives the
spectral-moment identity
\begin{multline}
\int_{-\infty}^{\infty}
\dd\omega'\,
\omega'\,
\bigl[S(\vb q,\omega')+S(-\vb q,\omega')\bigr]
\\
=
\frac{1}{\hbar^2}
\left\langle
\left[
\hat\varrho(\vb q),
\left[
\hat H_1,
\hat\varrho^\dagger(\vb q)
\right]
\right]
\right\rangle,
\label{eq:moment-form-final}
\end{multline}
since $\langle[\hat\varrho,[\hat H_1,\hat\varrho^\dagger]]\rangle$
equals the sum of the first moments of $\hat\varrho$ and
$\hat\varrho^\dagger=\hat\varrho(-\vb q)$. The one-sided moment follows
only if $S(\vb q)$ and $S(-\vb q)$ have equal first moments (e.g.\ by
inversion or time-reversal symmetry). The Floquet generalization
follows.

\paragraph*{Floquet spectral moment.}
For a periodically driven target,
\[
\hat H_1(t+T)=\hat H_1(t),
\qquad
T=\frac{2\pi}{\wL},
\]
let \(\hat U(t,t_0)\) denote the exact evolution operator,
\[
\iu\hbar\,\partial_t\hat U(t,t_0)
=
\hat H_1(t)\hat U(t,t_0),
\qquad
\hat U(t_0,t_0)=\mathbf 1 .
\]
Periodicity of the Hamiltonian implies
\begin{equation}
\hat U(t+T,t_0+T)
=
\hat U(t,t_0),
\label{eq:U-periodic}
\end{equation}
while the one-cycle operator
\(\hat U_F(t_0)=\hat U(t_0+T,t_0)\) defines the Floquet eigenstates
\(\ket{\Phi_a(t_0)}\) and quasienergies \(E_a\), modulo
\(\hbar\wL\).

We now cast the double commutator \pref{eq:doublecomm} into the
Heisenberg picture. The Hamiltonian in the Heisenberg representation 
is labeled by $\hat{\mathfrak H}_{1}(t)$
and the transformed density operator is
\begin{equation}
\hat{\mathfrak r}(\vb q,t)
\coloneqq
\hat U^\dagger(t,t_0)\,
\hat\varrho(\vb q)\,
\hat U(t,t_0).
\label{eq:rho-Heisenberg}
\end{equation}
Its matrix elements in the Floquet basis admit the decomposition
\begin{equation}
\bra{\Phi_a}
\hat{\mathfrak r}(\vb q,t)
\ket{\Phi_0}
=
e^{\iu(E_a-E_0)t/\hbar}
\sum_n
e^{\iu n\wL t}\,
\tilde\varrho_{a0}^{(n)}(\vb q),
\label{eq:rho-floquet-fourier}
\end{equation}
which defines the same Floquet density components used in the main
text.

Cycle averaging the double commutator on the initial Floquet state
gives
\begin{multline}
\overline{
\bra{\Phi_0}
\left[
\hat{\mathfrak r}(\vb q,t),
\left[
\hat{\mathfrak H}_{1}(t),
\hat{\mathfrak r}^\dagger(\vb q,t)
\right]
\right]
\ket{\Phi_0}
}
=
\frac{e^2\hbar^2q^2 N_\mathrm e}{\me} 
\label{eq:doublecomm-cycle}
\end{multline}
where $N_\mathrm e$ is the electron number and the overbar denotes
\[
\overline{X}
\coloneqq
\frac{1}{T}
\int_{t_0}^{t_0+T}
\dd t\,X(t).
\]
By \pref{eq:doublecomm}, the right-hand side is independent of the
drive amplitude and of the choice of the cycle-averaging window. The
drive enters the spectral decomposition, but not the equal-time
double commutator.

The Floquet moments follow from an exact identity. Let
$g_a(t)\coloneqq\bra{\Phi_a(t)}\hat\varrho(\vb q)\ket{\Phi_0(t)}$, which by
\pref{eq:rho-floquet-fourier} contains the frequencies
$\omega_{a0}+n\wL$ with amplitudes $\tilde\varrho^{(n)}_{a0}$. The
Schr\"odinger equation gives
$-\iu\dot g_a=\hbar^{-1}\bra{\Phi_a}[\hat H_1(t),\hat\varrho]\ket{\Phi_0}$.
Parseval's theorem and completeness of $\{\ket{\Phi_a(t)}\}$ at each $t$
then give, for
$M(\vb q)\coloneqq\sum_n\int\dd\omega'\,(\omega'+n\wL)
S_{\mathrm F}^{nn}(\vb q,\omega')$,
\begin{equation}
M(\vb q)=\frac{1}{\hbar^2}\,
\overline{\bra{\Phi_0(t)}\hat\varrho^\dagger(\vb q)
[\hat H_1(t),\hat\varrho(\vb q)]\ket{\Phi_0(t)}},
\label{eq:fsum-moment-identity}
\end{equation}
including the coherent term $a=0$. Since the cycle average of
$\langle[\hat H_1,\hat\varrho^\dagger\hat\varrho]\rangle
=-\iu\hbar\,\partial_t\langle\hat\varrho^\dagger\hat\varrho\rangle$
vanishes and $[\hat\varrho(\vb q),\hat\varrho(-\vb q)]=0$,
\pref{eq:doublecomm-cycle} yields
\begin{multline}
\frac12\sum_n
\int_{-\infty}^{\infty}
\dd\omega'\,
(\omega'+n\wL)\,
\bigl[s_{\mathrm F}^{nn}(\vb q,\omega')+s_{\mathrm F}^{nn}(-\vb q,\omega')\bigr]
\\
=
\frac{e^2q^2\bar n_{\mathrm e}}{2\me},
\label{eq:fsum-derived}
\end{multline}
i.e.\ $M(\vb q)+M(-\vb q)=e^2q^2N_{\mathrm e}/\me$ with
$s_{\mathrm F}=S_{\mathrm F}/V$ and $\bar n_{\mathrm e}=N_{\mathrm e}/V$.
This is the Floquet \(f\)-sum rule \pref{eq:fsum}.
The one-sided form $M(\vb q)=M(-\vb q)$ holds if $\hat H_1(t)$, the
density and the \emph{prepared state} share a generalized inversion
(inversion combined with a half-period shift) or a generalized time
reversal $\Theta$ with $\Theta\hat H_1(t)\Theta^{-1}=\hat H_1(t_0-t)$,
$\Theta\hat\varrho(\vb q)\Theta^{-1}=\hat\varrho(-\vb q)$ and
$\Theta\Phi_0(t)\propto\Phi_0(t_0-t)$: the Hermitian expectation
$\langle\hat\varrho^\dagger\hat H_1\hat\varrho\rangle_t$ then equals
$\langle\hat\varrho\hat H_1\hat\varrho^\dagger\rangle_{t_0-t}$, so the two
one-sided moments coincide after cycle averaging. Otherwise they need
not: for a single particle,
$M(\vb q)-M(-\vb q)=-(2Q^2/\hbar m)\,\vb q\cdot
\overline{\langle\hat{\boldsymbol\pi}\rangle}$ with the kinetic momentum
$\hat{\boldsymbol\pi}$, which is nonzero for Floquet states carrying a
cycle-averaged current (e.g.\ on a ring with static flux); in many-body
systems the difference is a density--current correlation. The shifted
frequency
\[
\omega'+n\wL
\]
includes the energy exchanged with the drive in Floquet sector \(n\).
The right-hand side depends only on the cycle-averaged electron density
and not on the distribution of spectral weight among the individual
Floquet sectors. Consequently, \pref{eq:fsum-derived} follows from the double commutator
alone, under the Hamiltonian assumptions stated above, for any preparation
of the target, and provides a direct consistency
check on approximate calculations of
\(S_{\mathrm F}^{nn}\).

\paragraph*{Coherent-sector contribution and scope.}
The moment in \pref{eq:fsum-derived} refers to the full correlator.
Using \pref{eq:SF-coherent-connected}, its coherent sector
contributes
\begin{equation}
\Delta_{\rm coh}(\vb q)
=
\frac{1}{\hbar}
\sum_n n\wL\,|R_n(\vb q)|^2
\label{eq:fsum-coherent}
\end{equation}
to the one-sided moment $M(\vb q)$: the factor \(\delta(\hbar\omega')\) removes the
\(\omega'\) part of the moment but leaves the Floquet energy
\(n\wL\). Hermiticity of the mean density,
$R_n(-\vb q)=R_{-n}(\vb q)^*$, gives $\Delta_{\rm coh}(-\vb q)=
-\Delta_{\rm coh}(\vb q)$. The coherent sector therefore cancels in the
symmetrized rule \pref{eq:fsum-derived} for any illumination, homogeneous
or structured, and the connected correlator alone has the same
symmetrized moment; a one-sided connected moment must subtract
$\Delta_{\rm coh}$ explicitly.
The derivation holds at $\vb q_1=\vb q_2=\vb q$; off-diagonal
bi-momentum moments involve non-$c$-number double commutators.

\paragraph*{Equilibrium limit.}
In the absence of target dressing,
\[
S_{\mathrm F}^{nn'}(\vb q,\omega)
\longrightarrow
\delta_{n0}\delta_{n'0}\,
S(\vb q,\omega),
\qquad
\bar n_{\mathrm e}
\longrightarrow
n_{\mathrm e}.
\]
Equation~\pref{eq:fsum-derived} then immediately reduces to
\begin{equation}
\int_{-\infty}^{\infty}
\dd\omega'\,
\omega'\,
s(\vb q,\omega')
=
\frac{e^2q^2n_{\mathrm e}}{2\me},
\label{eq:fsum-eq-extensive}
\end{equation}
in the intensive normalization used in the main text, for
equilibrium states with equal $\pm\vb q$ moments (inversion or
time-reversal symmetry). In an extensive
normalization, the right-hand side is multiplied by
\(\mathrm{Vol}\). Thus the textbook equilibrium \(f\)-sum rule is a
direct limiting case of the Floquet result rather than an independent
derivation.

\section{Relation to earlier scattering theories}
\label{app:limits}
The general Floquet--Born scattering theory admits three benchmark
specializations, distinguished by which projectile and target Floquet
degrees of freedom are retained. In the field-free Bethe limit, both
Floquet structures collapse. In the homogeneous-field
potential-scattering reduction, the target is replaced by a static
potential and only the projectile Volkov sidebands remain, yielding
the Bunkin--Fedorov result and, in the appropriate soft-photon regime,
the Kroll--Watson structure. Finally, specializing the driven target
to a single atom while retaining both projectile and target Floquet
dynamics recovers the laser-assisted atomic theory of Joachain and
coworkers. These cases are complementary specializations rather than
successive approximations.  A separate, cross-cutting decomposition is
the exact coherent/connected split
\pref{eq:SF-coherent-connected}: the coherent target-elastic branch
defines \(g_{\rm ind}\) and becomes the PINEM Bessel ladder after
amplitude-level resummation, whereas the three reductions below concern
how the projectile kernel and the connected target response are
specialized.

\paragraph*{Field-free Bethe limit.}
When $\vb A\to 0$ the Floquet expansions \pref{eq:FF} collapse onto a
single Fourier component each, $n_i=n_f=m_i=m_f=0$, and only the $N=0$
term contributes. With the momentum and energy transfers
\[
\vb q\coloneqq \vb k_f-\vb k_i,
\qquad
\hbar\omega\coloneqq \epsilon(\vb k_i)-\epsilon(\vb k_f),
\]
Equation~\pref{eq:dsigma} reduces, after summing over unresolved target final
states, to the textbook Bethe formula~\cite{Bethe1930,Inokuti1971,Joachain2012}
\begin{equation}
\frac{\dd^2 \sigma}{\dd \Omega \dd\epsilon} =
\qty(\frac{\me}{2\pi\hbar^2})^{2} \frac{k_f}{k_i}\,|v(\vb q)|^2\,
S(\vb q,\omega),
\label{eq:bethe}
\end{equation}
with the conventional dynamic structure factor
\(
S(\vb q,\omega) = \sum_f \abs{\bra{\Phi_f}\hat\varrho_1(\vb q)\ket{\Phi_i}}^{2}
\delta(\hbar\omega + E_i-E_f).
\)
The first-Born many-body Bethe theory of fast-electron energy-loss
spectroscopy is recovered as a special case.

\paragraph*{Equivalent dielectric representation.}
In the field-free translationally invariant limit, \pref{eq:bethe}
can also be written in the dielectric loss-function form reviewed in
Ref.~\cite{deAbajoRMP2010}. The identity \pref{eq:S-loss-function},
established for the metallic case in \secref{sec:plasmon-drude} but
following purely from the equilibrium density fluctuation--dissipation
theorem and the definition of the longitudinal dielectric function --
and hence equally valid here -- converts \pref{eq:bethe} directly into
\begin{equation}
\frac{\dd^2\sigma}{\dd\Omega\,\dd\epsilon}
=
\left(\frac{\me}{2\pi\hbar^2}\right)^{2}
\frac{k_f}{k_i}\,
|v(\vb q)|^2\,
\frac{q^{2}}{4\pi^{2}}\,
\mathrm{Im}\!\left[-\frac{1}{\varepsilon(\vb q,\omega)}
\right].
\label{eq:bethe-loss}
\end{equation}
This is the standard longitudinal EELS loss-function form. Unlike a
direct substitution of the projectile-dependent $v(\vb q)$ into the
dielectric identity itself -- which would incorrectly couple the
target's internal screening to the projectile charge $q_{\mathrm p}$,
and for an electron projectile would leave the wrong sign, since
$v(\vb q)<0$ -- the target's own Coulomb self-interaction
$V_{\mathrm C}(\vb q)=4\pi e^{2}/q^{2}$ enters only inside
\pref{eq:S-loss-function}, while the projectile coupling appears
solely through $|v(\vb q)|^2\ge 0$, exactly as in \pref{eq:bethe}.
The structure-factor representation \pref{eq:bethe} and the
loss-function representation \pref{eq:bethe-loss} are equivalent in
this equilibrium field-free limit; the Floquet generalization
developed in the main text is naturally expressed in the
structure-factor language $S\to S_{\mathrm F}^{nn'}$.

\paragraph*{Homogeneous-field potential scattering:
Bunkin--Fedorov reduction and Kroll--Watson structure.}
This specialization follows from the homogeneous projectile kernel of
\pref{eq:P-hom} after replacing the dynamical many-body target channel
by a static potential \(U(\vb r)\); all target Floquet indices are then
absent, while the projectile sideband index \(N\) remains.

The sideband-resolved angular weight \pref{eq:sideband-resolved-angular-weight}
is subject to elastic kinematics,
with the outgoing projectile wave number in the
\(N\)-th sideband fixed by
\begin{subequations}
\label{eq:kinematics}
\begin{equation}
\frac{\hbar^2 k_N^2}{2m_{\rm e}}
=
\frac{\hbar^2 k_i^2}{2m_{\rm e}}
-
N\hbar\omega_{\rm L},
\end{equation}
where
\begin{equation}
\mathbf Q_N=\mathbf k_N-\mathbf k_i,
\qquad
k_N=|\mathbf k_N|.
\end{equation}
\end{subequations}
The first-Born elastic sideband-resolved cross section, evaluated at
\pref{eq:kinematics}, is
\begin{equation}
\label{eq:cs-elastic}
\left(\frac{\dd\sigma}{\dd\Omega}\right)_{N}^{(0)}
=
\frac{k_N}{k_i}
\left(\frac{m_{\rm e}}{2\pi\hbar^2}\right)^2
\left|U(\mathbf Q_N)\right|^2 .
\end{equation}
The factor \(k_N/k_i\) is the usual outgoing-density-of-states and
incident-flux ratio evaluated after the energy-conservation condition
has fixed the final wave number.

The Volkov dressing then multiplies this elastic channel by
the \(N\)-th Fourier coefficient of the optical phase. Thus, for a
homogeneous, linearly polarized field,
\begin{equation}
\label{eq:BunkinFedorov}
\qty(\frac{\dd\sigma}{\dd\Omega})_N
=
J_N^2(\mathbf Q_N\cdot\boldsymbol{\alpha}_0)
\left(\frac{\dd\sigma}{\dd\Omega}\right)_{N}^{(0)}.
\end{equation}
Equation~\pref{eq:BunkinFedorov} is the Bunkin--Fedorov result obtained
within the first-Born scope of the present theory~\cite{BunkinFedorov1966}.
Kroll and Watson showed that, in the soft-photon regime, the same
sideband structure persists when the Born cross section is replaced by
the full field-free elastic cross section~\cite{KrollWatson1973}. The
latter is therefore a structural extension rather than a direct
reduction of the present first-Born formula. Because
\(J_N^2(\mathbf Q_N\cdot\boldsymbol{\alpha}_0)\) assumes a real quiver
amplitude, \pref{eq:BunkinFedorov} as written holds for a linearly
polarized homogeneous field; for general elliptic polarization,
\(\boldsymbol\alpha_0\) and its argument are replaced by the complex
coupling and phase convention of \secref{sec:applications}.

If the quantities \(k_N\), \(\Qv_N\), and the
field-free elastic cross section vary negligibly over the sideband range
with appreciable Bessel weight (a `soft-sideband limit'), then
\[
k_N\simeq k_i,\qquad
\Qv_N\simeq\Qv,\qquad
\left(\frac{\dd\sigma}{\dd\Omega}\right)_{N}^{(0)}
\simeq
\left(\frac{\dd\sigma}{\dd\Omega}\right)_{\rm el}^{(0)} ,
\]
and the Bessel identity \(\sum_N J_N^2(x)=1\) gives
\begin{equation}
\sum_N
\left(
\frac{\dd\sigma}{\dd\Omega}
\right)_N
\simeq
\left(
\frac{\dd\sigma}{\dd\Omega}
\right)_{\rm el}^{(0)} .
\label{eq:elastic-KW-closure}
\end{equation}
This is the elastic Kroll--Watson closure; its inelastic generalization is
presented in \secref{sec:KW-closure}.

\paragraph*{Single-atom Floquet specialization:
Joachain--D\"orr--Vu\v ci\'c--Potvliege theory.}
When the target is specialized to a single atom and the external field
is treated in the dipole approximation, \(\hat\varrho_1\) becomes the
atomic charge-density operator and
\(\ket{\Phi_{i/f}^{(m)}}\) are the Fourier components of laser-dressed
atomic Floquet states. The projectile remains described by homogeneous
Volkov states. Consequently, the photon-partition sum in
\pref{eq:MN} and the energy-conservation condition in
\pref{eq:S1B-Floquet} reproduce term by term the Born--Floquet
laser-assisted \((e,e')\) theory of
Refs.~\cite{Dorr1994,Vucic1995}. The present theory generalizes this
construction by replacing the atomic transition density with a
many-body Floquet density response.

The atomic calculations may incorporate ionization widths through
complex resonance quasienergies, whereas the formal development here
adopts the closed-target, unitary Floquet setting stated in
\secref{sec:scope}.

\paragraph*{Relation among the three specializations.}
The three cases retain different sectors of the general theory.
The Bethe limit removes both projectile and target dressing.
The Bunkin--Fedorov reduction retains projectile Volkov dressing but
replaces the target by a static potential. The Joachain specialization
retains both projectile and target Floquet dynamics but restricts the
target to a single atom. The present many-body theory occupies the
remaining general case: projectile dressing is retained together with
the Floquet density correlations of an interacting condensed-matter
target. In each case the coherent mean-density sector, when present,
must be separated before a connected-response reduction is applied.
Its weak-coupling contribution and PINEM resummation form the coherent
branch shown separately in \figref{fig:limits} (right), rather than a fourth
member of this specialization chain.

\section{Elastic scattering in an inhomogeneous field}
\label{app:elastic}
When the laser field is inhomogeneous, the elastic scattering
cross section involves a coherent sum of a zeroth and first-order
scattering amplitudes (\secref{sec:1Born}) in the spirit of the distorted-wave
Born approximation (DWBA) \cite{Taylor2006,Peng2004}.
In the present context, the zero-order term describes projectile scattering
in an inhomogeneous laser field; we call the associated amplitude `optical'. 
The first-order term is formulated here for 
elastic scattering from a potential $U(\mathbf r)$.
Thus we obtain a generalization of the Bunkin--Fedorov 
cross section \pref{eq:BunkinFedorov}
to the inhomogeneous laser drive. The appendix closes with the
far-field angular transform of the resummed coherent PINEM channel of
\secref{sec:pinem-bridge}, which follows from the same exit-plane
construction.

\paragraph*{Optical amplitude.}
The `zero-order' amplitude describes electron scattering in an inhomogeneous
laser field, \epref{eq:S0B}. The present subsection expresses
the amplitude $\braket{\chi_f}{\chi_i}$ in 
the paraxial localized-field regime.
Let the incident projectile follow the straight eikonal trajectory
\[
\vb r(s)=\vb r_\perp+\vb v_{\vb k_i}s,
\qquad
\vb r_\perp\cdot\vb v_{\vb k_i}=0,
\]
where \(s\) is the time parameter along the trajectory and
\(\vb r_\perp\) is the impact parameter in the plane transverse to the
incident beam. In the paraxial small-angle regime the same transverse
plane is used for the outgoing projection; corrections from the
difference between the incoming and outgoing transverse planes are
higher order in the scattering angle.

For a localized external 
field the eikonal--Volkov wavefunction \pref{eq:eikonal-eq} is used.
For the field-free projection it is convenient to use the
full-trajectory, impact-parameter form of the eikonal phase,
\[
\Theta_{\vb k_i}(\vb r_\perp,t_0)
=
\mathrm{Re}
\left[
\zeta_{\vb k_i}(\vb r_\perp)
e^{\iu(\omega_{\rm L}t_0+\varphi)}
\right],
\]
with
\[
\zeta_{\vb k_i}(\vb r_\perp)
=
-\frac{q_{\rm p}}{m_{\rm e}c}\,
\vb k_i\cdot
\int_{-\infty}^{\infty}
\dd s\,
\Ap(\vb r_\perp+\vb v_{\vb k_i}s)\,
e^{+\iu\omega_{\rm L}s}.
\]
This is the accumulated asymptotic coupling~\footnote{The 
sign of the trajectory harmonic follows the $e^{+\iu\omega_{\rm L}t}$ 
convention of \pref{eq:Acos}. The full-trajectory expression is the downstream
limit of the semi-infinite incoming coupling
\pref{eq:eikonal-zeta}.}.

Writing
\[
\zeta_{\vb k_i}(\vb r_\perp)
=
|\zeta_{\vb k_i}(\vb r_\perp)|
e^{\iu[\theta_{\vb k_i}(\vb r_\perp)-\pi/2]},
\]
one obtains
\[
\Theta_{\vb k_i}(\vb r_\perp,t_0)
=
|\zeta_{\vb k_i}(\vb r_\perp)|
\sin[
\omega_{\rm L}t_0+\varphi+\theta_{\vb k_i}(\vb r_\perp)
].
\]
The optical transmission factor has the Fourier expansion
\[
e^{-\iu\Theta_{\vb k_i}(\vb r_\perp,t_0)}
=
\sum_N
\mathcal F_{\vb k_i}^{(N)}(\vb r_\perp)
e^{+\iu N\omega_{\rm L}t_0},
\]
where
\[
\mathcal F_{\vb k_i}^{(N)}(\vb r_\perp)
=
J_{-N}(|\zeta_{\vb k_i}(\vb r_\perp)|)
e^{+\iu N[\varphi+\theta_{\vb k_i  }(\vb r_\perp)]}
\]
is the coefficient of electron energy loss $N\hbar\wL$ (a factor
$e^{-\iu N\wL t_0}$ would raise the electron energy), consistent with
the loss-$N$ kinematics \pref{eq:kinematics} used below and with the
main-text kernel \pref{eq:Graf-DWBA} at $\zeta^{(-)}=0$.
In the paraxial exit-plane representation for a spatially localized
laser field, the optical phase is accumulated along the single,
approximately undeflected incoming trajectory and evaluated on a plane
downstream of the field. Because the field vanishes on every future
trajectory segment starting at that plane, the outgoing channel is
field-free there and
\[
\zeta_{\vb k_N}^{(-)}(\vb r_{\rm exit})=0,
\]
where \(\vb r_{\rm exit}\) lies on the downstream exit plane. Thus the optical modulation on the exit plane is contained entirely in
the accumulated incoming coupling \(\zeta_{\vb k_i}(\vb r_\perp)\).
This vanishing follows from the localized field and the downstream
choice of exit plane, rather than from paraxiality alone.

For a fixed sideband index \(N\), energy conservation fixes the
magnitude \(k_N\) of the outgoing momentum.  We use
\(\vb k_\perp\) as the continuous transverse-momentum variable that
parametrizes the forward part of this energy shell.  At a selected
detector channel,
\begin{equation}
\vb k_\perp
\coloneqq
\vb k_N-
(\vb k_N\!\cdot\!\hat{\vb k}_i)\hat{\vb k}_i,
\qquad
\hat{\vb k}_i\coloneqq\vb k_i/k_i,
\label{eq:kperpN-def}
\end{equation}
and the corresponding forward on-shell momentum is
\begin{equation}
\vb k_N(\vb k_\perp)
\coloneqq
\vb k_\perp
+
\hat{\vb k}_i\sqrt{k_N^2-k_\perp^2}.
\label{eq:kN-transverse-shell}
\end{equation}
Thus \(N\) labels the outgoing energy shell, whereas
\(\vb k_\perp\) labels a continuous outgoing direction within that
shell.  The square root selects the forward branch and the propagating
domain is \(k_\perp<k_N\); in the paraxial approximation the
transverse Fourier integrals below may be extended to
\(\mathbb R^2\).

Throughout this appendix,
\(\chi_{\rm in}(\vb r_\perp)\) denotes the incident transverse wave
propagated field-free to the chosen downstream exit plane, in the
absence of both the optical modulation and the potential \(U\).  In
the thin-element paraxial approximation, the free propagation between
the entrance and exit planes is absorbed into this definition.  The
wave may represent beam shaping or focusing.

Projection onto a field-free outgoing transverse momentum state gives
the optical-only sideband amplitude
\begin{equation}
\label{eq:pinem-opt-ampl}
\mathcal A_N^{\rm opt}(\vb k_\perp)
\coloneqq
\int
\dd^2\vb r_\perp\,
e^{-\iu\vb k_\perp\cdot\vb r_\perp}
\mathcal F_{\vb k_i}^{(N)}(\vb r_\perp)
\chi_{\rm in}(\vb r_\perp).
\end{equation}

For \(N=0\) this expression contains the transmitted forward beam. If
one wants only the nontrivial optical diffraction part, the replacement
\[
\mathcal F_{\vb k_i}^{(N)}(\vb r_\perp)
\longrightarrow
\mathcal F_{\vb k_i}^{(N)}(\vb r_\perp)-\delta_{N0}
\]
removes the field-free identity contribution present only at \(N=0\),
where \(J_0\to1\) as the field switches off. For \(N\neq0\) no such
subtraction is needed: \(\mathcal F_{\vb k_i}^{(N)}(\vb r_\perp)\)
already vanishes identically in that limit, since \(J_N(0)=0\).

\paragraph*{First-Born amplitude.}
For an inhomogeneous optical field treated in the eikonal--Volkov
approximation, the potential \(U\) generates the first-Born
contribution to the
DWBA scattering amplitude,
\[
S^\text{1B}
=
-\frac{\iu}{\hbar}
\int_{-\infty}^{\infty}
\dd t
\int \dd^3\vb r\,
\chi_{\vb k_f}^{(-)*}(\vb r,t)\,
U(\vb r)\,
\chi_{\vb k_i}^{(+)}(\vb r,t),
\]
where \(\chi^{(+)}\) and \(\chi^{(-)}\) denote the incoming and outgoing
eikonal channel waves defined in \secref{sec:dsigma-inhom}. The 
$N$th Floquet component of the product
\(\chi_{\vb k_f}^{(-)*}\chi_{\vb k_i}^{(+)}\) defines the eikonal
Bessel kernel \pref{eq:Fkernel} after momentum-space Fourier
transformation. Factorizing the quasienergy conservation 
delta as in \pref{eq:MN} leads to the scattering amplitude
\begin{equation}
\mathcal M_N^{U}(\vb k_f,\vb k_i)
\coloneqq
\int\!\dd^3\vb q'\,
U(\vb q')\,
\tilde{\mathcal F}_{\vb k_f\vb k_i}^{(N)}(\vb Q -\vb q')
\label{eq:KrollWatsonEikonal}
\end{equation}
where $\Qv = \vb k_f -\vb k_i$.
The expression \pref{eq:KrollWatsonEikonal} follows also from our general
expression \pref{eq:MN} by replacing the target with the static
potential $U(\vb r)$, whose only nonzero Fourier component forces
$n=N$.

When the optical-only zero-order term can be neglected,
the amplitude \pref{eq:KrollWatsonEikonal} gives the sideband-resolved
cross section
\begin{equation}
\label{eq:pinem-U-cross}
\left(
\frac{\dd\sigma}{\dd\Omega}
\right)_N
\to
\left (\frac{\me}{2\pi\hbar^2}\right)^2
\frac{k_N}{k_i}
\left|
\mathcal M_N^{U}(\vb k_N,\vb k_i)
\right|^2,
\end{equation}
where the kinematics \pref{eq:kinematics} is imposed.
Comparing with \epref{eq:BunkinFedorov}, the inhomogeneous
drive introduces a momentum-space convolution of $U(\vb q)$
and the eikonal kernel.

\paragraph*{First-Born term in the exit-plane representation.}
Before adding \(\mathcal A_N^{\rm opt}\) and
\(\mathcal M_N^U\) coherently, one must place them in the same
outgoing-channel normalization and ensure that they refer to the same
incident transverse state.  The optical amplitude in
\epref{eq:pinem-opt-ampl} already contains the actual incident wave
\(\chi_{\rm in}\).  By contrast, the plane-wave Born expression
\(-m_{\rm e}U(\vb Q)/(2\pi\hbar^2)\) corresponds to one incident
momentum and must first be superposed over the angular spectrum of
\(\chi_{\rm in}\).

Define that angular spectrum by
\begin{equation}
a_{\rm in}(\vb k_\perp')
\coloneqq
\int\dd^2\vb r_\perp\,
e^{-\iu\vb k_\perp'\cdot\vb r_\perp}
\chi_{\rm in}(\vb r_\perp),
\label{eq:incident-angular-spectrum}
\end{equation}
so that
\begin{equation}
\chi_{\rm in}(\vb r_\perp)
=
\int\frac{\dd^2\vb k_\perp'}{(2\pi)^2}\,
e^{+\iu\vb k_\perp'\cdot\vb r_\perp}
a_{\rm in}(\vb k_\perp').
\label{eq:incident-angular-spectrum-inverse}
\end{equation}
For an incoming component on the incident energy shell, let
\begin{equation}
\vb k_i(\vb k_\perp')
\coloneqq
\vb k_\perp'
+
\hat{\vb k}_i\sqrt{k_i^2-k_\perp'^2}.
\label{eq:ki-transverse-shell}
\end{equation}
Although \(\chi_{\rm in}\) may describe beam shaping or focusing, the
use of the single central-momentum optical factor
\(\mathcal F_{\vb k_i}^{(N)}(\vb r_\perp)\) assumes a paraxially narrow
incident angular spectrum, such that
\[
\mathcal F_{\vb k_i(\vb k_\perp')}^{(N)}(\vb r_\perp)
\simeq
\mathcal F_{\vb k_i}^{(N)}(\vb r_\perp)
\]
for the relevant \(\vb r_\perp\) throughout the support of
\(a_{\rm in}(\vb k_\perp')\).

We define the momentum transfer between this incoming component and
an outgoing component on the \(N\)-th shell by
\begin{equation}
\vb Q_N(\vb k_\perp,\vb k_\perp')
\coloneqq
\vb k_N(\vb k_\perp)-\vb k_i(\vb k_\perp').
\label{eq:QN-two-transverse-momenta}
\end{equation}
The corresponding transverse plane-wave Born kernel is
\begin{equation}
\mathcal A_{{\rm B},N}^{U}
(\vb k_\perp,\vb k_\perp')
\coloneqq
-\frac{\me}{2\pi\hbar^2}
U\!\left[\vb Q_N(\vb k_\perp,\vb k_\perp')\right].
\label{eq:transverse-Born-kernel}
\end{equation}

The field-free amplitude scattered by \(U\) from the actual incident
wave is then
\begin{equation}
\mathcal A_{{\rm sc},N}^{U}[\chi_{\rm in}](\vb k_\perp)
\coloneqq
\int\frac{\dd^2\vb k_\perp'}{(2\pi)^2}\,
\mathcal A_{{\rm B},N}^{U}
(\vb k_\perp,\vb k_\perp')
a_{\rm in}(\vb k_\perp').
\label{eq:scattered-amplitude-input}
\end{equation}
For a beam-adapted incident plane wave,
\(a_{\rm in}(\vb k_\perp')=(2\pi)^2
\delta^{(2)}(\vb k_\perp')\), this reduces to
\begin{equation}
\mathcal A_{{\rm sc},N}^{U}[\chi_{\rm in}](\vb k_\perp)
=
-\frac{\me}{2\pi\hbar^2}
U\!\left(\vb Q_N(\vb k_\perp,\vb 0)\right),
\label{eq:passive-elastic-amplitude2}
\end{equation}
which is the plane-wave formula used previously.

We define the corresponding \(U\)-generated scattered wave at the
exit plane by the inverse transverse Fourier transform
\begin{equation}
\chi_{{\rm sc},N}^{U}[\chi_{\rm in}](\vb r_\perp)
\coloneqq
\int\frac{\dd^2\vb k_\perp}{(2\pi)^2}\,
e^{+\iu\vb k_\perp\cdot\vb r_\perp}
\mathcal A_{{\rm sc},N}^{U}[\chi_{\rm in}](\vb k_\perp).
\label{eq:passive-elastic-exit-wave}
\end{equation}
This is the scattered part only, not the complete field-free
exit-plane wave.  Because $\mathcal A^U_{{\rm sc},N}$ is a Born
scattering amplitude (dimension of length), the physical scattered exit
wave is $(2\pi\iu/k_N)\chi^U_{{\rm sc},N}$, by the paraxial relation
$f(\vb k_\perp)=(k/2\pi\iu)\int\dd^2\vb r_\perp\,
e^{-\iu\vb k_\perp\cdot\vb r_\perp}\,\delta\psi(\vb r_\perp)$ between an
exit-wave modification $\delta\psi$ and its scattering amplitude. The latter is
\(\chi_{\rm in}+(2\pi\iu/k_N)\chi_{{\rm sc},N}^{U}[\chi_{\rm in}]\).  Consequently,
\begin{equation}
\mathcal A_{{\rm sc},N}^{U}[\chi_{\rm in}](\vb k_\perp)
=
\int\dd^2\vb r_\perp\,
e^{-\iu\vb k_\perp\cdot\vb r_\perp}
\chi_{{\rm sc},N}^{U}[\chi_{\rm in}](\vb r_\perp).
\label{eq:passive-elastic-forward-transform}
\end{equation}
With this convention, the standard prefactor
\(-m_{\rm e}/(2\pi\hbar^2)\), the on-shell reduction to the
fixed-energy transverse shell, the actual incident angular spectrum,
and the usual incident-flux normalization are included in the
far-field amplitude associated with
\(\chi_{{\rm sc},N}^{U}[\chi_{\rm in}]\).

The corresponding optically modulated \(U\)-generated amplitude,
written as a scattering amplitude, is
\begin{equation}
\label{eq:pinem-U-exit}
\mathcal A_N^{U}(\vb k_\perp)
\coloneqq
\int
\dd^2\vb r_\perp\,
e^{-\iu\vb k_\perp\cdot\vb r_\perp}
\mathcal F_{\vb k_i}^{(N)}(\vb r_\perp)
\chi_{{\rm sc},N}^{U}[\chi_{\rm in}](\vb r_\perp).
\end{equation}
Thus both coherent contributions refer to the same incident transverse
state. The optical amplitude \pref{eq:pinem-opt-ampl} is a transverse
Fourier amplitude (dimension of area); in the same scattering-amplitude
normalization it reads $f_N^{\rm opt}\coloneqq(k_N/2\pi\iu)\,\mathcal
A_N^{\rm opt}$. Their sum can be written as the projection of the complete
sideband exit-plane wave,
\begin{multline}
f_N^{\rm meas}(\vb k_\perp)
\coloneqq
f_N^{\rm opt}(\vb k_\perp)
+
\mathcal A_N^{U}(\vb k_\perp)
\\
=
\frac{k_N}{2\pi\iu}\int\dd^2\vb r_\perp\,
e^{-\iu\vb k_\perp\cdot\vb r_\perp}
\mathcal F_{\vb k_i}^{(N)}(\vb r_\perp)
\\
\times\Bigl[
\chi_{\rm in}(\vb r_\perp)
+
\frac{2\pi\iu}{k_N}\chi_{{\rm sc},N}^{U}[\chi_{\rm in}](\vb r_\perp)
\Bigr].
\label{eq:pinem-total-exit-plane-amplitude}
\end{multline}

Multiplication of both field-free components by the same local factor
\(\mathcal F_{\vb k_i}^{(N)}(\vb r_\perp)\) is the paraxial,
localized-field, thin-element factorization used here.  It is the
exit-plane counterpart of the momentum-space convolution in
\pref{eq:KrollWatsonEikonal} after the on-shell reduction and the Born
prefactor have been absorbed into the scattered wave.  It should not
be read as an exact ordering when the optical and static interactions
overlap strongly in space.

\paragraph*{Differential cross section.}
With the zeroth and first-Born
amplitudes expressed in the same field-free outgoing-channel
normalization, the sideband-resolved elastic angular distribution is
coherent,
\begin{equation}
\label{eq:pinem-meas-coherent}
\left(
\frac{\dd\sigma}{\dd\Omega}
\right)_N^{\rm meas}
=
\frac{k_N}{k_i}
\left|
f_N^{\rm meas}(\vb k_\perp)
\right|^2 .
\end{equation}
(At $N=0$ the unscattered forward beam makes $f^{\rm opt}_0$
distributional at $\vb k_\perp=0$; a finite beam and its incident-flux
normalization are then required.)
Equivalently,
\begin{equation}
\label{eq:pinem-meas-expanded}
\left(
\frac{\dd\sigma}{\dd\Omega}
\right)_N^{\rm meas}
=
\left(
\frac{\dd\sigma}{\dd\Omega}
\right)_N^{\rm opt}
+
\left(
\frac{\dd\sigma}{\dd\Omega}
\right)_N^{\rm int}
+
\left(
\frac{\dd\sigma}{\dd\Omega}
\right)_N^{U},
\end{equation}
with
\begin{align*}
\left(
\frac{\dd\sigma}{\dd\Omega}
\right)_N^{\rm opt}
&=
\frac{k_N}{k_i}
\left|
f_N^{\rm opt}
\right|^2 ,\\
\left(
\frac{\dd\sigma}{\dd\Omega}
\right)_N^{\rm int}
&=
2\frac{k_N}{k_i}
\mathrm{Re}
\left[
(f_N^{\rm opt})^*
\mathcal A_N^{U}
\right]\\
&=-\frac{k_N^2}{\pi k_i}\,\mathrm{Im}\bigl[(\mathcal A_N^{\rm opt})^*\mathcal A_N^U\bigr],\\
\left(
\frac{\dd\sigma}{\dd\Omega}
\right)_N^{U}
&=
\frac{k_N}{k_i}
\left|
\mathcal A_N^{U}
\right|^2.
\end{align*}
In these expressions the common continuous argument
\(\vb k_\perp\) is suppressed; \(N\) fixes the outgoing energy shell,
whereas \(\vb k_\perp\) selects the detector direction on that shell.
The measured amplitude at a selected detector channel is obtained by
evaluating the preceding expressions at
\(\vb k_\perp=\vb k_\perp^{\rm det}\).  The two contributions are
given by \pref{eq:pinem-opt-ampl} and \pref{eq:pinem-U-exit}.
The interference term is present only when the
optical transmission amplitude and the \(U\)-generated amplitude are
collected in the same final projectile sideband, direction, and target
channel, and when the measurement preserves their relative phase.
The factor $1/\iu$ between Fourier and scattering amplitudes makes it
an imaginary-part (optical-theorem-type) term; without it the forward
flux would change at first order in $U$.
The interference term vanishes if the reference amplitude is absent
from the selected channel or if relative-phase averaging removes the
cross term. Subtraction of a reference-only intensity does not in
general remove the interference.

\paragraph*{Factorization of the Bessel kernel.} Let
\begin{equation}
\mathcal A_{{\rm ref},N}[\chi_{\rm in}](\vb k_\perp)
\coloneqq
\int
\dd^2\vb r_\perp\,
e^{-\iu\vb k_\perp\cdot\vb r_\perp}
\chi_{\rm in}(\vb r_\perp)
\label{eq:passive-transmitted-amplitude}
\end{equation}
denote the unscattered reference amplitude in the selected field-free
outgoing channel, and $f_{{\rm ref},N}\coloneqq(k_N/2\pi\iu)\mathcal
A_{{\rm ref},N}$ its scattering-amplitude form. The field-free amplitude scattered by \(U\) from
the same incident wave is
\(\mathcal A_{{\rm sc},N}^{U}[\chi_{\rm in}]\), defined in
\pref{eq:scattered-amplitude-input}. If
\[
\mathcal F_{\vb k_i}^{(N)}(\vb r_\perp)
\simeq
\mathcal F_{\vb k_i}^{(N)}(\vb r_{\perp0})
\]
over the transverse support of both field-free components, then,
suppressing the common argument \(\vb k_\perp\),
\begin{align*}
f_N^{\rm opt}
&\simeq
\mathcal F_{\vb k_i}^{(N)}(\vb r_{\perp0})
f_{{\rm ref},N}[\chi_{\rm in}],\\
\mathcal A_N^{U}
&\simeq
\mathcal F_{\vb k_i}^{(N)}(\vb r_{\perp0})
\mathcal A_{{\rm sc},N}^{U}[\chi_{\rm in}].
\end{align*}
The $\vb r_{\perp0}$ indicates the center of the common transverse support
of the propagated reference wave and the 
$U$-generated scattered wave. For a transversely-localized beam 
$\vb r_{\perp0}$ denotes the impact parameter.
Consequently the measured sideband amplitude factorizes as
\[
f_N^{\rm opt}
+
\mathcal A_N^{U}
\simeq
\mathcal F_{\vb k_i}^{(N)}(\vb r_{\perp0})
\left(
f_{{\rm ref},N}[\chi_{\rm in}]
+
\mathcal A_{{\rm sc},N}^{U}[\chi_{\rm in}]
\right).
\]
Squaring gives
\begin{multline*}
\left(
\frac{\dd\sigma}{\dd\Omega}
\right)_N^{\rm meas}
\simeq
\left|
\mathcal F_{\vb k_i}^{(N)}(\vb r_{\perp0})
\right|^2
\frac{k_N}{k_i}\\
\times\left|
f_{{\rm ref},N}[\chi_{\rm in}]
+
\mathcal A_{{\rm sc},N}^{U}[\chi_{\rm in}]
\right|^2 .
\end{multline*}
Thus
\begin{equation}
\label{eq:pinem-meas-factorized}
\left(
\frac{\dd\sigma}{\dd\Omega}
\right)_N^{\rm meas}
\simeq
P_N(\vb r_{\perp0})
\left(
\frac{\dd\sigma}{\dd\Omega}
\right)_{{\rm ff},N}^{\rm meas},
\end{equation}
where
\[
P_N(\vb r_{\perp0})
=
\left|
\mathcal F_{\vb k_i}^{(N)}(\vb r_{\perp0})
\right|^2
=
J_N^2(
|\zeta_{\vb k_i}(\vb r_{\perp0})|
)
\]
and
\begin{equation}
\label{eq:dsigma-passive-meas}
\left(
\frac{\dd\sigma}{\dd\Omega}
\right)_{{\rm ff},N}^{\rm meas}
=
\frac{k_N}{k_i}
\left|
f_{{\rm ref},N}[\chi_{\rm in}]
+
\mathcal A_{{\rm sc},N}^{U}[\chi_{\rm in}]
\right|^2
\end{equation}
where the label ff indicates a field-free measurement.
Equivalently,
\begin{multline*}
\left(
\frac{\dd\sigma}{\dd\Omega}
\right)_{{\rm ff},N}^{\rm meas}
=
\frac{k_N}{k_i}
\Bigl[
|f_{{\rm ref},N}[\chi_{\rm in}]|^2
+ \\
2\mathrm{Re}
\left(
\left(f_{{\rm ref},N}[\chi_{\rm in}]\right)^{*}
\mathcal A_{{\rm sc},N}^{U}[\chi_{\rm in}]
\right)
+
|\mathcal A_{{\rm sc},N}^{U}[\chi_{\rm in}]|^2
\Bigr].
\end{multline*}
Thus the interference between the unscattered reference amplitude and
the \(U\)-generated scattered amplitude is not neglected; it is included
inside the field-free reference-plus-scattered measurement channel.

In \epref{eq:pinem-meas-factorized}, the prefactor of the
sideband distribution $P_N = J_N^2$
formally also appears in the PINEM theory. To facilitate the comparison,
we define the coupling $g_0(\vb r_{\perp0})$ by
\begin{equation}
\label{eq:g0}
g_0(\vb r_{\perp0})\coloneqq\frac{\iu}{2}\,e^{-\iu\varphi}\,
\zeta_{\vb k_i}^*(\vb r_{\perp0}),\qquad
|\zeta_{\vb k_i}(\vb r_{\perp0})|
=
2|g_0(\vb r_{\perp0})|,
\end{equation}
so that $e^{-\iu\Theta}=\exp[g_0^*e^{\iu\wL t_0}-g_0e^{-\iu\wL t_0}]$
and $\mathcal F^{(N)}_{\vb k_i}=C_N(g_0)=e^{-\iu N\arg g_0}J_N(2|g_0|)$,
in the convention of \pref{eq:gind-field} and
\pref{eq:pinem-coherent-exponential}. Equivalently $g_0$ is the field
integral \pref{eq:gind-field} of the external longitudinal field
$\EE^{(+)}_{{\rm ext},z}=+\iu\wL A^{(+)*}_ze^{-\iu\varphi}/2c$ (the
coefficient of $e^{-\iu\wL t}$), with the same time origin and
trajectory as $g_{\rm ind}$;
leading to
\begin{equation}
\label{eq:pn}
P_N(\vb r_{\perp0})
=
J_N^2(2|g_0(\vb r_{\perp0})|).
\end{equation}
Here \(g_0\) denotes the localized projectile coupling already
prescribed in the channel Hamiltonian.  The pump-induced target mean
density supplies a second coherent coupling \(g_{\rm ind}\), defined
by \pref{eq:gind-field} or \pref{eq:gind-density}.  At strict first
Born order its one-photon amplitude is added coherently to the
optical reference and static-potential amplitudes above, including all
interference terms that reach the same target-elastic detector channel.
In the all-orders PINEM organization the coherent field entering
\(\mathcal F_{\vb k_i}^{(N)}\) is instead built from
\begin{equation}
g_{\rm tot}(\vb r_{\perp0})
=g_0(\vb r_{\perp0})+g_{\rm ind}(\vb r_{\perp0}),
\label{eq:gtot-elastic-app}
\end{equation}
so that \pref{eq:pn} becomes
\(P_N=J_N^2(2|g_{\rm tot}|)\).  The residual target interaction must
then be expressed through
\(\delta\hat\varrho_1=\hat\varrho_1-\bar\varrho\); adding the coherent
outer product from \pref{eq:SF-coherent-connected} again would double
count \(g_{\rm ind}\).

With a prescribed coupling identified with the sample-induced near
field, \epref{eq:pn} is the standard PINEM sideband distribution of
Park, Lin, and Zewail, obtained there from the high-energy time-dependent
Schrödinger evolution of an electron wave packet in the
nanostructure-scattered optical near field~\cite{Park2010}, and
derived by an independent quantum calculation by Garc\'{\i}a de
Abajo, Asenjo-Garcia, and Kociak~\cite{deAbajo2010}. We note that
$P_N(\vb r_{\perp0})$
becomes the normalized measured sideband distribution when the field-free channel factor
\pref{eq:dsigma-passive-meas}
is approximately independent of $N$. Namely,
\begin{equation}
\label{eq:pinem-meas-softsideband}
\left(
\frac{\dd\sigma}{\dd\Omega}
\right)_N^{\rm meas}
\simeq
P_N(\vb r_{\perp0})
\left(
\frac{\dd\sigma}{\dd\Omega}
\right)_{{\rm ff},0}^{\rm meas}.
\end{equation}

If the unscattered reference amplitude is removed from the selected channel, e.g.\ blocked
($f_{{\rm ref},N}[\chi_{\rm in}]=0$), the measured field-free
channel
reduces to the isolated
\(U\)-generated elastic contribution; subtracting a reference-only
intensity would leave the interference term. In that case
\[
\left(
\frac{\dd\sigma}{\dd\Omega}
\right)_{{\rm ff},N}^{\rm meas}
\longrightarrow
\left(
\frac{\dd\sigma}{\dd\Omega}
\right)_{N}^{(0)}
\]
with the field-free elastic cross section \pref{eq:cs-elastic} generated by the
static potential \(U\).
Equation~\pref{eq:pinem-meas-factorized} reduces to
\begin{equation}
\label{eq:pinem-U-factorized}
\left(
\frac{\dd\sigma}{\dd\Omega}
\right)_N^{U}
\simeq
P_N(\vb r_{\perp0})
\left(
\frac{\dd\sigma}{\dd\Omega}
\right)_{N}^{(0)}.
\end{equation}

The condition of validity of the local factorization is that both the
magnitude and the phase of
\(\mathcal F_{\vb k_i}^{(N)}(\vb r_\perp)\) vary weakly over the
transverse support selected by the field-free reference and scattered
components. If \(w_{\rm ch}\) is the characteristic
transverse width of this selected passive channel and \(L_{\rm opt}\)
is the length scale over which
\(\mathcal F_{\vb k_i}^{(N)}\) varies, one requires
\[
w_{\rm ch}
\left|
\nabla_\perp
\ln
\mathcal F_{\vb k_i}^{(N)}(\vb r_\perp)
\right|_{\vb r_{\perp0}}
\ll 1,
\]
or, more simply, \(w_{\rm ch}\ll L_{\rm opt}\). The approximation also
holds when the incoming electron wave packet itself is localized on a
scale much smaller than \(L_{\rm opt}\). It fails when the structured
external field imprints appreciable transverse phase or amplitude
variations across the region selected by the field-free
reference-plus-scattered channel.
In that case the optical field acts as a transverse grating, and
the full coherent diffraction integral must be retained, as in the
structured near-field diffraction problem studied by de Abajo,
Barwick, and Carbone~\cite{Abajo2016}.

\paragraph*{From local PINEM amplitudes to an angular distribution.}
The same exit-plane construction gives the far-field angular distribution
of the resummed coherent channel of \secref{sec:pinem-bridge}. Take
plane-wave incidence, omit the static potential $U$, and replace the
prescribed coupling $g_0(\vb R)$ by the total coherent coupling
$g_{\rm tot}(\vb R)=g_0(\vb R)+g_{\rm ind}(\vb R)$ [\pref{eq:g0},
\pref{eq:gind-field}], so that $\mathcal F_{\vb k_i}^{(N)}\to
C_N(g_{\rm tot})=e^{-\iu N\arg g_{\rm tot}}J_N(2|g_{\rm tot}|)$ of
\pref{eq:pinem-coherent-exponential}. With the impact parameter
$\vb R\equiv\vb r_\perp$, the transverse transfer $\vb Q_\perp\equiv\vb
k_\perp$, and the incident beam removed as in
\pref{eq:pinem-opt-ampl}, \pref{eq:pinem-total-exit-plane-amplitude}
becomes
\begin{equation}
\begin{aligned}
f_N^{\rm coh}(\vb Q_\perp)&=\frac{k}{2\pi\iu}\int\!\dd^2\vb R\,
e^{-\iu\vb Q_\perp\cdot\vb R}
\bigl[C_N(g_{\rm tot}(\vb R))-\delta_{N0}\bigr],\\
\frac{\dd\sigma_N^{\rm coh}}{\dd\Omega}&=\bigl|f_N^{\rm coh}(\vb Q_\perp)\bigr|^2,
\end{aligned}
\label{eq:pinem-angular}
\end{equation}
with $k$ the common longitudinal wave number ($k_N/k_i\to1$ at this
order). This is the far-field amplitude transform of
Ref.~\cite{Abajo2016}, including its erratum; its energy-gain
coefficients satisfy $f_\ell=C_{-\ell}$ in the present loss-index
convention. A transmitted-beam measurement restores the subtracted
incident beam together with its interference, as in
\pref{eq:pinem-meas-coherent}--\pref{eq:pinem-meas-expanded}, and a
static potential enters at amplitude level.

Two points matter for the interpretation. First, the local quantity
$P_N(\vb R)=|C_N(g_{\rm tot}(\vb R))|^2$ is a sideband probability along a
single straight trajectory, not an angular cross section: when the
coupling varies transversely, the Fourier transform acts on the
amplitude $C_N$, and the Bessel function stays inside the integral.
Second, \pref{eq:pinem-angular} is consistent with the first-Born
formalism. With $g_0=0$, the weak-coupling limit
$C_{+1}\simeq g_{\rm ind}^*$ and \pref{eq:gind-density} make
\pref{eq:pinem-angular} reproduce the one-photon Born amplitude
\pref{eq:weak-pinem-born-parent},
$f_{+1}=-(\me/2\pi\hbar^2)v(\Qv)R_{+1}(\Qv)$ at $Q_z=-\wL/v_{\rm e}$, using
$v_{\rm e}=\hbar k/\me$ in the nonrelativistic core. The PINEM
resummation of \secref{sec:pinem-bridge} is thus normalized and phased
consistently with the Born scattering amplitudes. It is an all-orders
reorganization of the coherent channel and is not used in
\secref{sec:two-level} or \figref{fig:eels-spectrum}, which are strict
first-Born results (\appref{app:two-level}).

\section{Details of the driven two-level model}
\label{app:two-level}
This appendix collects the algebra behind the driven two-level example of
\secref{sec:two-level}. We fix the phase convention of the Rabi coupling
and identify the one phase combination that survives in observables;
exhibit the two indistinguishable dressed-state transition paths,
$g\to e$ and $e\to g$, that produce the $J_N$--$J_{N+2}$ interference of
the connected, target-changing channel; derive the target-elastic
coherent contribution \pref{eq:two-level-coherent-cross-section}; and
list consistency checks. Everything here is evaluated at strict first-Born
order, as are \secref{sec:two-level} and \figref{fig:eels-spectrum}. The
all-orders PINEM reorganization of the coherent channel,
\pref{eq:pinem-angular} of \appref{app:elastic}, is not used: it would
absorb the induced coupling into the projectile channel, and adding
\pref{eq:two-level-coherent-cross-section} to it would count the coherent
mean field twice.

\paragraph*{Rabi coupling and phase convention.}
The coupling \pref{eq:two-level-coupling} follows from the length-gauge
form $-\hat{\vb D}\cdot\EEv=+\hat{\vb D}\cdot\dot{\vb A}/c$ of the
convention \pref{eq:H0}; the velocity gauge gives the same phase, with
$\omega_0$ in place of $\wL$. Hermiticity and
\pref{eq:two-level-density-model} give
$\varrho_{eg}(\vb q)=\varrho_{eg}(-\vb q)^*$: the transition density is
real in real space, and $\varrho_{eg}(\vb q)\simeq-\iu\vb q\cdot\vb D_{eg}$
with real $\vb D_{eg}$. The phase $\chi$ of $\Omega_c$ and the phases of
$\varrho_{eg}$ and $\varrho_{ge}$ each depend on the phase chosen for
$\ket{e}$. Only the combination
\begin{equation}
\Lambda\coloneqq-e^{2\iu\chi}\frac{\varrho_{ge}(\Qv)}{\varrho_{eg}(\Qv)}
=\frac{\vb D_{eg}\cdot\vb A_0^{(+)*}}{\vb D_{ge}\cdot\vb A_0^{(+)}}\,
\frac{\varrho_{ge}(\Qv)}{\varrho_{eg}(\Qv)}
\label{eq:two-level-kappa}
\end{equation}
is invariant under $\ket{e}\to e^{\iu\beta}\ket{e}$, and it is this
combination that fixes the relative sign and phase of the two
interfering paths below. For \pref{eq:two-level-density-model} with
linear polarization, $\Lambda=1$.

\paragraph*{Two-path interference in the connected channel.}
Write the Floquet states \pref{eq:two-level-states} through their periodic
parts $\ket{u_\pm(t)}\coloneqq e^{\iu E_\pm t/\hbar}\ket{\Phi_\pm(t)}$, with
$c=\cos(\theta/2)$, $s=\sin(\theta/2)$ and $\psi=\wL t+\varphi$. Direct
multiplication gives, for general density matrix elements,
\begin{align}
\bra{u_-}\hat\varrho_1(\vb q)\ket{u_+}
&=c^2\varrho_{eg}
-s^2e^{2\iu\chi}\varrho_{ge}\,e^{-2\iu\psi}
\nonumber\\
&\quad
+cs\,e^{\iu\chi}(\varrho_{ee}-\varrho_{gg})\,e^{-\iu\psi}.
\label{eq:two-level-path-matrix}
\end{align}
Each term is one route from $\Phi_+$ to $\Phi_-$:
\begin{itemize}
\item[(i)] $g\to e$: the density excites the ground component $c\ket{g}$
of the incoming state and projects onto the excited component
$c\ket{e}$ of the outgoing state. No optical phase is involved, so this
route is the Floquet harmonic $M=0$.
\item[(ii)] $e\to g$: the density de-excites the excited component of the
incoming state and projects onto the ground component of the outgoing
state. Each of the two states contributes one optical phase
$e^{-\iu\psi}$, so this route is the harmonic $M=-2$.
\item[(iii)] The intermediate harmonic $M=-1$ carries the diagonal
contrast $\varrho_{ee}-\varrho_{gg}$ and vanishes under
\pref{eq:two-level-density-model}.
\end{itemize}
In the cross section the target harmonic $M$ is combined with the
projectile coefficients $B_n$ of \pref{eq:dsigma-SF} at fixed
$N=n+M$, so route (i) pairs with $B_N$ and route (ii) with $B_{N+2}$.
In terms of \pref{eq:two-level-kappa},
\begin{equation}
\sum_n B_n\tilde\varrho_{-+}^{(N-n)}
=\varrho_{eg}\bigl[c^2B_N+\Lambda\,s^2e^{-2\iu\varphi}B_{N+2}\bigr],
\label{eq:two-level-path-general}
\end{equation}
and with $\varrho_{eg}=\varrho_{ge}=a\coloneqq q_{\rm dip}F$ of
\pref{eq:two-level-density-model},
\begin{align}
\sum_n B_n\tilde\varrho_{-+}^{(N-n)}
&=a\bigl[c^2B_N-s^2e^{2\iu(\chi-\varphi)}B_{N+2}\bigr]
\nonumber\\
&=a\,e^{\iu N(\varphi+\theta_{\Qv}^{\rm hom})}\\
&{\phantom =}\quad \times
\bigl[c^2J_N-s^2e^{2\iu(\chi+\theta_{\Qv}^{\rm hom})}J_{N+2}\bigr],
\label{eq:two-level-path-convolution}
\end{align}
with Bessel argument $|\Qv\cdot\alphav|$. This is \pref{eq:BN}. The two
routes are alternative amplitudes for the same initial and final Floquet
states and the same electron channel, so they interfere. Their harmonic
difference of two stems from the optical dressing of the states, not from
a two-photon Coulomb vertex. A consistent change of Floquet
representatives shifts the labels but not the interference.

\paragraph*{Target-elastic coherent channel.}
For $\Phi_+\to\Phi_+$ the mean-density harmonics
\pref{eq:two-level-coherent-R} enter instead. With $N=n+M$ as above,
\begin{align*}
\sum_nB_nR_{N-n}
&=\varrho_{gg}B_N
+a\,cs\,e^{\iu(\varphi-\chi)}B_{N-1}
\\
&\quad
+a\,cs\,e^{-\iu(\varphi-\chi)}B_{N+1},
\end{align*}
which gives \pref{eq:two-level-coherent-cross-section}. For linear
polarization, after removing the common phase $e^{\iu N\varphi}$ and using
$e^{\iu\chi}=-\iu\sigma$ and $J_{N-1}-J_{N+1}=2J_N'$, the bracket becomes
$\varrho_{gg}J_N(x)+\iu\sigma a\sin\theta\,J_N'(x)$. In the dipole limit
$a\simeq-\iu\Qv\cdot\vb D_{eg}$ this is
\begin{equation*}
\varrho_{gg}J_N(x)-(\Qv\cdot\vb P_0)\,J_N'(x),
\qquad
\vb P_0=-\sigma\sin\theta\,\vb D_{eg},
\end{equation*}
where $\vb P(t)=\vb P_0\sin\psi$ is the induced dipole of $\Phi_+$. For
linear polarization $R_{+1}e^{-\iu\varphi}=-R_{-1}e^{\iu\varphi}$, so this
dipole oscillates in phase or in antiphase with the electric field, not
in quadrature with it: a stationary Floquet state carries a reactive,
non-absorbing response. To first order in $\vb P_0$ the bracket is
$\varrho_{gg}J_N\bigl(x-\Qv\cdot\vb P_0/\varrho_{gg}\bigr)$, i.e.\ a
Bessel function of the electron quiver taken relative to the oscillating
target dipole. This induced-dipole term is what reshapes the
target-elastic (blue) comb of \figref{fig:eels-spectrum}(b) with respect
to the passive comb of panel~(a).

\paragraph*{Consistency checks.}
(i)~Quasienergies: $(E_--E_+)/\hbar=\omega_0+\Delta-\WR=\wL-\WR\to\omega_0$
for $\Omega\to0$ at fixed $\Delta>0$.
(ii)~Sideband closure at fixed transfer:
$\sum_N|\mathcal B_N|^2=c^4+s^4$ and
$\sum_N|\sum_nB_nR_{N-n}|^2=|\varrho_{gg}|^2+2|a|^2c^2s^2$.
(iii)~Dark line: $\mathcal B^{\rm lin}_{-1}=-\cos\theta\,J_1(x)$ vanishes
at maximal mixing.
(iv)~Born--PINEM matching: in weak coupling the angular transform
\pref{eq:pinem-angular} reproduces the one-photon coherent Born amplitude
(\appref{app:elastic}).
(v)~No double counting: when the coherent mean potential is resummed into
the projectile channel, the residual target-changing signal must be
computed from the connected correlator with the correspondingly dressed
projectile kernel, and \pref{eq:two-level-coherent-cross-section} is not
added again.
Finally, a change of laser intensity changes $\WR$, $\theta$, $x$ and
$|g^{\rm 2lev}_{\rm ind}|$ together; the fixed-$x$ comparison of
\figref{fig:two-level} is a controlled parameter illustration.

\section{Validity of the straight-line trajectory in the eikonal--Volkov construction}
\label{app:trajectory-bending}
The eikonal coupling parameter \pref{eq:eikonal-zeta} of
\secref{sec:projdyn-eikonal} is evaluated along the \emph{undeflected}
free-electron trajectory $\vb r(s)=\vb r-\vb v_{\vb k}s$. In the
inhomogeneous-field regime, the classical projectile is in fact bent by
the gradients of $\Ap(\vb r)$, and the straight-line assumption is
an approximation whose validity must be quantified. The purpose of this
appendix is to identify, at the level of scaling
estimates for a smooth localized field with a single characteristic
length, the dimensionless
small parameters that control this approximation and to give the
corresponding inequalities.

\paragraph*{Characteristic scales.}
At TEM/PINEM energies the projectile kinematics must be evaluated
relativistically; we therefore use the relativistic projectile
relations
\begin{equation}
p\equiv\hbar k=\gamma_{\mathrm p}m_{\mathrm e}v,\qquad
\epsilon_{\rm k}=(\gamma_{\mathrm p}-1)m_{\mathrm e}c^{2}
\label{eq:appC-relativistic-kinematics}
\end{equation}
in place of the nonrelativistic $v=\hbar k/m_{\mathrm e}$,
$\epsilon_{\rm k}=\tfrac{1}{2}m_{\mathrm e}v^{2}$, with
$\gamma_{\mathrm p}=1+\epsilon_{\rm k}/(m_{\mathrm e}c^{2})$ the projectile
Lorentz factor of \pref{eq:rel-mass-substitution}. The projectile
transverse mass in the eikonal equation of motion is
$m_{\mathrm p}^{\mathrm{eff}}=\gamma_{\mathrm p}m_{\mathrm e}$. With
these substitutions:
\begin{itemize}
\item the peak vector-potential amplitude
$A_{0}\equiv\sup_{\vb r}|\Ap(\vb r)|$, and the peak electric-field
scale $\EE_{0}\equiv\wL A_{0}/c$ generated by the harmonic time
dependence \pref{eq:Acos};
\item the field-inhomogeneity scale
$L^{-1} \coloneqq \bigl(\sup_{\vb r}|\nabla\Ap(\vb r)|\bigr)/A_0$, so that
$|\nabla\Ap(\vb r)|\le A_{0}/L$ pointwise;
\item the transit time across the inhomogeneity,
$\tau_{\rm tr}\equiv L/v$;
\item the classical (modulus of complex) quiver displacement
$\alpha_{0}\equiv |q_{\rm p}|A_{0}/(m_{\mathrm p}^{\mathrm{eff}}c\wL)
=|q_{\rm p}|A_{0}/(\gamma_{\mathrm p}m_{\mathrm e}c\wL)$ and the
ponderomotive energy
$U_{\rm p}\equiv q_{\rm p}^{2}A_{0}^{2}/(4 m_{\mathrm p}^{\mathrm{eff}}c^{2})
=\tfrac{1}{4}m_{\mathrm p}^{\mathrm{eff}}(\wL\alpha_{0})^{2}$.
\end{itemize}

Three independent dimensionless parameters characterize the regime of
validity of the straight-line eikonal--Volkov construction:
\begin{align}
\eta_{\rm eik} &\equiv \frac{1}{kL}, &
\eta_{\rm q}   &\equiv \frac{\alpha_{0}}{L}, &
\eta_{\rm cyc} &\equiv \frac{v}{\wL L}=(\wL\tau_{\rm tr})^{-1}.
\label{eq:trajbend-smallparams}
\end{align}
$\eta_{\rm eik}$ is the semiclassical (eikonal) parameter already
required by \pref{eq:eikonal-validity}. $\eta_{\rm q}$ measures the
projectile's quiver amplitude against the inhomogeneity scale.
$\eta_{\rm cyc}^{-1}=\wL\tau_{\rm tr}$ is the number of optical
radians accumulated during the transit and discriminates between the
cycle-averaged ($\eta_{\rm cyc}\ll 1$) and the sub-cycle
($\eta_{\rm cyc}\gtrsim 1$) regimes of the trajectory dynamics.

\paragraph*{Transverse impulse and deflection.}
The transverse Lorentz force on the projectile has an electric component
of amplitude $|q_{\rm p}|\EE_{0}$ oscillating at $\wL$, and a magnetic
component of amplitude $|q_{\rm p}|v A_{0}/(cL)$ arising from
$\vb B=\nabla\times\vb A$. The maximum (uncanceled) transverse impulse
acquired during the transit is therefore estimated by
\begin{equation}
\frac{|\Delta p_{\perp}|}{p}\lesssim
\frac{|q_{\rm p}|\EE_{0}}{p}\,
\min\!\bigl(\tau_{\rm tr},\,\wL^{-1}\bigr)
+\frac{|q_{\rm p}|A_{0}}{m_{\mathrm p}^{\mathrm{eff}}c}\,
\frac{\tau_{\rm tr}}{L},
\label{eq:trajbend-impulse}
\end{equation}
where the $\min$ encodes the cancellation of the oscillating part of
the electric force over each full cycle. In the cycle-averaged regime
$\eta_{\rm cyc}\ll 1$ the secular contribution reduces to the
ponderomotive force $\langle\vb F\rangle=-\nabla U_{\rm p}(\vb r)$; after cycle
averaging, the secular contribution is of order $U_{\rm p}/(pv)$, while
oscillatory motion during traversal remains relevant to the local-angle
and phase estimates below.

The corresponding transverse displacement during the transit follows, to
leading order in deflection and nonrecoil, from the in-transit kinetic
transverse momentum $\delta\vb p_\perp(t)$,
\begin{equation}
\begin{gathered}
\Delta\vb r_{\perp}(t)=\frac{1}{m_{\mathrm p}^{\mathrm{eff}}}
\int_{t_{\rm in}}^{t}\dd t'\,\delta\vb p_{\perp}(t'),\\
\frac{|\Delta\vb r_{\perp}|}{L}\lesssim\theta_{*}\equiv
\sup_{\rm transit}\frac{|\delta\vb p_{\perp}|}{p},
\end{gathered}
\label{eq:trajbend-displacement}
\end{equation}
where $p=\gamma_{\mathrm p}m_{\mathrm e}v=m_{\mathrm p}^{\mathrm{eff}}v$
is the relativistic projectile momentum
\pref{eq:appC-relativistic-kinematics} and $\tau_{\rm tr}\sim L/v$, so
$\theta_{*}\ll1$ is sufficient for both a small in-transit angle and a small
transverse displacement. The
exit kick alone does not bound $\theta_{*}$: a transverse field component
can deflect the projectile during traversal and return it to its initial
direction.

\paragraph*{Three distinct deflection mechanisms.}
\begin{enumerate}
\item \emph{Sub-cycle direct impulse} ($\eta_{\rm cyc}\gtrsim 1$). The
projectile crosses the field-localization region in less than one
optical cycle, so the oscillating electric force does not average out.
The transverse impulse is linear in the field amplitude, giving
\begin{equation}
\eta_{\rm def}^{\rm sub}
\;\equiv\;\frac{|q_{\rm p}|\EE_{0}L}{pv},
\label{eq:eta-def-sub}
\end{equation}
which reduces to $|q_{\rm p}|\EE_{0}L/(2\epsilon_{\rm k})$ nonrelativistically
($pv\to 2\epsilon_{\rm k}$).
\item \emph{Many-cycle residual oscillatory impulse}
($\eta_{\rm cyc}\ll 1$). The projectile crosses many optical cycles
during the transit and the oscillating part of the electric force
returns the projectile to within a single-cycle excursion of the
unperturbed path. The residual transverse drift is bounded by
\begin{equation}
\eta_{\rm def}^{\rm osc}
\;\equiv\;\frac{|q_{\rm p}|\EE_{0}}{p\wL}
\;=\;\frac{|q_{\rm p}|A_{0}}{m_{\mathrm p}^{\mathrm{eff}}c\,v},
\label{eq:eta-def-osc}
\end{equation}
linear in $A_{0}$ but independent of $L$.
\item \emph{Many-cycle secular ponderomotive deflection}
($\eta_{\rm cyc}\ll 1$). The cycle-averaged force is the ponderomotive
gradient $\langle\vb F\rangle=-\nabla U_{\rm p}(\vb r)$, giving a
quadratic-in-$A_{0}$ secular displacement
\begin{equation}
\eta_{\rm def}^{\rm pond}
\;\equiv\;\frac{U_{\rm p}}{\tfrac{1}{2}pv}
\;=\;\frac{2U_{\rm p}}{pv},
\label{eq:eta-def-pond}
\end{equation}
which reduces to $U_{\rm p}/\epsilon_{\rm k}$ nonrelativistically.
\end{enumerate}
The magnetic term of \pref{eq:trajbend-impulse} is not a separate entry of
this list. With $\tau_{\rm tr}\sim L/v$ it has the scale
$|q_{\rm p}|A_{0}/(m_{\mathrm p}^{\mathrm{eff}}c\,v)=\eta_{\rm def}^{\rm osc}$ in either
regime, so that the conservative in-transit estimate has the scale
$\eta_{\rm def}^{\rm osc}[1+\min(1,\eta_{\rm cyc}^{-1})]$, up to shape factors. For
sub-cycle transit this scale exceeds the electric estimate
$\eta_{\rm def}^{\rm sub}$ by the factor $\eta_{\rm cyc}$; in a given geometry the
two contributions may add or cancel. At leading eikonal order, with
field-free endpoints, the net exit kick is the transverse gradient of the complete trajectory phase,
$\Delta\vb p_\perp=-\hbar\nabla_\perp\Theta_{\vb k}$; it does not determine the
angle during traversal.
These three parameters are not independent: with
$p=m_{\mathrm p}^{\mathrm{eff}}v$ and \pref{eq:trajbend-smallparams},
\begin{align}
\eta_{\rm def}^{\rm osc}
&=\frac{\eta_{\rm q}}{\eta_{\rm cyc}}
=\frac{\wL\alpha_{0}}{v},\nonumber\\
\eta_{\rm def}^{\rm sub}
&=\frac{\eta_{\rm def}^{\rm osc}}{\eta_{\rm cyc}},\nonumber\\
\eta_{\rm def}^{\rm pond}
&=\tfrac{1}{2}\bigl(\eta_{\rm def}^{\rm osc}\bigr)^{2},
\label{eq:eta-def-relations}
\end{align}
so that $\eta_{\rm def}^{\rm osc}=v_{\rm quiver}/v$ with quiver
velocity $v_{\rm quiver}=\wL\alpha_{0}$, and all three follow from
$\eta_{\rm q}$ and $\eta_{\rm cyc}$ alone. The mechanisms remain
physically distinct -- they scale differently in $A_{0}$ and $L$, and
$\eta_{\rm def}^{\rm sub}\neq\eta_{\rm def}^{\rm pond}\eta_{\rm cyc}^{-1}$
-- but their relative sizes are fixed. Writing $\eta_{\rm def}$ for the
characteristic scale,
$\eta_{\rm def}=\eta_{\rm def}^{\rm osc}$ in either regime (for
$\eta_{\rm cyc}\gtrsim1$ through the magnetic term, for $\eta_{\rm cyc}\ll1$
through the residual oscillatory impulse), at scaling accuracy with
geometry-dependent constants suppressed, the
ponderomotive contribution being automatically subdominant whenever
$\eta_{\rm def}^{\rm osc}\ll1$.

\paragraph*{Regime of validity.}
The joint inequalities that delimit the validity of the straight-line
eikonal--Volkov ansatz at the level of the classical trajectory
geometry are
\begin{subequations}
\label{eq:trajbend-validity}
\begin{align}
\eta_{\rm eik}=\frac{1}{kL}&\ll 1,
\label{eq:trajbend-validity-eik}\\[2pt]
\eta_{\rm q}=\frac{\alpha_{0}}{L}&\ll 1,
\label{eq:trajbend-validity-quiver}\\[2pt]
\eta_{\rm def}^{\rm osc}=\frac{\wL\alpha_{0}}{v}&\ll 1.
\label{eq:trajbend-validity-def}
\end{align}
\end{subequations}
Equation~\pref{eq:trajbend-validity-eik} guarantees the eikonal expansion
underlying \pref{eq:eikonal-ansatz}--\pref{eq:eikonal-eq};
\epref{eq:trajbend-validity-quiver} ensures that the quiver motion does
not sample regions where $\vb A(\vb r)$ differs significantly from its
value on the undeflected path; \epref{eq:trajbend-validity-def}
guarantees that the secular trajectory deflection is parametrically
small both as a fraction of the longitudinal momentum and as a fraction
of $L$, and by \pref{eq:eta-def-relations} it is the single sufficient
condition covering both regimes, implying
$\eta_{\rm def},\eta_{\rm def}^{\rm pond}\ll1$. Corrections to the trajectory coupling
$\zeta_{\vb k}(\vb r)$ must be assessed in absolute size; a relative
estimate additionally requires that the unperturbed coupling not be
suppressed by cancellations. These corrections modify the projectile
kernel in \pref{eq:dsigma-inhom}, not the intrinsic target structure
factor $\SF$.

\paragraph*{Phase fidelity.}
The conditions \pref{eq:trajbend-validity} only control the
\emph{classical} trajectory geometry. In addition to
trajectory geometry, we estimate the geometric phase contribution
associated with the extra path length of the bent trajectory, using the
reference wave number $k$. Changes in the field sampled along the path
require separate control.
Writing $z$ for the longitudinal coordinate along the undeflected path
and
\[
\theta_{\rm d}(z)
\equiv
\left|
\frac{\dd\Delta\vb r_{\perp}}{\dd z}
\right|
\simeq
\frac{|\delta\vb p_{\perp}(z)|}{p}
\]
for the local deflection angle, the path-length element is
\[
\dd\ell
=
\dd z\,\sqrt{1+\theta_{\rm d}^{2}(z)}
\simeq
\dd z\left[1+\frac{\theta_{\rm d}^{2}(z)}{2}\right].
\]
The resulting path-length phase contribution is therefore
\begin{equation}
\delta\phi_{\rm bend}
=
k(\ell-L)
\simeq
\frac{k}{2}\int_{0}^{L}\dd z\,\theta_{\rm d}^{2}(z)
=
\frac{kL}{2}\theta_{\rm d,rms}^{2},
\label{eq:trajbend-phasefidelity}
\end{equation}
where
\[
\theta_{\rm d,rms}^{2}
\equiv
\frac{1}{L}\int_{0}^{L}\dd z\,\theta_{\rm d}^{2}(z).
\]
Phase fidelity thus requires
\[
kL\,\theta_{\rm d,rms}^{2}\ll1.
\]
Using the operative deflection parameter $\eta_{\rm def}$ of
\pref{eq:eta-def-relations} as a characteristic upper bound on the
in-transit trajectory angle gives the conservative scaling condition,
with geometry-dependent constants suppressed,
\[
kL\,\eta_{\rm def}^{\,2}\ll1,
\quad\text{or equivalently}\quad
\eta_{\rm def}\ll\frac{1}{\sqrt{kL}}
=\sqrt{\eta_{\rm eik}}.
\]

\paragraph*{Numerical estimates for UTEM/PINEM parameters.}
For representative UTEM/PINEM parameters
($\epsilon_{\rm k}=200\,\mathrm{keV}$, $\gamma_{\mathrm p}\approx 1.39$,
$v\approx 0.7\,c$, $pv\approx 343\,\mathrm{keV}$,
$\hbar\wL\approx 1.5\,\mathrm{eV}$ so $\wL\approx 2.3\times 10^{15}\,
\mathrm{s}^{-1}$, $L\approx 30\,\mathrm{nm}$, peak optical field
$\EE_{0}\approx 3.3\times10^{4}\,\mathrm{statV\,cm^{-1}}$ ($=1\,\mathrm{GV\,m^{-1}}$, the value used in the numbers below), and a 200~keV de Broglie
wavelength $\lambda_{\rm dB}\approx 2.5\,\mathrm{pm}$ giving
$k\approx 2.5\times 10^{12}\,\mathrm{m}^{-1}$, $kL\approx 7.5\times 10^{4}$),
the relativistic projectile kinematics
\pref{eq:appC-relativistic-kinematics} are used throughout; the
dimensionless numbers come out as
\begin{align}
\eta_{\rm cyc}&\;\approx\;3.05, &
\eta_{\rm eik}&\;\approx\;1.33\times 10^{-5}, \nonumber\\
\eta_{\rm q}&\;\approx\;8.11\times 10^{-4}, &
\eta_{\rm def}^{\rm pond}&\;\approx\;3.54\times 10^{-8},
\label{eq:trajbend-numerics-1}\\
\eta_{\rm def}^{\rm sub}&\;\approx\;8.72\times 10^{-5}, &
\eta_{\rm def}^{\rm osc}&\;\approx\;2.66\times 10^{-4}.
\end{align}
These values satisfy \pref{eq:eta-def-relations} identically. At these
parameters $\eta_{\rm cyc}\approx3$, so the transit is
\emph{sub-cycle}, not many-cycle, and the operative deflection
parameter is $\eta_{\rm def}=\eta_{\rm def}^{\rm osc}$,
through the magnetic term (the electric sub-cycle impulse is the smaller
$\eta_{\rm def}^{\rm sub}$); the
trajectory-geometry inequalities \pref{eq:trajbend-validity} are
comfortably satisfied
($\eta_{\rm q},\eta_{\rm def}^{\rm osc}\lesssim 10^{-3}$).
Sub-cycle transit also means $v/(\wL L)=\eta_{\rm cyc}>1$: for these
parameters the local Volkov/SVE replacement
$\zeta_{\vb k}^{(\pm)}\to-\iu\vb k\cdot\alphav(\vb r)$ of
\secref{sec:projdyn-eikonal} does not apply, and the full trajectory
kernel must be used (\secref{sec:dsigma-inhom}); the trajectory-bending
estimates of this appendix are unaffected.
The two phase-fidelity products are both much smaller than unity:
\begin{equation}
kL\bigl(\eta_{\rm def}^{\rm sub}\bigr)^2
\approx 5.72\times10^{-4},
\qquad
kL\bigl(\eta_{\rm def}^{\rm osc}\bigr)^2
\approx 5.32\times10^{-3}.
\label{eq:trajbend-numerics-2}
\end{equation}
The projectile Lorentz factor $\gamma_{\mathrm p}\approx1.39$ enters
through $m_{\mathrm p}^{\mathrm{eff}}=\gamma_{\mathrm p}m_{\mathrm e}$
in $\alpha_{0}$ and through $pv\approx0.86\times2\epsilon_{\rm k}$; by
\pref{eq:eta-def-relations} this fixes the relativistic scaling of all
three deflection parameters. Because the electric sub-cycle phase-fidelity
product scales as
\[
kL\bigl(\eta_{\rm def}^{\rm sub}\bigr)^{2}
=
\frac{k|q_{\rm p}|^{2}\EE_{0}^{2}L^{3}}{(pv)^{2}},
\]
its relativistic rescaling involves both $k$ and $pv$. Relative to a
nonrelativistic evaluation at the same kinetic energy,
\[
\frac{
 \bigl[kL(\eta_{\rm def}^{\rm sub})^{2}\bigr]_{\rm rel}
}{
 \bigl[kL(\eta_{\rm def}^{\rm sub})^{2}\bigr]_{\rm nr}
}
=
\frac{k_{\rm rel}}{k_{\rm nr}}
\left(\frac{2\epsilon_{\rm k}}{pv}\right)^{2}
\approx1.5
\]
at $\epsilon_{\rm k}=200\,\mathrm{keV}$. The operative product
$kL(\eta_{\rm def}^{\rm osc})^{2}=k|q_{\rm p}|^{2}\EE_{0}^{2}L/(p\wL)^{2}$ scales as
$1/p$ at fixed field, so its rescaling is $p_{\rm nr}/p_{\rm rel}\approx0.91$.
This order-unity correction does
not alter the regime classification: both the trajectory-geometry
inequalities and the corrected phase-fidelity condition remain
comfortably satisfied.
The straight-line trajectory therefore reproduces the classical
momentum geometry to better than a part in $10^{3}$ and, for the
quoted parameters, also preserves the bending-induced eikonal phase.
The distinction between observables nevertheless remains useful in
general. The inequalities
\pref{eq:trajbend-validity} control trajectory geometry. Accuracy of the
sideband amplitudes also requires that the induced changes of the complex
coupling be small at the coupling strength and sideband under
consideration; relative errors are not uniform near spectral zeros.
Phase-sensitive comparisons additionally require control of the
accumulated phase, \pref{eq:trajbend-phasefidelity}. The trajectory-geometry and path-length phase conditions are
comfortably satisfied for the numerical example above. In regimes that violate
\pref{eq:trajbend-validity-quiver} -- tightly localized laser fields
with $L\lesssim\alpha_{0}$ -- or that drive
$\eta_{\rm def}\gtrsim1$ -- strong-field drives with
$|q_{\rm p}|\EE_{0}\gtrsim p\,\wL$ -- an explicit deflected-trajectory
eikonal construction or a canonical-momentum
(Floquet--Lagrangian) treatment is required. Such a treatment is also
needed for phase-sensitive observables when
$kL\,\eta_{\rm def}^{\,2}\not\ll1$, even if the classical deflection
angle itself remains small.

\paragraph*{Controlled phase-fidelity window.}
Rewriting the phase-fidelity condition
\pref{eq:trajbend-phasefidelity} explicitly as a bound on the peak
optical field gives
\begin{equation}
\EE_{0} \;\ll\; \frac{p\,\wL}{|q_{\rm p}|\sqrt{kL}},
\label{eq:phase-fidelity-window}
\end{equation}
which
follows from $kL\,(\eta_{\rm def}^{\rm osc})^{2}\ll1$ under the single-scale,
smooth-profile assumptions stated above.
Equation~\pref{eq:phase-fidelity-window} is a conservative scaling
window for this geometric bending-phase contribution under the
assumptions above. For the same parameters as above,
\epref{eq:phase-fidelity-window} translates into the characteristic
field scales
$7.9\times10^{5}$$\,\mathrm{statV\,cm^{-1}}$,
$4.6\times10^{5}$$\,\mathrm{statV\,cm^{-1}}$, and
$2.5\times10^{5}$$\,\mathrm{statV\,cm^{-1}}$ at
$L=10$, $30$, and $100\,\mathrm{nm}$, respectively. Using
$I=c\EE_{0}^{2}/(8\pi)$, these correspond to
characteristic intensity scales of approximately
$7.5\times10^{20}$, $2.5\times10^{20}$, and
$7.5\times10^{19}$$\,\mathrm{erg\,s^{-1}cm^{-2}}$.

The bending-induced phase fidelity is therefore comfortably
controlled for typical PINEM fields
$\EE_{0}\sim3.3\times10^{3}\,\mathrm{statV\,cm^{-1}}$ at the standard localization scale
$L\sim30\,\mathrm{nm}$. It is likewise controlled for the
strong-field example
\(\EE_{0}\sim3.3\times10^{4}\,\mathrm{statV\,cm^{-1}}\) used in
\pref{eq:trajbend-numerics-1}--\pref{eq:trajbend-numerics-2}, for
which
$kL(\eta_{\rm def}^{\rm osc})^{2}\approx5.32\times10^{-3}$.
The example therefore passes the trajectory-geometry and
bending-phase estimates; the coupling-accuracy qualification above remains
applicable to particular spectra. These field scales refer only to the bending-induced
phase correction; other validity requirements, in particular
$U_{\rm p}\ll\hbar\omega_{\rm L}$, can become restrictive at
substantially lower fields.

\section{Validity of the slowly-varying-envelope reduction of the eikonal--Volkov cross section}
\label{app:validity}
\label{app:SVE-reduction}

This appendix establishes the parametric conditions under which the
kernel $\tilde{\mathcal F}^{(n)}_{\vb k_f\vb k_i}(\Qv-\vb q')$ in
\pref{eq:dsigma-inhom} is sufficiently peaked on $\vb q'=\Qv$ that the
Coulomb factors $v(\vb q'_{1,2})$ can be pulled outside the momentum
integral and evaluated at the kinematic transfer,
$v(\vb q'_{1,2})\to v(\Qv)$. This is the parametric content of
regime~(i) of \secref{sec:dsigma-inhom}.

\paragraph*{Characteristic length scales.}
The projectile geometry and the two target sectors introduce the
following scales in the convolution \pref{eq:dsigma-inhom}:
\begin{itemize}
  \item the kinematic momentum transfer $Q=|\Qv|$;
  \item the spatial envelope of the optical drive,
        $L$, defined operationally over the interaction region (a
        global supremum of $|\nabla\ln\vb A|$ is unbounded in the tails of
        a Gaussian profile),
        which controls the transverse support of
        $\tilde{\mathcal F}^{(n)}_{\vb k_f\vb k_i}(\vb K)$;
  \item the projectile de Broglie wavelength $1/k$, already constrained
        by the eikonal condition $kL\gg 1$ of
        \pref{eq:eikonal-validity};
  \item the intrinsic connected-response momentum scale
        \begin{equation}
        \Delta q_{S,c}\;\equiv\;
        \Bigl(\sup_{\vb q}\bigl|\nabla_{\vb q}\ln
        S_{\mathrm{F,c}}^{nn'}(\vb q,\omega_N)\bigr|\Bigr)^{-1},
        \label{eq:DeltaqS-def}
        \end{equation}
        at fixed Floquet indices and energy, with the complex logarithm
        (phases included), and the corresponding
        real-space correlation length
        $\xi_{S,c}\equiv 1/\Delta q_{S,c}$ on which the target's
        connected diagonal density response varies. (The logarithmic
        gradient has the dimensions of a length and its inverse those
        of a momentum, so $\Delta q_{S,c}$ is a momentum scale.)
  \item at \(\omega_N=0\), the independent coherent mean-density scale
        \[
        \begin{aligned}
        \Delta q_R\equiv
        \Bigl(\max_{j\in\{n,n'\}}\sup_{\vb q}\bigl|\nabla_{\vb q}\ln
        R_j(\vb q)\bigr|\Bigr)^{-1},
        \\
        \xi_R\equiv1/\Delta q_R .
        \end{aligned}
        \]
        defined on the sampled nonzero support of the relevant harmonic
        pair; each harmonic is differentiated separately, since a
        translation phase cancels in $R_n(\vb q)R_{n'}^*(\vb q)$.
        There is no reason for \(\Delta q_R\) to equal
        \(\Delta q_{S,c}\). For translational-symmetry-broken targets we further
        introduce an inhomogeneity length $\xi_T$ that sets the
        off-diagonal width of
        the sector under consideration in the
        relative-momentum coordinate $\vb q_1-\vb q_2$ -- equivalently,
        a relative-momentum scale $\Delta q_T\equiv 1/\xi_T$.
\end{itemize}
These scales must be evaluated for the complete complex factors,
including their phases, on the actually sampled support: a target
displaced by $\vb r_0$ has $R_n(\vb q)=R^{(0)}_{n}(\vb q)e^{-\iu\vb q\cdot\vb r_0}$,
whose phase gradient $\vb r_0$ is invisible in $|R_n|$. With a kernel of
width $\Delta K$ the pull-out of the coherent sector requires
$\Delta K\max_j\sup|\nabla_{\vb q}\ln R_j|\ll1$,
i.e.\ $|\nabla_{\vb q}\ln R_j|/L\ll1$ for $\Delta K\sim1/L$ ($r_0/L\ll1$ for a
pure translation); a translational-symmetry-broken connected response
must be bounded as a complex function of both momentum arguments. Near zeros of the leading result, absolute bounds
replace logarithmic ones.

\paragraph*{Support of $\tilde{\mathcal F}^{(n)}_{\vb k_f\vb k_i}(\vb K)$.}
By \pref{eq:Fkernel}, $\tilde{\mathcal F}^{(n)}_{\vb k_f\vb k_i}(\vb K)$ is the
Fourier transform of $\mathcal F^{(n)}_{\vb k_f\vb k_i}(\vb r)$, whose magnitude
$J_n(|\delta\zeta_{fi}(\vb r)|)$ and phase
$n[\varphi+\theta_{fi}(\vb r)]$ vary smoothly with $\vb r$ on the
field-envelope scale $L$ in the local SVE limit and at moderate
dressing, and by the chain rule
\begin{multline}
\bigl|\nabla_{\vb r}\mathcal F^{(n)}_{\vb k_f\vb k_i}(\vb r)\bigr|
\le\bigl|J_n'(|\delta\zeta_{fi}|)\bigr|\,\bigl|\nabla_{\vb r}|\delta\zeta_{fi}|\bigr|\\
+\bigl|nJ_n(|\delta\zeta_{fi}|)\bigr|\,\bigl|\nabla_{\vb r}\theta_{fi}\bigr| .
\label{eq:F-r-derivative}
\end{multline}
Standard repeated-integration-by-parts (Paley--Wiener) estimates then
give the decay
\begin{equation}
\bigl|\tilde{\mathcal F}^{(n)}_{\vb k_f\vb k_i}(\vb K)\bigr|
\;\lesssim\;C_n\,L^{3}\,\bigl(1+|\vb K|L\bigr)^{-M},
\quad M\ge 1,
\label{eq:F-tilde-decay}
\end{equation}
with $C_n$ set by derivative norms of the
specified profile and regulator, not by a single local value of $J_n$,
and $M$ as large as the smoothness of $\vb A(\vb r)$ permits. This decay estimate, and
the normalized second moment defined next, presuppose that
$\mathcal F^{(n)}_{\vb k_f\vb k_i}(\vb r)$ itself vanishes outside the
localized field. This holds for every $n\neq0$ of the local SVE
kernel, since
$J_n(0)=0$. It does not hold for the full two-sided kernel, whose
coupling difference persists downstream along the trajectory
(\secref{sec:dsigma-inhom}): there $\mathcal F^{(n)}\propto
e^{-\iu n\wL z/v}$ and $\tilde{\mathcal F}^{(n)}$ is centered at
$K_z=-n\wL/v$. The estimates of this appendix therefore refer to the
local Volkov/SVE kernel, which requires in addition
$v/(\wL L_\parallel)\ll1$. At $n=0$, however, $J_0(0)=1$, so
$\mathcal F^{(0)}_{\vb k_f\vb k_i}(\vb r)\to1$ (not $0$) outside the
field, and its Fourier transform contains the distributional identity
term $\delta^{(3)}(\vb K)$ [in the normalization \pref{eq:Fkernel}] -- exactly the contribution isolated
by the $\mathcal F^{(N)}\to\mathcal F^{(N)}-\delta_{N0}$ subtraction of
\appref{app:elastic}. The decay \pref{eq:F-tilde-decay} and the
normalized second moment below are meaningful for $n=0$ only after
this identity term is first subtracted; unless stated otherwise, the
$n=0$ estimates below refer to the subtracted kernel. The subtraction
serves the estimate only: the identity and its interference terms are
part of the full $n=0$ observable and must be reassembled with a finite
regulator. Equivalently, defining
the (normalized) second moment
\begin{equation}
\langle K^{2}\rangle_{\tilde{\mathcal F}}\;\equiv\;
\frac{\int\!\dd^{3}\vb K\,K^{2}\,
|\tilde{\mathcal F}^{(n)}_{\vb k_f\vb k_i}(\vb K)|^{2}}
     {\int\!\dd^{3}\vb K\,
|\tilde{\mathcal F}^{(n)}_{\vb k_f\vb k_i}(\vb K)|^{2}}
\;=\;\frac{c_{n}}{L^{2}},
\label{eq:K2-moment}
\end{equation}
the kernel is concentrated on $|\vb K|\lesssim 1/L$, with
$c_{n}=\order{1}$ depending on the profile of $\vb A(\vb r)$ at moderate
dressing. The coefficient is not uniform in the coupling strength: for
$\mathcal F_1=J_1(a\,e^{-x^2/2L^2})$ one finds $\langle
K^2\rangle L^2=0.49,\,0.40,\,7.5,\,28$ for $a=0.5,\,2,\,8,\,20$, since strong
dressing sweeps through Bessel zeros.

\paragraph*{Variation of $v(\vb q')$ over the kernel support.}
Writing $\vb q'_{1,2}=\Qv+\vb K_{1,2}$ in \pref{eq:dsigma-inhom} and
Taylor-expanding the Coulomb kernel
$v(\vb q')=4\pi q_{\rm p}/|\vb q'|^{2}$ around $\vb K=0$,
\begin{multline}
\frac{v(\Qv+\vb K)}{v(\Qv)}
\;=\;1
\;-\;\frac{2\,\Qv\cdot\vb K}{Q^{2}}\\
\;+\;\frac{4(\Qv\cdot\vb K)^{2}-Q^{2}\,K^{2}}{Q^{4}}
\;+\;\order{(K/Q)^{3}}.
\label{eq:v-Taylor}
\end{multline}
The linear term integrates to zero against any kernel that is even
under $\vb K\to-\vb K$, a property inherited by the magnitude
$|\tilde{\mathcal F}^{(n)}_{\vb k_f\vb k_i}|^{2}$ from a real-valued, parity-symmetric
field envelope $\vb A(\vb r)$ in the local SVE limit; in general the
relevant first moments of the complete kernel contraction, including
the complex $n\neq n'$ cross terms, must vanish; for asymmetric profiles it is bounded
by $|\langle\vb K\rangle_{\tilde{\mathcal F}}|/Q\lesssim 1/(QL)$. The
quadratic term contributes at relative order
$\langle K^{2}\rangle_{\tilde{\mathcal F}}/Q^{2}=c_{n}/(QL)^{2}$, by
\pref{eq:K2-moment}. The pull-out
$v(\vb q'_{1,2})\to v(\Qv)$ is therefore controlled by
\begin{equation}
QL\;\gg\;1,
\label{eq:SVE-QL}
\end{equation}
for $c_n=\order{1}$, with leading relative error $\order{(QL)^{-2}}$ when
the first moments vanish and
$\order{(QL)^{-1}}$ otherwise.

\paragraph*{Variation of $\SF^{nn'}$ over the kernel
support.}
Before estimating gradients, the target factor must be split as
\begin{multline*}
\SF^{nn'}(\vb q_1,\vb q_2;\omega_N)
=
R_n(\vb q_1)R_{n'}^*(\vb q_2)\delta(\hbar\omega_N)
{}\\
{}+\SFc^{nn'}(\vb q_1,\vb q_2;\omega_N).
\end{multline*}
The gradient expansion is then applied sector by sector, because the
coherent outer product and the connected response need not share a
momentum scale.  Each sector has two distinct non-trivial dependences
on the convolution variables $\vb q'_{1,2}$:
\begin{itemize}
  \item \emph{Diagonal momentum dependence.} For translationally
        invariant targets, the connected sector obeys
        $\SFc^{nn'}(\vb q_1,\vb q_2;\omega_N)
        =(2\pi)^3\delta^{(3)}(\vb q_1-\vb q_2)\,
        s_{\mathrm{F,c}}^{nn'}(\vb q_1,\omega_N)$ [cf.\ \pref{eq:bimomentum-translation}]. The diagonal
        $S_{\mathrm{F,c}}^{nn'}(\Qv+\vb K,\omega_N)$ varies on the intrinsic
        target momentum scale $\Delta q_{S,c}$ of \pref{eq:DeltaqS-def};
        pulling it out at $\vb q=\Qv$ is controlled by the same Taylor
        expansion as for $v$, with smallness parameter
        \begin{equation}
        \frac{1}{L\,\Delta q_{S,c}}\;\ll\;1
        \quad\Longleftrightarrow\quad
        L\;\gg\;\xi_{S,c},
        \label{eq:SVE-Ls}
        \end{equation}
        and leading relative correction
        $\order{(L\Delta q_{S,c})^{-2}}
        =\order{(L/\xi_{S,c})^{-2}}$.  At \(\omega_N=0\), a
        strict first-Born treatment of the full correlator requires
        the additional condition \(L\Delta q_R\gg1\) before the
        coherent factor \(R_nR_{n'}^*\) can be pulled out.
  \item \emph{Off-diagonal coherence.} For
        translational-symmetry-broken targets,
        either \(\SFc^{nn'}\) or the coherent
        outer product can have nonzero support away from
        $\vb q_1=\vb q_2$ with width
        $\sim 1/\xi_T$ in $\vb q_1-\vb q_2$. The kernel pair
        $\tilde{\mathcal F}^{(n)}_{\vb k_f\vb k_i}(\Qv-\vb q'_{1})\,
        \tilde{\mathcal F}^{(n')\,*}_{\vb k_f\vb k_i}(\Qv-\vb q'_{2})$ restricts
        the relative-momentum coordinate $\vb q'_{1}-\vb q'_{2}$ to
        width $\sim 1/L$. The condition
        \begin{equation}
        L\;\gg\;\xi_T,
        \label{eq:SVE-xiT}
        \end{equation}
        does \emph{not} suppress the existence of the
        target's off-diagonal coherences; it ensures only that the
        projectile-side kernel samples them within a narrow
        neighborhood of $\vb q_1=\vb q_2$. The value of the full bi-momentum
        response in that neighborhood contributes at leading order; only
        its variation over the sampled region can be expanded, at relative
        order $\order{(L/\xi_T)^{-2}}$ when the first moments vanish.
        Narrow sampling does not replace that response by a momentum delta
        function. Genuine $\vb q_1\neq\vb q_2$ target weight
        is recovered in the opposite, strongly localized-field regime.
\end{itemize}
In the PINEM-resummed organization of
\secref{sec:pinem-bridge}, the coherent mean potential is moved into
the projectile channel through \(g_{\rm tot}=g_0+g_{\rm ind}\).
Only \(\SFc\) then remains in the residual target
contraction, so \(\Delta q_R\) controls the projectile-side near-field
profile rather than a second target-response contraction.

\paragraph*{Joint validity hierarchy.}
The four base inequalities
\begin{subequations}
\label{eq:SVE-validity}
\begin{align}
QL\;&\gg\;1, &\text{(Coulomb)}
\label{eq:SVE-validity-QL}\\[2pt]
L\,\Delta q_{S,c}\;&\gg\;1,
&\text{(connected response)}
\label{eq:SVE-validity-Ls}\\[2pt]
L\;&\gg\;\xi_T, &\text{(target coherence)}
\label{eq:SVE-validity-xiT}\\[2pt]
kL\;&\gg\;1, &\text{(eikonal)}
\label{eq:SVE-validity-kL}
\end{align}
\end{subequations}
delimit the parametric regime in which the Coulomb factors can be
pulled outside the momentum integral in \pref{eq:dsigma-inhom}.
Equation~\pref{eq:SVE-validity-QL} controls the variation of $v(\vb q')$ across
the kernel; \epref{eq:SVE-validity-Ls} controls the variation of the
diagonal connected target response; \epref{eq:SVE-validity-xiT}
restricts
the sampled momentum difference (it does not remove the target's
coherences);
\epref{eq:SVE-validity-kL} guarantees the eikonal--Volkov construction
itself. The local Volkov/SVE specialization of the kernel used in these
estimates requires, in addition, $v/(\wL L_\parallel)\ll1$; this is a
condition on that specialization, not on the full trajectory theory,
whose kernel remains available outside this limit
(\appref{app:trajectory-bending}).  If the coherent first-Born sector is retained explicitly at
\(\omega_N=0\), the hierarchy must be supplemented by
\(L\Delta q_R\gg1\); this additional condition is absent from the
residual target contraction in the PINEM-resummed organization.

\paragraph*{Reduced cross section.}
The following reduction additionally restricts to a translationally
invariant target and, for the quoted quadratic scaling, to
vanishing first moments of the complete
kernel contraction; both restrictions are stated explicitly
because neither follows automatically from
\pref{eq:SVE-validity}. Translational invariance of the target is a
separate assumption from the SVE conditions above: even when
$L\gg\xi_T$ holds, a genuinely translational-symmetry-broken target
retains bi-momentum structure in $\SF^{nn'}$, and the
convolution in \pref{eq:dsigma-inhom} then yields a narrow \emph{sampling}
of $\SF^{nn'}(\vb q_1,\vb q_2;\omega_N)$ near
$\vb q_1\approx\vb q_2\approx\Qv$, not automatically the single-momentum
$S_{\mathrm F}^{nn'}(\Qv,\omega_N)$; the latter requires the target's own
translational invariance in addition. Under
\pref{eq:SVE-validity}, the additional \(L\Delta q_R\gg1\) condition
whenever the strict first-Born coherent sector is retained, and this
target restriction, the inhomogeneous-field cross section
\pref{eq:dsigma-inhom} reduces to
\begin{multline}
\frac1V\frac{\dd^{2}\sigma}{\dd\Omega\dd\epsilon}\bigg|_{\mathrm{SVE}}
\simeq\qty(\frac{\me}{2\pi\hbar^{2}})^{2}|v(\Qv)|^{2}\\
\times\sum_N\frac{k_f}{k_i}\sum_{n,n'}\mathcal K^{nn'}_N(\Qv)\\
\times s_{\mathrm F}^{(N-n)(N-n')}(\Qv,\omega_N),
\label{eq:dsigma-SVE}
\end{multline}
Equation~\pref{eq:dsigma-SVE}, up to the relative corrections estimated
below, is the full strict-first-Born result:
its \(S_{\mathrm F}\) must be decomposed into coherent and connected
parts before their gradient errors are estimated.  In the
PINEM-resummed organization the same formula applies to the residual
target-changing signal with \(S_{\mathrm F}\) replaced by
\(S_{\mathrm{F,c}}\) and with the coherent mean potential already
included in the projectile kernel through \(g_{\rm tot}\).
Here the projectile-side harmonic-overlap weight is
\begin{equation}
\mathcal K^{nn'}_N(\Qv)\;\coloneqq\;
\frac{(2\pi)^3}{V}\!\int\!\dd^{3}\vb K\,
\tilde{\mathcal F}^{(n)}_{\vb k_f\vb k_i}(-\vb K)\,
\tilde{\mathcal F}^{(n')\,*}_{\vb k_f\vb k_i}(-\vb K).
\label{eq:K-kernel-SVE}
\end{equation}
Equation~\pref{eq:dsigma-SVE} is the strict-homogeneous Bethe--Floquet
cross section \pref{eq:dsigma-SF} with the Bessel products replaced by the
projectile harmonic-overlap weight \pref{eq:K-kernel-SVE}; no momentum
convolution remains once $v$ and $s_{\mathrm F}$ are evaluated at $\Qv$.
By Parseval, $\mathcal K^{nn'}_N=V^{-1}\int_V\dd^3\vb r\,\mathcal
F^{(n)}\mathcal F^{(n')*}$ for a specified sample window; it is not
normalized to one for each sideband. It recovers \pref{eq:dsigma-SF} in the
limit $L\to\infty$, $\mathcal K^{nn'}_N(\Qv)\to
J_n J_{n'}\,e^{\iu(n-n')(\varphi+\theta_{\Qv}^{\rm hom})}$.

\paragraph*{Leading corrections.}
When the first
moments of the complete kernel contraction vanish (for $n=n'$ in the
local SVE limit a parity-symmetric real envelope suffices), the linear
contributions to \pref{eq:v-Taylor} and to the target-side expansions
vanish, and the relative corrections to \pref{eq:dsigma-SVE}, meaningful
away from zeros of the leading result, are the quadratic scaling
estimates
\begin{align}
\delta\sigma_{\rm Coulomb}\;&\sim\;\order{(QL)^{-2}},\\
\delta\sigma_{\rm target,c}\;&\sim\;
\order{(L\Delta q_{S,c})^{-2}}
=\order{(L/\xi_{S,c})^{-2}},\\
\delta\sigma_{\rm target,coh}\;&\sim\;
\order{(L\Delta q_R)^{-2}}
=\order{(L/\xi_R)^{-2}},\\
\delta\sigma_{\rm offdiag}\;&\sim\;\order{(L/\xi_T)^{-2}}.
\end{align}
The coherent estimate applies only to a strict first-Born calculation
at \(\omega_N=0\); it is absorbed into the projectile-kernel accuracy
once the coherent mean potential is PINEM-resummed.
Otherwise the linear term in \pref{eq:v-Taylor} survives and
$\delta\sigma_{\rm Coulomb}$ degrades to $\order{(QL)^{-1}}$; the target-side
estimates $\delta\sigma_{\rm target,c}$,
$\delta\sigma_{\rm target,coh}$, and
$\delta\sigma_{\rm offdiag}$ are
governed by the analogous sectorwise Taylor expansions of
$S_{\mathrm{F,c}}^{nn'}$, \(R_nR_{n'}^*\), and
$\SFc^{nn'}$ and degrade in the same way. These are
scaling estimates for the regime in which the SVE reduction is
parametrically justified, not a general theorem valid for arbitrary
field profiles; they make explicit the parametric content of
regime~(i) of \secref{sec:dsigma-inhom}.
Operationally, for a specified sample and beam window one evaluates
the centroid $\boldsymbol\mu_n=\langle\vb K\rangle_n$ and width $\Delta
K_n$ of each relevant kernel (and the complex moments of the kernel pairs
for $n\neq n'$), and bounds the variation of the complete target factors
on that support. The pull-out is controlled when $|\boldsymbol\mu_n|/Q$,
$\Delta K_n/Q$ and their target-response counterparts are small, with
negligible weight near the Coulomb singularity or other unresolved sharp
structure; vanishing first moments give quadratic leading corrections,
otherwise the leading correction is linear. For fixed moderate dressing
and smooth local SVE profiles these conditions reduce to the inequalities
\pref{eq:SVE-validity}.

\bibliography{references}
\end{document}

%% file: figures/fig_geometry.tex
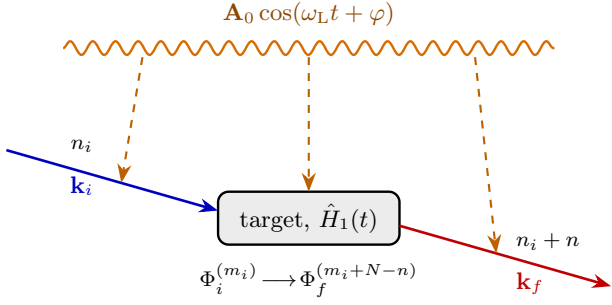
\begin{figure}[t]
\centering
\begin{tikzpicture}[>={Stealth[length=2.5mm]},thick,scale=1.0]
  \draw[orange!85!black,decorate,
        decoration={snake,amplitude=0.9mm,segment length=3mm},
        line width=0.9pt]
    (-3.25,2.25) -- (3.25,2.25);
  \node[above=2pt,font=\small,orange!55!black] at (0,2.35)
    {$\vb A_0\cos(\wL t+\varphi)$};

  \draw[fill=gray!15,rounded corners=4pt,line width=0.9pt]
    (-1.20,-0.35) rectangle (1.20,0.35);
  \node[font=\small] at (0,0) {target, $\hat H_1(t)$};

  \node[font=\footnotesize] at (0,-0.82)
    {$\Phi_i^{(m_i)}\!\longrightarrow\!\Phi_f^{(m_i+N-n)}$};

  \draw[->,blue!75!black,line width=1.0pt]
    (-4.00,0.90) -- (-1.20,0.10);
  \node[font=\footnotesize] at (-3.00,0.96) {$n_i$};
  \node[blue!75!black,font=\small] at (-3.00,0.43) {$\vb k_i$};

  \draw[->,red!75!black,line width=1.0pt]
    (1.20,-0.10) -- (4.00,-0.90);
  \node[font=\footnotesize,anchor=west] at (2.62,-0.31) {$n_i+n$};
  \node[red!75!black,font=\small,anchor=west] at (2.62,-0.86) {$\vb k_f$};

  \draw[->,dashed,orange!75!black,line width=0.8pt]
    (-2.20,2.13) -- (-2.48,0.466);
  \draw[->,dashed,orange!75!black,line width=0.8pt]
    (0,2.13) -- (0,0.35);
  \draw[->,dashed,orange!75!black,line width=0.8pt]
    (2.20,2.13) -- (2.48,-0.466);

\end{tikzpicture}
\caption{Schematic representation of the scattering geometry and
of a term in the inelastic
scattering amplitude~\pref{eq:MN}. The incoming projectile with wave
vector $\vb k_i$ scatters into an outgoing projectile with wave vector
$\vb k_f$, while the target undergoes a transition from an initial
state $i$ to a final state $f$. The dashed arrows indicate that the
same prescribed periodic field dresses the projectile
and the target; they do not
represent photon exchange at three localized events. The corresponding
Floquet--Fourier components carry projectile indices $n_i$ and
$n_i+n$ and target indices $m_i$ and $m_i+N-n$. Accordingly,
the projectile
Fourier index changes by $n$ while $N$ is the externally resolved
Floquet sideband index of the combined projectile--target
transition.
}
\label{fig:geometry}
\end{figure}

%% file: figures/fig_limits.tex
\begin{figure*}[t]
\centering
\begin{tikzpicture}[
  >={Stealth[length=2.2mm]},
  every node/.style={font=\footnotesize},
  block/.style={draw, rounded corners=2pt, align=center,
                inner sep=4pt, minimum width=3.2cm, minimum height=1.5cm},
  pblock/.style={draw, dashed, rounded corners=2pt, align=center,
                inner sep=4pt, minimum width=3.2cm, minimum height=1.5cm},
  centerblock/.style={draw, thick, rounded corners=3pt, align=center,
                inner sep=6pt, fill=gray!12, minimum width=9.4cm,
                minimum height=1.8cm},
  oplbl/.style={font=\scriptsize, fill=white, inner sep=1.5pt, align=center},
]
  \node[centerblock] (center) at (0,5.9)
    {\textbf{Present work: exact first-Born Floquet scattering}, \epref{eq:dsigma-bimomentum}\\[2pt]
     bi-momentum target response
     $\SF^{nn'}(\vb q_1,\vb q_2;\omega)$
     $\;+\;$ exact projectile kernel
     };

  \draw (0,5.0) -- (0,4.6);
  \draw (-6.5,4.6) -- (6.5,4.6);

  \node[pblock] (eik)     at (-6.5,3.2)
    {eikonal--Volkov\\ projectile kernel\\[1pt] \epref{eq:dsigma-inhom}};
  \node[pblock] (homvolk) at (-2.2,3.2)
    {homogeneous Volkov\\ projectile kernel};
  \node[block]  (momdiag) at ( 2.2,3.2)
    {momentum-diagonal\\ target response\\[1pt]
     $(2\pi)^3\delta^{(3)}(\vb q_1\!-\!\vb q_2)\,s_{\mathrm F}^{nn'}$};
  \node[block]  (bethe)   at ( 6.5,3.2)
    {Bethe EELS\\[1pt] $|v(\Qv)|^{2}S(\Qv,\omega)$};

  \draw[->,dashed] (-6.5,4.6) -- node[oplbl,right]{eikonal approx.} (eik.north);
  \draw[->] ( 2.2,4.6) -- node[oplbl,right]{target transl.\\ invariance} (momdiag.north);
  \draw[->] ( 6.5,4.6) -- node[oplbl,left]{$\vb A_0\!\to\!0$} (bethe.north);
  \draw[->,dashed] (eik.east) -- node[oplbl]{homog.\\ field} (homvolk.west);

  \node[block] (hbf)  at (-4.8,0.4)
    {homogeneous\\ Bethe--Floquet};
  \node[block, minimum width=4.4cm] (bfkw) at (-0.1,0.4)
    {{Bunkin--Fedorov} \\[1pt]
     laser-assisted\\ elastic scattering};

  \draw[->,dashed] (homvolk.south) -- (hbf.north east);
  \draw[->] (homvolk.south) -- node[oplbl,right,pos=0.55]{$+$ static elastic $U(\vb r)$}
        (bfkw.north);

  \node[block, minimum width=4.4cm, draw=black, text=black] (coh) at (4.75,0.25)
    {exact coherent mean-density sector\\[1pt]
     $R_M(\vb q_1)R_{M'}^{*}(\vb q_2)\,\delta(\hbar\omega)$, \pref{eq:SF-coherent-connected}\\[2pt]
     fundamental harmonic $\to g_{\rm ind}$, \pref{eq:gind-density}\\[1pt]
     after coherent resummation: local\\
     $P_N(\vb R)=J_N^2(2|g_0+g_{\rm ind}|)$};
  \draw[->] (4.35,4.6) -- node[oplbl, right, pos=0.80]
    {coherent/connected\\split \pref{eq:SF-coherent-connected}} (4.35,0 |- coh.north);
\end{tikzpicture}
\caption{Independent reductions of the general first-Born
Floquet--Fourier cross-section \pref{eq:dsigma-bimomentum}.
Dashed arrows denote projectile-kernel specializations; solid arrows
denote target-side or full-theory limits. Target translational invariance makes the bi-momentum
response momentum diagonal and reduces the double momentum integral to
a single-$\vb q$ convolution. The eikonal--Volkov construction
\pref{eq:dsigma-inhom} is a projectile-kernel specialization; in a
homogeneous field it reduces to the exact Volkov kernel, whose momentum
delta functions yield the homogeneous Bethe--Floquet expression.
Combining that kernel with the projectile-only
approximation to the connected target response gives the
Bessel-weighted equilibrium result $J_N^{2}S_{\mathrm c}(\Qv,\omega_N)$ (not
shown). The right branch starts from the exact coherent mean-density
split \pref{eq:SF-coherent-connected}. Resumming its fundamental oscillatory
component along a straight, nonrecoiling trajectory gives the local PINEM
probabilities $P_N(\vb R)=J_N^2(2|g_0+g_{\rm ind}|)$, with $g_{\rm ind}$ from
\pref{eq:gind-density}. The paraxial angular amplitude follows from
\pref{eq:pinem-angular}; static scattering and any collected reference
enter at amplitude level. Switching off the drive gives field-free Bethe
EELS~\cite{Bethe1930,Egerton2011,Inokuti1971}. Combining the
homogeneous Volkov kernel with a static elastic target gives the
Bunkin--Fedorov result, the first-Born homogeneous-field elastic
sideband formula; its soft-photon Kroll--Watson extension retains the
same sideband structure but replaces the Born cross-section by the full
field-free elastic cross-section and therefore reaches beyond first
Born~\cite{BunkinFedorov1966,KrollWatson1973}. The atomic
Born--Floquet precursor of the many-body case was developed
in Refs.~\cite{Dorr1994,Vucic1995,Joachain2012}.
\label{fig:limits}}
\end{figure*}
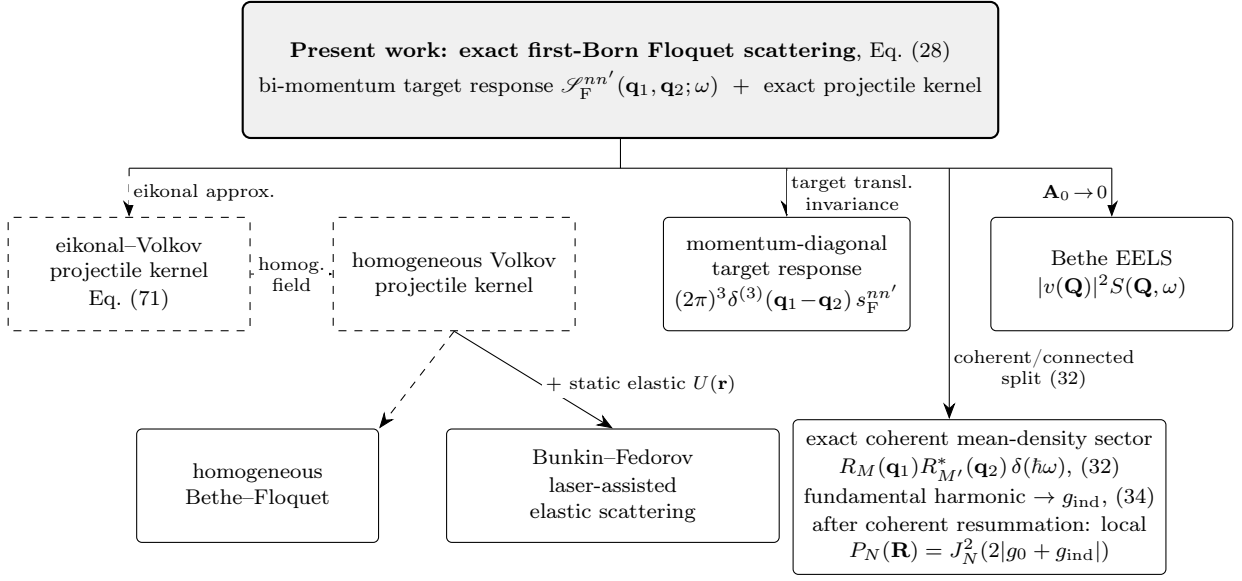

%% file: figures/fig_metal_spectra.tex
\begin{figure*}[t]
\centering
\begin{tikzpicture}
\begin{axis}[
  name=panela,
  width=0.43\linewidth, height=4.0cm,
  ybar, bar width=2.2pt,
  xlabel={$N$}, ylabel={$J_N^{2}(|\Qv\!\cdot\!\alphav|)$},
  xmin=-4.5, xmax=4.5,
  ymin=0, ymax=1.0,
  xtick={-4,-3,-2,-1,0,1,2,3,4},
  enlarge x limits=0.02,
  tick label style={font=\scriptsize},
  label style={font=\footnotesize},
  legend cell align=left,
  legend style={font=\tiny, at={(0.98,0.98)}, anchor=north east,
                draw=none, fill=none, inner sep=1pt, row sep=-2pt},
  title={(a) Drude plasmon comb},
  title style={font=\footnotesize, yshift=-2pt},
]
\addplot[fill=blue!30, draw=blue!60!black]
  coordinates {(-4,0.0000)(-3,0.0000)(-2,0.0009)(-1,0.0587)(0,0.8807)
               (1,0.0587)(2,0.0009)(3,0.0000)(4,0.0000)};
\addlegendentry{$|\Qv\!\cdot\!\alphav|=0.5$}
\addplot[fill=orange!40, draw=orange!70!black]
  coordinates {(-4,0.0001)(-3,0.0037)(-2,0.0539)(-1,0.3113)(0,0.2620)
               (1,0.3113)(2,0.0539)(3,0.0037)(4,0.0001)};
\addlegendentry{$|\Qv\!\cdot\!\alphav|=1.5$}
\addplot[fill=red!40, draw=red!70!black]
  coordinates {(-4,0.0054)(-3,0.0469)(-2,0.1990)(-1,0.2471)(0,0.0023)
               (1,0.2471)(2,0.1990)(3,0.0469)(4,0.0054)};
\addlegendentry{$|\Qv\!\cdot\!\alphav|=2.5$}
\end{axis}
\begin{axis}[
  at={(panela.east)}, anchor=west, xshift=1.7cm,
  width=0.50\linewidth, height=4.0cm,
  xlabel={$\omega_{\rm loss}/\wL$},
  ylabel={$\dd^{2}\sigma/\dd\Omega\,\dd \epsilon$ (arb.~units)},
  xmin=-1, xmax=5,
  ymin=0, ymax=2.2,
  tick label style={font=\scriptsize},
  label style={font=\footnotesize},
  legend cell align=left,
  legend style={font=\tiny, at={(0.98,0.98)}, anchor=north east,
                draw=none, fill=none, inner sep=1pt, row sep=-2pt},
  title={(b) Diffusive Floquet bands ($|\Qv\!\cdot\!\alphav|=1.5$)},
  title style={font=\footnotesize, yshift=-2pt},
]
\addplot[blue!70!black, thick] coordinates {
  (-1.000,0.055)(-0.975,0.786)(-0.950,1.298)(-0.925,1.546)(-0.900,1.607)
  (-0.875,1.568)(-0.850,1.485)(-0.825,1.388)(-0.800,1.291)(-0.775,1.201)
  (-0.750,1.118)(-0.725,1.043)(-0.700,0.977)(-0.675,0.917)(-0.650,0.864)
  (-0.625,0.816)(-0.600,0.772)(-0.575,0.733)(-0.550,0.698)(-0.525,0.665)
  (-0.500,0.636)(-0.475,0.609)(-0.450,0.584)(-0.425,0.561)(-0.400,0.540)
  (-0.375,0.520)(-0.350,0.502)(-0.325,0.485)(-0.300,0.469)(-0.275,0.454)
  (-0.250,0.440)(-0.225,0.427)(-0.200,0.414)(-0.175,0.403)(-0.150,0.392)
  (-0.125,0.381)(-0.100,0.371)(-0.075,0.362)(-0.050,0.353)(-0.025,0.344)
  (0.000,0.336)(0.025,0.945)(0.050,1.369)(0.075,1.572)(0.100,1.617)
  (0.125,1.579)(0.150,1.504)(0.175,1.417)(0.200,1.331)(0.225,1.250)
  (0.250,1.176)(0.275,1.109)(0.300,1.048)(0.325,0.994)(0.350,0.945)
  (0.375,0.901)(0.400,0.861)(0.425,0.825)(0.450,0.791)(0.475,0.761)
  (0.500,0.733)(0.525,0.707)(0.550,0.683)(0.575,0.661)(0.600,0.640)
  (0.625,0.621)(0.650,0.603)(0.675,0.586)(0.700,0.570)(0.725,0.555)
  (0.750,0.541)(0.775,0.528)(0.800,0.515)(0.825,0.503)(0.850,0.492)
  (0.875,0.481)(0.900,0.470)(0.925,0.461)(0.950,0.451)(0.975,0.442)
  (1.000,0.434)(1.025,1.158)(1.050,1.662)(1.075,1.904)(1.100,1.959)
  (1.125,1.914)(1.150,1.825)(1.175,1.723)(1.200,1.621)(1.225,1.525)
  (1.250,1.437)(1.275,1.358)(1.300,1.286)(1.325,1.222)(1.350,1.164)
  (1.375,1.112)(1.400,1.065)(1.425,1.022)(1.450,0.982)(1.475,0.946)
  (1.500,0.913)(1.525,0.882)(1.550,0.854)(1.575,0.828)(1.600,0.803)
  (1.625,0.780)(1.650,0.759)(1.675,0.739)(1.700,0.720)(1.725,0.702)
  (1.750,0.685)(1.775,0.669)(1.800,0.654)(1.825,0.640)(1.850,0.626)
  (1.875,0.613)(1.900,0.601)(1.925,0.589)(1.950,0.578)(1.975,0.567)
  (2.000,0.557)(2.025,0.674)(2.050,0.753)(2.075,0.787)(2.100,0.789)
  (2.125,0.774)(2.150,0.751)(2.175,0.727)(2.200,0.703)(2.225,0.680)
  (2.250,0.658)(2.275,0.639)(2.300,0.621)(2.325,0.604)(2.350,0.589)
  (2.375,0.575)(2.400,0.561)(2.425,0.549)(2.450,0.537)(2.475,0.527)
  (2.500,0.516)(2.525,0.507)(2.550,0.498)(2.575,0.489)(2.600,0.481)
  (2.625,0.473)(2.650,0.465)(2.675,0.458)(2.700,0.451)(2.725,0.444)
  (2.750,0.438)(2.775,0.432)(2.800,0.426)(2.825,0.420)(2.850,0.415)
  (2.875,0.409)(2.900,0.404)(2.925,0.399)(2.950,0.394)(2.975,0.390)
  (3.000,0.385)(3.025,0.389)(3.050,0.391)(3.075,0.390)(3.100,0.387)
  (3.125,0.382)(3.150,0.377)(3.175,0.372)(3.200,0.367)(3.225,0.363)
  (3.250,0.358)(3.275,0.354)(3.300,0.350)(3.325,0.346)(3.350,0.342)
  (3.375,0.338)(3.400,0.335)(3.425,0.331)(3.450,0.328)(3.475,0.325)
  (3.500,0.322)(3.525,0.318)(3.550,0.315)(3.575,0.313)(3.600,0.310)
  (3.625,0.307)(3.650,0.304)(3.675,0.301)(3.700,0.299)(3.725,0.296)
  (3.750,0.294)(3.775,0.291)(3.800,0.289)(3.825,0.287)(3.850,0.284)
  (3.875,0.282)(3.900,0.280)(3.925,0.278)(3.950,0.276)(3.975,0.273)
  (4.000,0.271)(4.025,0.270)(4.050,0.268)(4.075,0.266)(4.100,0.264)
  (4.125,0.262)(4.150,0.260)(4.175,0.258)(4.200,0.257)(4.225,0.255)
  (4.250,0.253)(4.275,0.251)(4.300,0.250)(4.325,0.248)(4.350,0.246)
  (4.375,0.245)(4.400,0.243)(4.425,0.241)(4.450,0.240)(4.475,0.238)
  (4.500,0.237)(4.525,0.235)(4.550,0.234)(4.575,0.232)(4.600,0.231)
  (4.625,0.229)(4.650,0.228)(4.675,0.227)(4.700,0.225)(4.725,0.224)
  (4.750,0.223)(4.775,0.221)(4.800,0.220)(4.825,0.219)(4.850,0.217)
  (4.875,0.216)(4.900,0.215)(4.925,0.214)(4.950,0.212)(4.975,0.211)
  (5.000,0.210)};
\addlegendentry{$\xi_{\rm scr}=D(Q^{2}+\kappa^{2})/\wL=0.1$ (comb resolved)}
\addplot[red!80!black, thick, dashed] coordinates {
  (-1.000,0.028)(-0.975,0.036)(-0.950,0.044)(-0.925,0.052)(-0.900,0.059)
  (-0.875,0.067)(-0.850,0.074)(-0.825,0.081)(-0.800,0.088)(-0.775,0.094)
  (-0.750,0.101)(-0.725,0.107)(-0.700,0.113)(-0.675,0.119)(-0.650,0.124)
  (-0.625,0.129)(-0.600,0.134)(-0.575,0.139)(-0.550,0.143)(-0.525,0.147)
  (-0.500,0.151)(-0.475,0.154)(-0.450,0.157)(-0.425,0.160)(-0.400,0.163)
  (-0.375,0.165)(-0.350,0.167)(-0.325,0.169)(-0.300,0.171)(-0.275,0.173)
  (-0.250,0.174)(-0.225,0.175)(-0.200,0.176)(-0.175,0.177)(-0.150,0.177)
  (-0.125,0.178)(-0.100,0.178)(-0.075,0.178)(-0.050,0.178)(-0.025,0.178)
  (0.000,0.178)(0.025,0.185)(0.050,0.191)(0.075,0.197)(0.100,0.203)
  (0.125,0.209)(0.150,0.214)(0.175,0.220)(0.200,0.225)(0.225,0.230)
  (0.250,0.235)(0.275,0.239)(0.300,0.243)(0.325,0.247)(0.350,0.251)
  (0.375,0.255)(0.400,0.258)(0.425,0.261)(0.450,0.263)(0.475,0.266)
  (0.500,0.268)(0.525,0.270)(0.550,0.272)(0.575,0.273)(0.600,0.275)
  (0.625,0.276)(0.650,0.276)(0.675,0.277)(0.700,0.278)(0.725,0.278)
  (0.750,0.278)(0.775,0.278)(0.800,0.278)(0.825,0.278)(0.850,0.277)
  (0.875,0.277)(0.900,0.276)(0.925,0.275)(0.950,0.274)(0.975,0.274)
  (1.000,0.273)(1.025,0.279)(1.050,0.286)(1.075,0.292)(1.100,0.299)
  (1.125,0.305)(1.150,0.311)(1.175,0.317)(1.200,0.322)(1.225,0.327)
  (1.250,0.333)(1.275,0.337)(1.300,0.342)(1.325,0.346)(1.350,0.350)
  (1.375,0.354)(1.400,0.357)(1.425,0.360)(1.450,0.363)(1.475,0.366)
  (1.500,0.368)(1.525,0.370)(1.550,0.371)(1.575,0.373)(1.600,0.374)
  (1.625,0.375)(1.650,0.376)(1.675,0.376)(1.700,0.376)(1.725,0.377)
  (1.750,0.376)(1.775,0.376)(1.800,0.376)(1.825,0.375)(1.850,0.374)
  (1.875,0.373)(1.900,0.372)(1.925,0.371)(1.950,0.370)(1.975,0.369)
  (2.000,0.367)(2.025,0.367)(2.050,0.367)(2.075,0.366)(2.100,0.366)
  (2.125,0.366)(2.150,0.365)(2.175,0.364)(2.200,0.364)(2.225,0.363)
  (2.250,0.362)(2.275,0.361)(2.300,0.360)(2.325,0.359)(2.350,0.358)
  (2.375,0.357)(2.400,0.355)(2.425,0.354)(2.450,0.353)(2.475,0.351)
  (2.500,0.350)(2.525,0.348)(2.550,0.347)(2.575,0.345)(2.600,0.344)
  (2.625,0.342)(2.650,0.340)(2.675,0.338)(2.700,0.337)(2.725,0.335)
  (2.750,0.333)(2.775,0.331)(2.800,0.329)(2.825,0.328)(2.850,0.326)
  (2.875,0.324)(2.900,0.322)(2.925,0.320)(2.950,0.318)(2.975,0.316)
  (3.000,0.314)(3.025,0.312)(3.050,0.311)(3.075,0.309)(3.100,0.307)
  (3.125,0.305)(3.150,0.303)(3.175,0.302)(3.200,0.300)(3.225,0.298)
  (3.250,0.296)(3.275,0.294)(3.300,0.293)(3.325,0.291)(3.350,0.289)
  (3.375,0.287)(3.400,0.286)(3.425,0.284)(3.450,0.282)(3.475,0.281)
  (3.500,0.279)(3.525,0.277)(3.550,0.276)(3.575,0.274)(3.600,0.272)
  (3.625,0.271)(3.650,0.269)(3.675,0.267)(3.700,0.266)(3.725,0.264)
  (3.750,0.263)(3.775,0.261)(3.800,0.259)(3.825,0.258)(3.850,0.256)
  (3.875,0.255)(3.900,0.253)(3.925,0.252)(3.950,0.250)(3.975,0.249)
  (4.000,0.248)(4.025,0.246)(4.050,0.245)(4.075,0.243)(4.100,0.242)
  (4.125,0.241)(4.150,0.239)(4.175,0.238)(4.200,0.237)(4.225,0.235)
  (4.250,0.234)(4.275,0.233)(4.300,0.231)(4.325,0.230)(4.350,0.229)
  (4.375,0.228)(4.400,0.226)(4.425,0.225)(4.450,0.224)(4.475,0.223)
  (4.500,0.221)(4.525,0.220)(4.550,0.219)(4.575,0.218)(4.600,0.217)
  (4.625,0.216)(4.650,0.215)(4.675,0.213)(4.700,0.212)(4.725,0.211)
  (4.750,0.210)(4.775,0.209)(4.800,0.208)(4.825,0.207)(4.850,0.206)
  (4.875,0.205)(4.900,0.204)(4.925,0.203)(4.950,0.202)(4.975,0.201)
  (5.000,0.200)};
\addlegendentry{$\xi_{\rm scr}=D(Q^{2}+\kappa^{2})/\wL=1.0$ (bands overlap)}
\end{axis}
\end{tikzpicture}
\caption{Cross-sections for the metallic worked examples of
\secref{sec:plasmon}, at representative dimensionless parameters.
(a)~Bessel-weight distribution $J_N^{2}(|\Qv\cdot\alphav|)$
entering the Drude plasmon-loss cross-section
\pref{eq:dsigma-Drude}, shown at three couplings:
$|\Qv\!\cdot\!\alphav|=0.5$ (weak drive, $N=0$ dominates),
$1.5$ (broad $N=\pm1$ comb), and $2.5$ (zero of $J_0^{2}$,
comb dominated by $N=\pm1,\pm2$).
(b)~Doubly differential cross-section \pref{eq:dsigma-diff-sum},
evaluated with the screened diffusive response
\pref{eq:S-diff-screened}, for the diffusive metal at
$|\Qv\!\cdot\!\alphav|=1.5$ and two values of the screened
sideband-overlap parameter
$\xi_{\rm scr}\equiv D(Q^{2}+\kappa^{2})/\wL$.
For $\xi_{\rm scr}=0.1$ the diffusive sidebands are narrow and the
Bessel comb is clearly resolved; for $\xi_{\rm scr}=1.0$ the bands
overlap and the comb washes out into a smooth Bessel-weighted
continuum. The two values are illustrative dimensionless choices
spanning the resolved and overlapping regimes, not estimates for a
specified metal; they are understood to satisfy the additional
diffusive consistency condition $D(Q^{2}+\kappa^{2})\tau\ll1$ stated
in \secref{sec:plasmon-diffusive}.}
\label{fig:metal-spectra}
\end{figure*}
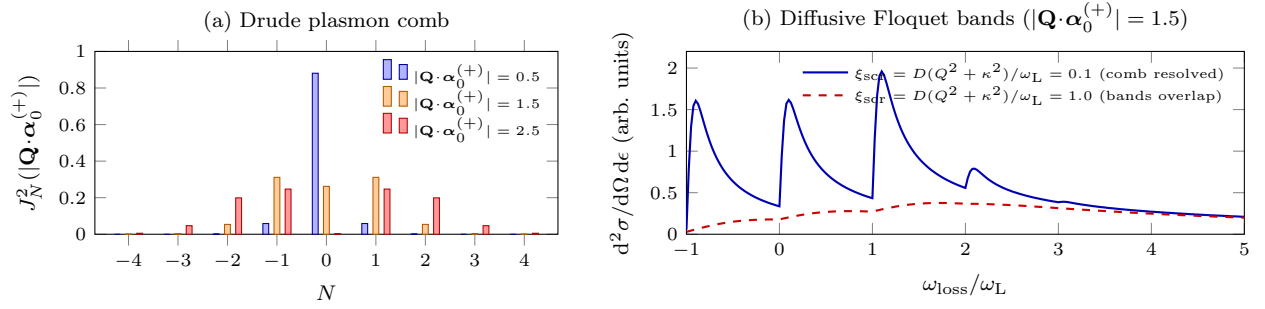

%% file: figures/fig_two_level.tex
\begin{figure}[t]
\centering
\begin{tikzpicture}
\begin{axis}[
  width=1.0\linewidth, height=4.0cm,
  ybar, bar width=4pt,
  xlabel={Sideband index $N$},
  ylabel={$W_N^{\rm lin}=|\mathcal B_N^{\rm lin}|^{2}$},
  xmin=-4.5, xmax=4.5,
  ymin=0, ymax=0.4,
  xtick={-4,-3,-2,-1,0,1,2,3,4},
  enlarge x limits=0.04,
  tick label style={font=\scriptsize},
  label style={font=\footnotesize},
  legend cell align=left,
  legend style={font=\tiny, at={(0.98,0.98)}, anchor=north east,
                draw=none, fill=none, inner sep=1pt, row sep=-2pt},
]
\addplot[pattern=north east lines, pattern color=blue!70!black, draw=blue!60!black]
  coordinates {(-4,0.0004)(-3,0.0064)(-2,0.0589)(-1,0.2657)(0,0.2512)
               (1,0.2905)(2,0.0500)(3,0.0034)(4,0.0001)};
\addlegendentry{$\theta=\pi/8$ (weak mixing, hatched)}
\addplot[pattern=crosshatch, pattern color=red!70!black, draw=red!70!black]
  coordinates {(-4,0.0149)(-3,0.0958)(-2,0.1384)(-1,0.0000)(0,0.1384)
               (1,0.0958)(2,0.0149)(3,0.0010)(4,0.0000)};
\addlegendentry{$\theta=\pi/2$ (maximal mixing, cross-hatched)}
\end{axis}
\end{tikzpicture}
\caption{Connected two-level sideband weights for a linearly polarized drive.
The plotted quantity is $W_N^{\rm lin}=|\mathcal B_N^{\rm lin}|^2$,
Eq.~\pref{eq:bessel-diff}, which multiplies the connected cross-section
\pref{eq:dsigma-2lev}, at $x=\Qv\cdot\boldsymbol\alpha_0=1.5$. The angle
$\theta$ mixes the bare ground and excited components of the Floquet
states \pref{eq:two-level-states} ($c=\cos(\theta/2)$,
$s=\sin(\theta/2)$); it is distinct from the projectile polarization
phase $\theta_{\Qv}^{\rm hom}$. At weak mixing ($\theta=\pi/8$) the
$c^2J_N$ route of the two-path picture in the text dominates. At maximal
mixing ($\theta=\pi/2$) the two routes cancel in the connected $N=-1$
channel, producing the dark line predicted by the linearly polarized RWA
model; other sidebands are redistributed according to
Eq.~\pref{eq:bessel-diff}. The target-elastic coherent contribution,
Eq.~\pref{eq:two-level-coherent-cross-section}, is not included.}
\label{fig:two-level}
\end{figure}
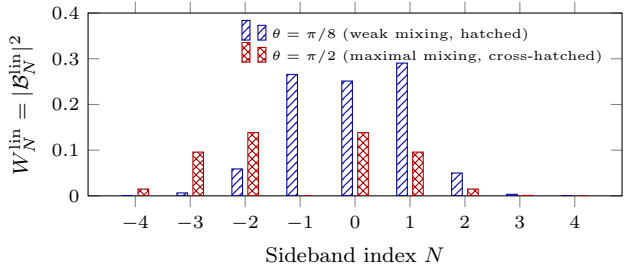

%% file: references.bib
@string{physrev = "Phys. Rev."}

@string{jpca = "J. Phys. Chem. A"}

@string{jpcm = "J. Phys.: Condens. Matter"}

@string{nc = "Nature Commun."}

@string{nat = "Nature"}

@string{njp = "New J. Phys."}

@string{nl = "Nano Lett."}

@string{nphoton = "Nature Photon."}

@string{acsphoton = "ACS Photonics"}

@string{pra =  "Phys. Rev. A"}

@string{prb =  "Phys. Rev. B"}

@string{prl = "Phys. Rev. Lett."}

@string{rmp = "Rev. Mod. Phys."}

@string{ap = "Ann. Phys."}

@string{zp = "Z. Phys."}

@article{Vucic1995,
  title={Inelastic fast-electron--hydrogen-atom collision in a laser field},
  author={Vu{\v{c}}i{\'c}, Svetlana},
  journal=pra,
  volume={51},
  number={6},
  pages={4754},
  year={1995},
  publisher={APS}
}

@book{Joachain2012,
  title={Atoms in intense laser fields},
  author={Joachain, Charles J and Kylstra, Niels J and Potvliege, Robert M},
  year={2012},
  publisher={Cambridge University Press}
}

@article{Dorr1994,
  title="Born-{Floquet} theory of laser-assisted electron-atom collisions",
  author={D{\"o}rr, M and Joachain, CJ and Potvliege, RM and Vu{\v{c}}i{\'c}, S},
  journal=pra,
  volume={49},
  number={6},
  pages={4852},
  year={1994},
  publisher={APS}
}

@article{Bethe1930,
  title={Zur {Theorie} des {Durchgangs} schneller {Korpuskularstrahlen} durch {Materie}},
  author={Bethe, H},
  journal=ap,
  volume={397},
  number={3},
  pages={325--400},
  year={1930}
}

@article{Volkov1935,
  title={{\"U}ber eine {Klasse} von {L{\"o}sungen} der {Diracschen} {Gleichung}},
  author={Volkov, D M},
  journal=zp,
  volume={94},
  pages={250--260},
  year={1935}
}

@book{Egerton2011,
  title={Electron Energy-Loss Spectroscopy in the Electron Microscope},
  author={Egerton, R F},
  edition={3},
  year={2011},
  publisher={Springer}
}

@article{Barwick2009,
  title={Photon-induced near-field electron microscopy},
  author={Barwick, B and Flannigan, D J and Zewail, A H},
  journal=nat,
  volume={462},
  pages={902--906},
  year={2009}
}

@article{BunkinFedorov1966,
  author = {Bunkin, F. V. and Fedorov, M. V.},
  title = {Bremsstrahlung in a strong radiation field},
  journal = {Sov. Phys. JETP},
  volume = {22},
  pages = {844--848},
  year = {1966},
  note = {[Zh. Eksp. Teor. Fiz. 49, 1215--1221 (1965)]}
}

@article{Inokuti1971,
  title={Inelastic Collisions of Fast Charged Particles with Atoms and Molecules---The {Bethe} Theory Revisited},
  author={Inokuti, M},
  journal=rmp,
  volume={43},
  pages={297--347},
  year={1971}
}

@article{KrollWatson1973,
  title={Charged-Particle Scattering in the Presence of a Strong Electromagnetic Wave},
  author={Kroll, N M and Watson, K M},
  journal=pra,
  volume={8},
  pages={804--809},
  year={1973}
}

@article{deAbajo2010,
  title={Multiphoton Absorption and Emission by Interaction of Swift Electrons with Evanescent Light Fields},
  author={Garc{\'\i}a de Abajo, F J and Asenjo-Garcia, A and Kociak, M},
  journal=nl,
  volume={10},
  pages={1859--1863},
  year={2010}
}

@article{Abajo2016,
  title = {Electron diffraction by plasmon waves},
  author = {Garc{\'i}a de Abajo, F. J. and Barwick, B. and Carbone, F.},
  journal = prb,
  volume = {94},
  pages = {041404(R)},
  year = {2016},
  doi = {10.1103/PhysRevB.94.041404}
}

@book{Pines1999,
  title={Elementary Excitations in Solids},
  author={Pines, D},
  year={1999},
  edition={reprint},
  publisher={Westview Press},
  note={Original Benjamin, New York, 1963.}
}

@article{RitchieRitchie1957,
  title={Plasma Losses by Fast Electrons in Thin Films},
  author={Ritchie, R H},
  journal=physrev,
  volume={106},
  pages={874--881},
  year={1957}
}

@incollection{AltshulerAronov1985,
  title={Electron--Electron Interactions in Disordered Systems},
  author={Altshuler, B L and Aronov, A G},
  booktitle={Electron--Electron Interactions in Disordered Systems},
  editor={Efros, A L and Pollak, M},
  series={Modern Problems in Condensed Matter Sciences},
  volume={10},
  pages={1--153},
  year={1985},
  publisher={North-Holland}
}

@book{Akkermans2007,
  title={Mesoscopic Physics of Electrons and Photons},
  author={Akkermans, E and Montambaux, G},
  year={2007},
  publisher={Cambridge University Press}
}

@article{Muller2024,
  title={Inelastic electron-light scattering at dielectric thin films},
  author={M{\"u}ller, N and Vosse, G and Evers, F and Sch{\"a}fer, S},
  journal={arXiv preprint arXiv:2405.11495},
  year={2024}
}

@article{Park2010,
  title={Photon-induced near-field electron microscopy ({PINEM}): theoretical and experimental},
  author={Park, S T and Lin, M and Zewail, A H},
  journal=njp,
  volume={12},
  pages={123028},
  year={2010}
}

@article{ParkZewail2012,
  author  = {Park, Sang Tae and Zewail, Ahmed H.},
  title   = {Relativistic Effects in Photon-Induced Near Field
             Electron Microscopy},
  journal = jpca,
  volume  = {116},
  number  = {46},
  pages   = {11128--11133},
  year    = {2012},
  doi     = {10.1021/jp304534n}
}

@article{ParkZewail2014,
  title={Photon-induced near-field electron microscopy: Mathematical formulation of the relation between the experimental observables and the optically driven charge density of nanoparticles},
  author={Park, S T and Zewail, A H},
  journal=pra,
  volume={89},
  pages={013851},
  year={2014}
}

@article{Gaida2024,
  title={Attosecond electron microscopy by free-electron homodyne detection},
  author={Gaida, J H and Louren{\c{c}}o-Martins, H and Sivis, M and {Garc{\'\i}a de Abajo}, F J and Ropers, C and others},
  journal=nphoton,
  volume={18},
  pages={509--515},
  year={2024}
}

@article{deAbajoRoadmap2025,
  title={Roadmap for Quantum Nanophotonics with Free Electrons},
  author={{Garc{\'\i}a de Abajo}, F J and others},
  journal=acsphoton,
  volume={12},
  pages={4760--4817},
  year={2025}
}

@article{deAbajoRMP2010,
  title={Optical excitations in electron microscopy},
  author={Garc{\'\i}a de Abajo, F J},
  journal=rmp,
  volume={82},
  pages={209--275},
  year={2010}
}

@article{Feist2015,
  title={Quantum coherent optical phase modulation in an ultrafast transmission electron microscope},
  author={Feist, A and Echternkamp, K E and Schauss, J and Yalunin, S V and Sch{\"a}fer, S and Ropers, C},
  journal=nat,
  volume={521},
  pages={200--203},
  year={2015}
}

@article{Vanacore2018,
  title={Attosecond coherent control of free-electron wave functions using semi-infinite light fields},
  author={Vanacore, G M and Madan, I and Berruto, G and Wang, K and Pomarico, E and Lamb, R J and McGrouther, D and Kaminer, I and Barwick, B and Garc{\'\i}a de Abajo, F J and Carbone, F},
  journal=nc,
  volume={9},
  pages={2694},
  year={2018}
}

@article{Kfir2019,
  title={Entanglements of Electrons and Cavity Photons in the Strong-Coupling Regime},
  author={Kfir, O},
  journal=prl,
  volume={123},
  pages={103602},
  year={2019}
}

@book{Watson1944,
  title={A Treatise on the Theory of {Bessel} Functions},
  author={Watson, G N},
  edition={2},
  year={1944},
  publisher={Cambridge University Press}
}

@article{VanHove1954,
  title={Correlations in Space and Time and {Born} Approximation Scattering in Systems of Interacting Particles},
  author={Van Hove, L},
  journal=physrev,
  volume={95},
  pages={249--262},
  year={1954}
}

@book{Squires2012,
  title={Introduction to the Theory of Thermal Neutron Scattering},
  author={Squires, G L},
  edition={3},
  year={2012},
  publisher={Cambridge University Press}
}

@article{Shirley1965,
  title={Solution of the {Schr{\"o}dinger} Equation with a {Hamiltonian} Periodic in Time},
  author={Shirley, Jon H},
  journal=physrev,
  volume={138},
  pages={B979--B987},
  year={1965}
}

@article{Sambe1973,
  title={Steady States and Quasienergies of a Quantum-Mechanical System in an Oscillating Field},
  author={Sambe, Hideo},
  journal=pra,
  volume={7},
  pages={2203--2213},
  year={1973}
}

@article{Peskin1994,
  title={Time-Independent Scattering Theory for Time-Periodic {Hamiltonians}: Formulation and Complex-Scaling Calculations of Above-Threshold-Ionization Spectra},
  author={Peskin, U and Moiseyev, N},
  journal=pra,
  volume={49},
  pages={3712--3725},
  year={1994}
}

@article{ArqueLopez2022,
  title={Atomic {Floquet} Physics Revealed by Free Electrons},
  author={Arqu{\'e} L{\'o}pez, Eduardo and Di Giulio, Valerio and {Garc{\'\i}a de Abajo}, F J},
  journal={Phys. Rev. Research},
  volume={4},
  pages={013241},
  year={2022}
}

@misc{DLMF,
  title={{NIST} Digital Library of Mathematical Functions},
  author={Olver, F W J and Daalhuis, A B Olde and Lozier, D W and Schneider, B I and Boisvert, R F and Clark, C W and Miller, B R and Saunders, B V and Cohl, H S and McClain  eds., M A},
  howpublished={\texttt{http://dlmf.nist.gov/}},
  note={Release 1.2.4 of 2025-03-15}
}

@book{Taylor2006,
  title={Scattering theory: the quantum theory of nonrelativistic collisions},
  author={Taylor, John R},
  year={2006},
  publisher={Dover Publications}
}

@article{Bilitewski2015,
  title = "Scattering theory for {Floquet}-{Bloch} states",
  author = {Bilitewski, Thomas and Cooper, Nigel R.},
  journal = pra,
  volume = {91},
  issue = {3},
  pages = {033601},
  numpages = {20},
  year = {2015},
  month = {Mar},
  publisher = {American Physical Society},
  doi = {10.1103/PhysRevA.91.033601},
  url = {https://link.aps.org/doi/10.1103/PhysRevA.91.033601}
}

@Article{Meuret2024,
author="Meuret, Sophie
and Louren{\c{c}}o-Martins, Hugo
and Weber, S{\'e}bastien
and Houdellier, Florent
and Arbouet, Arnaud",
title="Photon-Induced Near-Field Electron Microscopy of Nanostructured Metallic Films and Membranes",
journal=acsphoton,
year="2024",
month="Mar",
day="20",
publisher="American Chemical Society",
volume="11",
number="3",
pages="977--984",
doi="10.1021/acsphotonics.3c01223",
url="https://doi.org/10.1021/acsphotonics.3c01223"
}

@article{Asmar2022,
doi = {10.1088/1361-648X/ac709d},
url = {https://doi.org/10.1088/1361-648X/ac709d},
year = {2022},
month = {jun},
publisher = {IOP Publishing},
volume = {34},
number = {31},
pages = {315602},
author = {Asmar, Mahmoud M and Tse, Wang-Kong},
title = {Impurity screening and Friedel oscillations in {Floquet}-driven two-dimensional metals},
journal = jpcm,
}

@Article{Neufeld2019,
author="Neufeld, Ofer
and Podolsky, Daniel
and Cohen, Oren",
title="Floquet group theory and its application to selection rules in harmonic generation",
journal=nc,
year="2019",
month="Jan",
day="24",
volume="10",
number="1",
pages="405",
issn="2041-1723",
doi="10.1038/s41467-018-07935-y",
url="https://doi.org/10.1038/s41467-018-07935-y"
}

@article{Kumar2020,
  title = {Linear response theory and optical conductivity of {Floquet} topological insulators},
  author = {Kumar, Abhishek and Rodriguez-Vega, M. and Pereg-Barnea, T. and Seradjeh, B.},
  journal = prb,
  volume = {101},
  pages = {174314},
  year = {2020},
  doi = {10.1103/PhysRevB.101.174314}
}

@book{Peng2004,
  title={High energy electron diffraction and microscopy},
  author={Peng, Lian-Mao and Dudarev, Sergiej Leonidovi{\v{c}} and Whelan, Michael J},
  volume={61},
  year={2004},
  publisher={Oxford University Press}
}

@misc{fn:rel-catalogue,
  note={It is useful to distinguish three
relativistic ingredients that enter at TEM energies.
(1)~\emph{Kinematics}: the relation between acceleration voltage,
projectile momentum, scattering angle, and energy loss.
(2)~Projectile \emph{dressing}: replacement of the scalar
eikonal--Volkov phase by its Dirac--Volkov spinor extension, with
spin, recoil and radiation-reaction corrections. (3)~\emph{Interaction
kernel}: replacement of the instantaneous scalar Coulomb potential by
a retarded electromagnetic propagator coupling to both charge and
current response. For scalar PINEM only~(1) is typically relevant;
quantitative EELS at $100$--$300\,\mathrm{keV}$ generally requires~(1)
and~(3) and in some settings elements 
of~(2)\cite{Egerton2011,Inokuti1971,Volkov1935,deAbajoRMP2010}.}
}
